\documentclass[aps,prd,reprint,superscriptaddress,nofootinbib,amsfonts,amssymb,amsmath]{revtex4-1}

\usepackage{dsfont}
\usepackage{graphicx}
\usepackage[colorlinks]{hyperref}
\usepackage{xcolor}
\usepackage{bbm}

\hypersetup{
colorlinks=true,
citecolor=blue,
citebordercolor=red,
linktoc=all,
linkcolor=blue
}

\allowdisplaybreaks

 \graphicspath{{./figures/}}
\def\CC{{\mathcal C}}

\def\CN{{\mathcal N}}

\def\llangle{\left\langle}
\def\rrangle{\right\rangle}
\def\0#1#2{\frac{#1}{#2}}

\def\E{\text{e}}
\DeclareMathOperator{\tr}{Tr}
\newcommand{\EH}{\text{EH}}
\newcommand{\SM}{\text{SM}}
\newcommand{\TT}{\text{tt}}
\newcommand{\Tr}{\text{tr}}
\newcommand{\Flow}{\text{Flow}}
\newcommand{\eff}{\text{eff}}
\newcommand{\start}{\text{start}}
\def\NI{{\rm NI}}
\def\mNI{{\rm mNI}}

\DeclareMathOperator*{\sumint}{%
\mathchoice%
  {\ooalign{\scalebox{.85}{$\displaystyle\sum$}\cr\hidewidth$\displaystyle\int$\hidewidth\cr}}
  {\ooalign{\raisebox{.22\height}{\scalebox{.6}{$\textstyle\sum$}}\cr\hidewidth$\textstyle\int$\hidewidth\cr}}
  {\ooalign{\raisebox{.2\height}{\scalebox{.5}{$\scriptstyle\sum$}}\cr$\scriptstyle\int$\cr}}
  {\ooalign{\raisebox{.2\height}{\scalebox{.5}{$\scriptstyle\sum$}}\cr$\scriptstyle\int$\cr}}
}

\begin{document}

\title{Curvature dependence of quantum gravity}

% LIST_OF_AUTHORS.TEX         
%
\author{Nicolai~Christiansen}
%\email[]{nic.christiansen@googlemail.com}
\affiliation{Institut f\"ur Theoretische Physik, Universit\"at Heidelberg,
Philosophenweg 16, 69120 Heidelberg, Germany}
\author{Kevin~Falls}
%\email[]{kevin.g.falls@gmail.com}
\affiliation{Institut f\"ur Theoretische Physik, Universit\"at Heidelberg,
Philosophenweg 16, 69120 Heidelberg, Germany}
\author{Jan~M.~Pawlowski}
%\email[]{J.Pawlowski@thphys.uni-heidelberg.de}
\affiliation{Institut f\"ur Theoretische Physik, Universit\"at Heidelberg,
Philosophenweg 16, 69120 Heidelberg, Germany}
\affiliation{ExtreMe Matter Institute EMMI, GSI Helmholtzzentrum f\"ur
Schwerionenforschung mbH, Planckstr.\ 1, 64291 Darmstadt, Germany}
\author{Manuel~Reichert}
%\email[]{reichert@thphys.uni-heidelberg.de}
\affiliation{Institut f\"ur Theoretische Physik, Universit\"at Heidelberg,
Philosophenweg 16, 69120 Heidelberg, Germany}
%
%end                                                                          

\begin{abstract}
  We investigate the phase diagram of quantum gravity with a vertex
  expansion about constantly-curved backgrounds. The
  graviton two- and three-point function are evaluated with a spectral sum on a
  sphere. We obtain, for the first time, curvature-dependent UV fixed point functions of
  the dynamical fluctuation couplings $g^*(R)$, $\mu^*(R)$, and $\lambda_3^*(R)$,
  and the background $f(R)$-potential.

  Based on these fixed-point functions we compute solutions to the
  quantum and the background equation of motion with and without
  Standard Model matter. We have checked that the
  solutions are robust against changes of the truncation.
\end{abstract}

\maketitle

\section{Introduction}
Modern theoretical physics is built upon two pillars, namely quantum field 
theory and general relativity. Theories of quantum gravity aim at the 
unification of gravity with quantum dynamics. 
A candidate for a quantum theory of gravity is the asymptotic safety scenario, 
which goes back to Weinberg's idea in 1976 \cite{Weinberg:1980gg}. Its 
construction is based on a non-trivial ultraviolet (UV) fixed point in the 
renormalisation group flow. The fixed point of asymptotic safety implies 
coupling constants that are finite at arbitrarily high energy scales, while 
they depend only on a finite number of free parameters. Hence, an 
asymptotically safe quantum field theory does not necessarily have a scale of 
maximal validity and thus can potentially describe physical interactions 
at the most fundamental level.
The possibility of an interacting UV fixed point in quantum gravity 
attracted increasing attention over the last two decades. Beginning with the 
pioneering work by Reuter \cite{Reuter:1996cp}, good evidence for its 
existence was found in pure gravity setups as well as in systems with 
gravity coupled to gauge and matter fields  
\cite{Christiansen:2012rx,Codello:2013fpa,Christiansen:2014raa,Christiansen:2015rva,
  Meibohm:2015twa,Meibohm:2016mkp,Henz:2016aoh,Christiansen:2016sjn,Denz:2016qks,
  Knorr:2017fus,Christiansen:2017cxa,Knorr:2017mhu,
  Donkin:2012ud,Morris:2016spn,Percacci:2016arh,
  Manrique:2009uh,Manrique:2010mq,Manrique:2010am,Becker:2014qya,
  Falkenberg:1996bq,Reuter:2001ag,Lauscher:2001ya,Lauscher:2002sq,Litim:2003vp,
  Fischer:2006fz,Codello:2006in,Machado:2007ea,Codello:2008vh,Eichhorn:2009ah,
  Benedetti:2009rx,Eichhorn:2010tb,Groh:2010ta,Manrique:2011jc,Benedetti:2012dx,
  Dietz:2012ic,Falls:2013bv,Falls:2014tra,Falls:2015qga,Eichhorn:2015bna,Falls:2015cta,
  Demmel:2015oqa,Gies:2015tca,Gies:2016con,Biemans:2016rvp,Falls:2017cze,Hamada:2017rvn,Gonzalez-Martin:2017gza,
  Becker:2017tcx,Dou:1997fg,Percacci:2002ie,Folkerts:2011jz,Harst:2011zx,
  Eichhorn:2011pc,Eichhorn:2012va,Dona:2012am,Henz:2013oxa,Dona:2013qba,
  Percacci:2015wwa,Oda:2015sma,Dona:2015tnf,Eichhorn:2016esv,Eichhorn:2016vvy,
  Christiansen:2017gtg,Eichhorn:2017eht,Biemans:2017zca,
  Eichhorn:2017ylw,Eichhorn:2017lry,Eichhorn:2017sok}.
For reviews see 
\cite{Niedermaier:2006wt,Percacci:2007sz,Litim:2011cp,
  Reuter:2012id,Bonanno:2017pkg,Eichhorn:2017egq}.
  
Most studies on asymptotically safe quantum gravity are based on the
functional renormalisation group (FRG), \cite{Wetterich:1992yh} 
and \cite{Ellwanger:1993mw,Morris:1993qb}. In its modern form as a flow
equation for the effective action $\Gamma[\phi]$ of the theory, it
constitutes a powerful method for non-perturbative calculations in
continuum quantum field theory. Here $\phi$ is a super-field that
comprises all fields in the theory. This formulation, as all
formulations based on metric correlation functions, demands the
introduction of a background metric $\bar g_{\mu\nu}$ and a
corresponding fluctuation field $h_{\mu\nu}$. Inevitably, correlation
functions, as well as the effective action, depend separately on these
fields. Note, however, that it is the correlation functions of the
fluctuation field that carry the dynamics of the system. Indeed, the
flow equation for the effective action is directly proportional to the
two-point function (propagator) of the fluctuation field in a generic
background. Phrased differently, the solution of the flow equation
requires the knowledge of two-point and higher correlation functions
of the fluctuation field. This already entails that the correlation
functions of the background field and mixed correlations of background
and fluctuation can only be constructed based on the pure
fluctuation field correlations. More details on these important
relations and a brief overview of the current state are provided in
\autoref{sec:Motivation}.

Our setup is detailed in \autoref{sec:FRG} and allows for the
computation and the distinction of the background and quantum equation
of motion (EoM).  We argue that these equations have a common solution
at a vanishing infrared (IR) FRG cutoff scale $k=0$ due to background
independence. In turn, the solutions to the background and quantum EoM do
not agree at a finite cutoff scale $k\neq 0$, which signals the loss
of background independence in the presence of the FRG-regulator. This
is also seen in our explicit computations at the UV fixed
point. We further argue that the quantum EoM, and
\emph{not} the background EoM, should be used to
determine the self-consistent background at finite $k$.

Solutions to the background EoM appear as a minimum in
the background potential $f(R)/R^2$, which we compute for the
first time from the dynamical background-dependent fluctuation
couplings without a background field approximation. In the present
work, we compute the UV fixed point background potential
$f^*(R)$. Interestingly, in the pure quantum gravity setting, we do not
find a solution to the background EoM, while a solution
appears at small positive curvature for Standard Model (SM) matter
content. The quantum EoM, on the other hand, has a
solution also in the pure quantum gravity setting.

This work is organised as follows: In \autoref{sec:Motivation} we
discuss the importance of background independence and its
manifestation in the current framework. This includes a brief overview
and description of the results obtained in the literature. In \autoref{sec:FRG}
we introduce the FRG with a particular focus on the background and quantum 
EoM and the Nielsen identity that relates them. We furthermore introduce the vertex
expansion used in this work. In \autoref{sec:curvature} we construct
an approximate momentum space on spherical backgrounds. This
allows us to use previously developed techniques that were based on
running correlation functions in momentum space. In
\autoref{sec:results} we present our results, which include the
non-trivial UV fixed point functions for the dynamical
couplings as well as a detailed discussion of the background and
quantum EoM. In \autoref{sec:summary} we summarise our
results. The technical details are specified in the appendices.

\section{Background independence in quantum gravity}
\label{sec:Motivation}
Most applications of the FRG to quantum gravity to date do not resolve
the difference between background and fluctuation field and employ the
background field approximation. There only one metric
$g_{\mu\nu}=\bar g_{\mu\nu}+h_{\mu\nu}$ is used in the effective
action. However, the non-trivial interplay of the metric fluctuations
with the background plays a decisive r\^ole for background
independence of the theory. These non-trivial relations are governed
by non-trivial split-Ward or Nielsen identities (NIs), see e.g.\
\cite{Litim:2002ce,Litim:2002hj,Pawlowski:2003sk,Pawlowski:2005xe,
  Donkin:2012ud,Safari:2015dva,Bridle:2013sra,Dietz:2015owa,Morris:2016spn,
  Percacci:2016arh,Safari:2016gtj,Nieto:2017ddk} for formal progress
and applications in scalar theories, gauge theories and
gravity. Accordingly, the background field approximation violates the
NIs, which leads to the seemingly contradictory situation that it is
at odds with background independence even though it only features one
metric. In the past decade, quite some progress has been made in
overcoming the background field approximation, see
\cite{Pawlowski:2003sk,Pawlowski:2005xe,Manrique:2009uh,
  Manrique:2010mq,Manrique:2010am,Donkin:2012ud,Christiansen:2012rx,
  Codello:2013fpa,Christiansen:2014raa,Becker:2014qya,Christiansen:2015rva,
  Meibohm:2015twa,Meibohm:2016mkp,Henz:2016aoh,Dietz:2015owa,Safari:2015dva,
  Morris:2016spn,Percacci:2016arh,Christiansen:2016sjn,Denz:2016qks,Knorr:2017fus,Christiansen:2017cxa,
  Nieto:2017ddk,Knorr:2017mhu,Litim:2002ce,Litim:2002hj,Bridle:2013sra,
  Safari:2016gtj}.

\subsection{Approaches to fluctuation and background correlation functions}
All these works
should be seen in the context of gaining background independence and
physical diffeomorphism invariance in asymptotically safe gravity.
Here we briefly summarise the state of the art within the different
approaches. \\[-2ex]

{\it (1)} One approach utilises the fact that the NIs relate background metric
correlations to fluctuation ones. This leaves us with a system of one
type of correlations and it is possible to solve the system of flow
equations for fluctuation correlation functions either directly or
implicitly. This strategy has been set up and pursued in
\cite{Litim:2002ce,Litim:2002hj,Pawlowski:2003sk,Pawlowski:2005xe,
  Donkin:2012ud,Safari:2015dva,Bridle:2013sra,Dietz:2015owa,Morris:2016spn,
  Percacci:2016arh,Safari:2016gtj,Nieto:2017ddk} for generic theories
within the background field approach. At present, applications in
gravity still utilise the background field approximation beyond either the
first-order, or the second-order in the fluctuation field \cite{Donkin:2012ud}.
Such a closure of the flow equation with the
background field approximation is mandatory and all approaches aim at
introducing this approximation on a high order of the fluctuation
field.  Note in this context that it is only the second and higher
order $n$-point functions of the fluctuation field that drive the flow. \\[-2ex]

{\it (2a)} A second approach utilises the fact that the dynamics of the system is
carried by the correlation functions of the fluctuation field.
This is also reflected by the fact that the system of
flow equations for the fluctuation correlations is
closed. Consequently one may solve these flows for a specific
background metric that facilitates the computation, e.g.\ the flat
background. Then, background correlations are computed within an
expansion or extension about the flat background to access
the physical background that solves the quantum EoM.
This strategy has been set up and pursued in
\cite{Christiansen:2012rx,Codello:2013fpa,Christiansen:2014raa,
  Christiansen:2015rva,Meibohm:2015twa,Meibohm:2016mkp,
  Christiansen:2016sjn,Denz:2016qks,Knorr:2017fus,
  Christiansen:2017cxa,Knorr:2017mhu} for gravity, also guided by
successful applications in non-Abelian gauge theories, see e.g.\
\cite{Braun:2007bx,Braun:2009gm,Braun:2010cy,Fister:2013bh,Reinosa:2014ooa}. 
At present, fluctuation correlations up to the four-point function
have been included \cite{Denz:2016qks}, as well as a full fluctuation
effective potential \cite{Knorr:2017mhu}.  First results in a Taylor
expansion of the background about a flat one have been presented in
\cite{Knorr:2017fus}.\\[-2ex]

{\it (2b)} A third approach avoids the latter step of extending the results to 
physical backgrounds by computing instantly the flow equations for the
fluctuation correlation functions for general backgrounds. This has
been investigated in
\cite{Manrique:2009uh,Manrique:2010mq,Manrique:2010am,Becker:2014qya}.
As in the other approaches, the background field approximation has been used
for higher correlation functions. At present, this holds for all
correlation functions beyond the one-point function of the fluctuation
field.

\section{General framework}
\label{sec:genFrame}
In the present work, we develop an approach in the class {\it (2b)}. 
The present work does a qualitative step
towards background independence and diffeomorphism invariance in
asymptotically safe gravity by computing fluctuation correlation
functions up to the three-point function as well as the full
$f(R)$-potential of the background field. As already mentioned in the
introduction, we compute the fixed point potential $f^*(R)$
for $k\to\infty$ but the present approach also allows for its
computation in the physical limit $k\to0$. This potential certainly
has interesting applications in cosmology. The interplay of
asymptotically safe gravity and cosmology is investigated in e.g.~
\cite{Bonanno:2001xi,Bonanno:2001hi,Bentivegna:2003rr,Reuter:2005kb,
  Bonanno:2007wg,Weinberg:2009wa,Bonanno:2009nj,Bonanno:2010mk,Koch:2010nn,
  Casadio:2010fw,Contillo:2010ju,Bonanno:2010bt,Hindmarsh:2011hx,Cai:2011kd,
  Bonanno:2012jy,Hindmarsh:2012rc,Bonanno:2013dja,Copeland:2013vva,
  Becker:2014jua,Saltas:2015vsc,Nielsen:2015una,Bonanno:2015fga,Falls:2016wsa,Bonanno:2017pkg},
and we hope to add to this in the near future.

The present approach is built on the vertex expansion setup to quantum
gravity put forward in
\cite{Christiansen:2012rx,Codello:2013fpa,Christiansen:2014raa,
  Christiansen:2015rva,Meibohm:2015twa,Meibohm:2016mkp,Christiansen:2016sjn,
  Denz:2016qks,Knorr:2017fus,Christiansen:2017cxa,Knorr:2017mhu}.
However, instead of expanding about the flat background, we consider
for the first time coupling constants of the dynamical graviton field
as arbitrary functions of the background curvature.  We restrict
ourselves to spherical backgrounds.  A key point for this is the
construction of an approximate momentum space, which allows us to
utilise the previously developed techniques of running metric
correlators in momentum space. With the resulting curvature-dependent
dynamical couplings we find viable UV fixed-point functions
for all curvatures of the spherical background considered.
Interestingly these fixed-point functions of the effective couplings
are almost curvature independent: the couplings try to counterbalance
the explicit curvature dependence and thus try to keep the fixed point
curvature independent.  The fixed point functions provide further
evidence in favour of the asymptotic safety scenario.

\subsection{FRG and Nielsen identities for gravity}
\label{sec:FRG}
To compute correlation functions in quantum gravity, we
utilise the FRG approach to gravity \cite{Reuter:1996cp}.  In this
approach, the functional integral involves a momentum dependent mass
function $R_k$, which acts as an IR regulator suppressing
momenta $p^2 \lesssim k^2$ relative to the cutoff scale $k$.  This
leads to a scale-dependent effective action $\Gamma_k[\bar g,\phi]$,
which includes contributions from high momentum fluctuations.  Here
the dynamical metric
$g_{\mu\nu}= \bar g_{\mu\nu} + \sqrt{Z_h G_N} h_{\mu\nu}$ is expanded
around a non-dynamical background metric $\bar g$ with the
fluctuations $h$.  The fluctuation field is rescaled with Newton's
coupling such that it has the standard mass-dimension one of a bosonic
field.  In this work, we utilise a linear metric split and we restrict
$\bar g$ to spherical backgrounds.  Combined with ghost fields
$c, \bar c$ we denote the fluctuation super-field $\phi=(h,c,\bar c)$.
The scale-dependence of $\Gamma_k$ is then dictated by the flow
equation \cite{Wetterich:1992yh, Ellwanger:1993mw,Morris:1993qb},
\begin{align}
  \partial_t \Gamma_k &= \frac{1}{2} \tr\big[ G_{hh,k}\, \partial_t R_{h,k}\big]  
  - \tr\big[ G_{\bar c c,k}\,\partial_t R_{c,k}\big]\,, 
\label{eq:gen_flow_eq}
\end{align}
with the graviton and ghost regulators $R_{h,k}$ and $R_{c,k}$
respectively. The regulator terms are diagonal (symplectic) in field
space, hence the diagonal graviton and (symplectic) ghost propagators,
$G_{hh,k}$ and $G_{\bar c c,k}$, read
\begin{align}
  \label{eq:prop} 
  G_k= \left(\Gamma_k^{(0,2)}+R_{k}\right)^{-1}\,,
\end{align}
with the general
one-particle irreducible correlation functions given as derivatives of
the effective action,
\begin{align}
  \label{eq:Gnm}
  \Gamma^{(n,m)}[\bar g,h]=
  \0{\delta\Gamma[\bar g,h]}{\delta\bar g^n\delta h^m}\,.
\end{align}
In \eqref{eq:gen_flow_eq} we have introduced the derivative 
with respect to the RG time $t = \log k/k_{\text{in}}$ where
 $k_{\text{in}}$ is a reference scale, usually taken to be the
initial scale. The trace implies integrals over continuous and sums
over discrete indices.

An important issue in quantum gravity is the background independence
of physical observables. They are expectation values of diffeomorphism
invariant operators, and hence do not depend on the gauge
fixing. Examples for such observables are correlations of the
curvature scalar. Another relevant example is the free energy
of the theory, $-\log Z[\bar g, J=0]$, with
$\delta Z[\bar g, J=0]/\delta \bar g=0$. These observables cannot
depend on the choice of the background metric, which only enters via
the gauge fixing. The latter fact is encoded in the NI 
for the effective action: The difference between
background derivatives and fluctuation derivatives is proportional to
derivatives of the gauge fixing sector,
\begin{align}\nonumber 
  \NI=&\,\0{\delta\Gamma}{\delta \bar g_{\mu\nu}} -\0{\delta\Gamma}{\delta h_{\mu\nu}} \\[1ex] 
  &\, -\llangle \left[\0{\delta}{\delta \bar g_{\mu\nu}}-\0{\delta}{\delta \hat h_{\mu\nu}} \right] 
  ( S_{\text{gf}}+S_{\text{gh}} )\rrangle=0\,,
  \label{eq:Nielsen}
\end{align} 
where $S_{\text{gf}}$ is the gauge fixing term and
$S_{\text{gh}} $ is the corresponding ghost term, and
$h_{\mu\nu}=\langle \hat h_{\mu\nu}\rangle$. Note that
\eqref{eq:Nielsen} is nothing but the Dyson-Schwinger equation for the
difference of derivatives w.r.t.\ $\bar g$ and $h$. For the fully
diffeomorphism-invariant Vilkovisky-deWitt or geometrical effective
action the relation \eqref{eq:Nielsen} is even more concise: the split
is not linear and we have $ g = \bar g+f(\bar g,h)$, where
$f(\bar g,h)=\sqrt{G_N} h+O(h^2)$ depends on the Vilkovisky
connection. The NI then reads
\begin{align}
  \label{eq:geoNielsen} 
  \NI_{\text{geo}}=\0{\delta\Gamma_{\text{geo}}}{\delta 
  \bar g_{\mu\nu}} -\CC(\bar g,h) 
  \0{\delta\Gamma_{\text{geo}}}{\delta h_{\mu\nu}}  = 0\,, 
\end{align}
where $\CC(\bar g,h)$ is the expectation value of the (covariant)
derivative of $h(\bar g,g)$, for a discussion in the present FRG
setting see
\cite{Pawlowski:2003sk,Pawlowski:2005xe,Donkin:2012ud,Safari:2015dva}.

The NIs, \eqref{eq:Nielsen} and \eqref{eq:geoNielsen}, entail that in
both cases the effective action is not a function of $g=\bar g + h$ or
$g=\bar g + f(\bar g, h)$ respectively. This property holds for
general splits and prevents the simple expansion of the effective
action in terms of diffeomorphism invariants. Apart from this
disappointing consequence of the NIs, it also entails good news: the
effective action only depends on one field as background and
fluctuation derivatives are connected.

An important property that follows from background
independence is the fact that a solution of the background 
equation of motion (EoM)
\begin{align}
 \label{eq:backgr-eom}
  \0{\delta \Gamma[\bar g, h]}{\delta \bar g_{\mu\nu}}
  \bigg|_{\bar g = \bar g_{\rm eom},h=0} = 0\,,
\end{align}
is also one of the quantum EoM, 
\begin{align}
 \label{eq:quantum-eom}
  \0{\delta \Gamma[\bar g, h]}{\delta  h_{\mu\nu}}
  \bigg|_{\bar g = \bar g_{\rm eom},h=0} = 0\,.
\end{align}
see, e.g., \cite{Reinosa:2014ooa} for a discussion of this in
Yang-Mills theories. In \eqref{eq:backgr-eom} and
\eqref{eq:quantum-eom} we have already taken the standard choice $h=0$
but the statement hold for general combinations
$\bar g_\text{EoM}(h)$ that solves either of the equations.
The concise form \eqref{eq:geoNielsen} for the geometrical effective
action makes it apparent that a solution of either EoM,
\eqref{eq:backgr-eom} or \eqref{eq:quantum-eom}, also entails a
solution of the other one. Note that at $h=0$ we have
$\CC(\bar g,0)=\mathds{1}$.

Even though less apparent, the same holds true for the effective
action in the linear split: to that end we solve the quantum EoM
\eqref{eq:quantum-eom} as an equation for $\bar g_{\text{eom}}(h)$. As
the current $J$ in the generating functional simply is
$J=\delta\Gamma/\delta h$, the quantum EoM implies the vanishing of
$J$ and the effective action is given by
$\Gamma[\bar g_{\text{eom}}(0),0]=-\log Z[\bar g, J=0]$, the free
energy.  However, we have already discussed that $\log Z[\bar g,0]$ is
background-independent and it follows that \eqref{eq:backgr-eom}
holds.

The above properties and relations are a cornerstone of the background
formalism as they encode background independence of observables. The
NIs also link background diffeomorphism invariance to
the Slavnov-Taylor identities (STIs) that hold for diffeomorphism
transformations of the fluctuation field: the quantum deformation of
classical diffeomorphism symmetry is either encoded in the expectation
value of the gauge fixing sector or in the expectation value $\CC(\bar g,h)$.

At finite $k$, the regulator term introduces a genuine dependence on
the background field. Then $\log Z_k[\bar g, 0]$ is not background
independent. Consequently the STIs turns into modified STIs (mSTIs)
and the NIs turn into modified NIs (mNIs). For the linear split, the
mNI reads
\begin{align}
\mNI = \NI-  \012 \text{Tr}\left[\01{\sqrt{\bar g}} \0{\delta \sqrt{\bar g} 
  R_k[\bar g]}{\delta \bar g_{\mu\nu}} G_k  \right]=0 \,,
 \label{eq:Nielsen-R-term}
\end{align}
see \cite{Litim:2002ce,Litim:2002hj} for details 
and \cite{Eichhorn:2018akn} for an application
to quantum gravity. Importantly the right-hand side of
\eqref{eq:Nielsen-R-term} signals the loss of background
independence. It is proportional to the regulator and vanishes for
$k\to 0$ where background independence is restored. A similar
violation of background independence linear in the regulator is
present in the geometrical approach, see
\cite{Pawlowski:2003sk,Pawlowski:2005xe,Donkin:2012ud,Safari:2015dva}. 

In summary, this leaves us with non-equivalent solutions to the
EoMs in the presence of the regulator: a solution of
the quantum EoM \eqref{eq:quantum-eom} does not solve the
background EoM \eqref{eq:backgr-eom}. However, typically the
asymptotically safe UV regime of quantum gravity is accessed in the
limit $k\to\infty$ as this already encodes the important scaling
information in this regime. In the present work, we also follow this
strategy and hence we have to deal with different solutions of
background and quantum EoMs, if they exist at
all. Note that the right-hand side of the mNI is simply the
expectation value of the background derivative of the regulator
term. Accordingly, it is the background EoM that is deformed
directly by the presence of the regulator while the quantum EoM
feels its influence only indirectly. Therefore it is suggestive to
estimate the physical UV-limit of the EoM in the limit $k\to 0$ by
the quantum EoM in the limit $k\to\infty$.

Studies in asymptotically safe quantum gravity have focused so far on
finding solutions to \eqref{eq:backgr-eom}.  For instance in
\cite{Falls:2016wsa} they didn't find a solution to
\eqref{eq:backgr-eom} in a polynomial expansion with the background
field approximation.  Other approaches with the background field
approximation found a solution with the exponential parameterisation
\cite{Ohta:2015fcu,Ohta:2015efa} and within the geometrical approach
\cite{Demmel:2015oqa,Gonzalez-Martin:2017gza}.  In this work we are
for the first time able to disentangle \eqref{eq:backgr-eom} and
\eqref{eq:quantum-eom} in a quantum gravity setting and look for
separate solutions to the EoMs.

\begin{figure}[t]
\includegraphics[width=\linewidth]{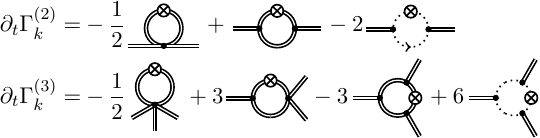}
\caption{Displayed are the diagrammatic representations of the flows of the 
  graviton two- and three-point functions. Double and dashed lines represent 
  dressed graviton and ghost propagators respectively, while filled circles
  denote dressed vertices. Crossed circles stand for regulator insertions.
  All quantities are explicit background curvature dependent and carry further 
  background curvature dependence via the spectral value of the respective 
  vertex/propagator.}
\label{fig:flow_n_point}
\end{figure}

We disentangle the background and fluctuation field by expanding the 
scale dependent effective action around a background according to 
\begin{align}
  \label{vertex_exp}
  \Gamma_k[\bar{g},h] =\sum_{n=0}^{\infty} \frac{1}{n!} \Gamma^{(0,n)}_k[\bar{g},h=0]\, h^n \,.
\end{align} 
The flow equations that govern the scale-dependence of the vertex functions 
are obtained by $n$ field derivatives of the flow equation for the effective 
action \eqref{eq:gen_flow_eq}. They are depicted in a diagrammatic language 
in \autoref{fig:flow_n_point} for cases $n=2$ and $n=3$. These flow equations
are familiar from computations on a flat background 
\cite{Christiansen:2012rx,Christiansen:2014raa,Christiansen:2015rva,Denz:2016qks}, 
here however all propagators and vertices depend non-trivially on the background.

From here on we drop the index $k$ to improve readability,
the scale dependence of the couplings, correlation functions and 
wave function renormalisations is implicitly understood.

\subsection{Background independence in non-perturbative expansion schemes}
\label{eq:FunRel}
It is important to discuss the relations of the approaches described
in \autoref{sec:Motivation} in particular for future developments and
the full resolution of {\it physical} background independence. This
chapter extends a similar discussion from \cite{Denz:2016qks} in the
context of modified STIs for diffeomorphism
transformations to NIs.  Despite its importance one may
skip this chapter for a first reading as its results are not necessary
for the derivations and computations presented in this work.

We have technically very different options to access
physical background independence of quantum gravity. Seemingly they
have different advantages and disadvantages. For example, approach
{\it (1)} via the NIs has the charm of directly
implementing background independence. In turn, the results of 
{\it (2b)} may apparently not satisfy the NIs. 

For resolving this issue it is instructive to discuss the approach
{\it (2a)}. There the fluctuation correlation functions are computed
for a specific background.  Results for general backgrounds have then
to be obtained with an expansion/extension of the results for the
specific background.  This could be done via the NIs in
which case background independence is guaranteed. This procedure for
guaranteeing STIs and NIs
has been discussed in detail in \cite{Fischer:2008uz} in the context
of non-Abelian gauge theories, and in \cite{Denz:2016qks} for gravity. We briefly 
repeat and extend the structural argument presented there: First, we notice that the 
functional equations for all correlation functions can be cast in the form
\begin{align}
  \label{eq:fun}
  \Gamma^{(n,m)}[\bar g,h]= {\rm FRG}_{n,m}[\{\Gamma^{(i\leq n\,,\,2\leq j\leq m+2)}
  [\bar g,h]\},\bar g]\,.
\end{align}
Eq.~\eqref{eq:fun} follows from integrating the FRG equations for $\Gamma^{(n,m)}$, which have
precisely the same structure for all theories: the flows of
$\Gamma^{(n,m)}$ are given by one-loop diagrams with full propagators
and full vertices.  The latter are given in terms of the correlation
functions $\{\Gamma^{(i\leq n\,,\,2\leq j\leq m+2)}\}$, see, e.g.,
\cite{Pawlowski:2005xe,Fischer:2008uz}. This also entails that the
lowest fluctuation correlation function that contributes to the
diagrams is the two-point function, i.e., the propagator.

In gravity \eqref{eq:fun} follows straightforwardly from
\eqref{eq:gen_flow_eq} by integrating the flow equation and taking
$\bar g$- and $h$-derivatives.  As a side remark, we note that the
order of derivatives on the right-hand side is different within other
functional approaches. For example, for Dyson-Schwinger equations (DSE) 
the right-hand side ${\rm DSE}_{n,m}$ 
for the $\Gamma^{( n,m)}$ depends on
$\{\Gamma^{(i\leq n,j\leq m+r-2)}\}$ and contain up to $r-2$-loop
diagrams. Here $r$ is the highest order of the field in the classical
action, see e.g. \cite{Pawlowski:2005xe}. In typical examples of
renormalisable theories, we have $r=3, 4$, but in gravity, we have
$r=\infty$. This singles out the flow equation for gravity as the only
functional approach that only connects a finite order of correlation
functions in each equation. The coupling of the whole tower of
equations then comes from the highest order correlation functions on
the right-hand side. In turn, each DSE already contains all orders on
the right-hand side of \eqref{eq:fun}, that is $2\leq j$ without upper
bound. Similar statements as for the DSE hold for 2PI or $n$PI
hierarchies.

Importantly, for all functional approaches the right-hand side of
\eqref{eq:fun} goes only up to the same order of background metric
derivatives, $i\leq n$. This allows us to view \eqref{eq:fun}
as functional relations for the highest order background metric
correlation functions that have as an input
$\{\Gamma^{(n-1,m)}\}$. Moreover, the NI relates a
derivative w.r.t.\ $\bar g$ to one w.r.t.\ $h$. For emphasising the 
similarities to the functional relations \eqref{eq:fun} we rewrite the NI. 
For simplicity we use the linear split NI, \eqref{eq:Nielsen} and \eqref{eq:Nielsen-R-term},  
\begin{align} 
 \label{eq:sym}
  \Gamma^{(n,m)}[\bar g,h]={}& \Gamma^{(n-1,m+1)}[\bar g,h] \\[1ex] 
  &+ {\CN}_{n,m}[G,\{\Gamma^{(i\leq n-1,j\leq m+1)}
  [\bar g,h]\},\bar g]\,,  \notag
\end{align}
where $\CN$ stands for the expectation value in \eqref{eq:Nielsen}, and
additionally for the regulator loop in \eqref{eq:Nielsen-R-term}, and we
have singled out the propagator $G$ for elucidating the
orders of the correlation functions on both sides. 
Importantly, \eqref{eq:sym}
makes the fact apparent that for the NI, \eqref{eq:Nielsen} and \eqref{eq:Nielsen-R-term},
the order of background derivatives is at most
$n-1$. Note also that \eqref{eq:sym} is nothing but the difference of
the Dyson-Schwinger equation for $h$ and $\bar g$ derivatives. In this
difference, the terms with the higher vertices with $j\geq m+2$ drop
out.

In summary, this leaves us with two towers of functional
relations. While the first one, \eqref{eq:fun} describes the full set of
correlation functions, the second one, \eqref{eq:sym} can be used to
iteratively solve the tower of mixed fluctuation-background
correlations on the basis of the fluctuating correlation functions $\{\Gamma^{(0,m)}\}$.
In both cases, we can solve 
the system for the higher-order correlations of the background on the
basis of the lower order correlations. If we use \eqref{eq:sym} with an iteration
starting with the results from the flow equation for
$\{\Gamma^{(0,m)}[\bar g_{\text{sp}}, h]\}$ for a specific
background $\bar g_{\text{sp}}$, this closure of the system
automatically satisfies the NI. Accordingly, {\it any}
set of fluctuation correlation functions
$\{\Gamma^{(0,m)}[\bar g_{\text{sp}}, h]\}$ can be iteratively
extended to a full set of fluctuation-background correlation functions
in an iterative procedure. Note that this procedure can be also
applied to the case {\it (2b)}.

While this seems to indicate that satisfying the symmetry identities
is not relevant (it can be done for all inputs), it points at a more
intricate structure already known from non-Abelian gauge theories. To
that end let us assume we have derived a globally unique solution of all
correlation functions within this iterative procedure starting from
the fluctuations correlation functions. If no approximation is
involved, this solution automatically would satisfy the full set of
functional relations for $\{\Gamma^{(n,m)}\}$ that can be derived from the
flow equation. However, in the presence of approximations these
additional functional relations represent infinite many additional
constraints on the iterative solution. These constraints are bound to
fail in generic non-perturbative approximation schemes as any
functional relation triggers specific resummations in given
approximations. It is a priori not clear which of the functional
relations are more important. Note also that typically the iterative
solutions of the symmetry identities are bound to violate the locality
constraints of local quantum field theories that are tightly connected
to the unitarity of the theory. In conclusion, it is fair to say that
only a combination of all approaches is likely to provide a final
resolution of {\it physical} background independence and
diffeomorphism invariance in combination with unitarity.

\section{Vertices in curved backgrounds}
\label{sec:curvature}
This section contains technical details about the construction of an approximate 
momentum space and the vertex flow equations on curved backgrounds. 
If one is not interested in these details, one 
may proceed to \autoref{sec:results}.

\subsection{Spectral decomposition}
\label{sec:spectral_decomposition}
We extend our previous expansion schemes about the flat Euclidean
background to one that allows for arbitrary constant curvatures. To
that end we first discuss the procedure at the example of the
propagators: propagators for non-trivial metrics $\bar g$ with constant
curvature can be written in terms of the scalar Laplacian $\Delta_{\bar g} = -\bar{\nabla}^2$
and curvature terms proportional to the background scalar curvature $\bar{R}$,
\begin{align}\label{eq:GDeltaR}
G=G(\Delta_{\bar g},\bar{R})\,.
\end{align}
For the flat metric \eqref{eq:GDeltaR} reduces to $G(p^2,0)$, where $p^2$
are the continuous spectral values of the flat scalar Laplacian. In a
spectral basis the propagator is diagonal and reads for general
curvatures
\begin{align}\label{eq:GDeltaRspec}
  \left.\langle \varphi_{\lambda}|
    G|\varphi_{\lambda}\rangle\right|_{\lambda=p^2} =G(p^2,\bar{R})\,,
\end{align}
and $\lambda=p^2$ are the discrete or continuous eigenvalues for the
given metric, and $\{|\varphi_{\lambda=p^2}\rangle \}$ is the
orthonormal complete basis of eigenfunctions of the scalar Laplacian
\begin{align}\label{eq:spec}
  \Delta_{\bar g} |\varphi_{\lambda}\rangle = \lambda
  |\varphi_{\lambda=p^2}\rangle \,,
\end{align}
see App.~\ref{app:propagator} for explicit expression for the propagator.
The tricky part in this representation are the vertices, which are operators 
that map $n$ vectors onto the real numbers. For example
the three-point function can be written in a spectral representation in terms 
of an expansion in
the tensor basis with eigenfunctions of $\Delta_{\bar{g}}$, 
\begin{align}\label{eq:Gamma3lambda}
  \Gamma^{(3)} = \!\!\sumint_{\lambda_1,\lambda_2,\lambda_3}
  \!\!\Gamma^{(3)}(\lambda_1, \lambda_2, \lambda_3,\bar{R})
  \langle \varphi_{\lambda_1}| \otimes
  \langle \varphi_{\lambda_2}| \otimes \langle \varphi_{\lambda_3}| \,, 
\end{align}
where the spectral values in general also depend on the curvature and
$\sumint$ runs over discrete or continuous spectral values. Also,
$\sumint$ may also include a non-trivial spectral measure weight
$\mu(\lambda)$. The representation of the higher $n$-point functions
follows straightforwardly from \eqref{eq:Gamma3lambda}. Inserting this
into the flow equation of the inverse propagator, we arrive at
\begin{align} 
\partial_t \Gamma^{(2)}(\lambda,\bar{R}) 
  = & -\012\sumint_{\lambda_1} 
  \Gamma^{(4)}(\lambda,\lambda,\lambda_1,\lambda_1,\bar{R}) (G \dot{R}_k
  G)(\lambda_1,\bar{R}) \notag\\[1ex]  
& + \sumint_{\lambda_2,\lambda_3} \! \Gamma^{(3)}(\lambda, \lambda_2,
  \lambda_3,\bar{R})G(\lambda_2,\bar{R})\notag\\[1ex] 
& \times (G \dot{R}_k G)(\lambda_3,\bar R)
  \Gamma^{(3)}(\lambda_3, \lambda_2,\lambda,\bar R)\,,
\label{eq:flow3}
\end{align}
where we denoted $\dot R_k = \partial_t R_k$.
The vertex functions $ \Gamma^{(n)}$ are complicated functions
of $\lambda_i$. 

On a flat background, the eigenfunctions of the Laplace operator are
also eigenfunctions of the partial derivatives and the representation
of the vertex functions follows trivially. On a curved background,
however, the covariant derivatives do not commute with the Laplace
operator and the representation of uncontracted covariant derivatives
on the set of functions $\{|\varphi_{\lambda=p^2}\rangle \}$ is
complicated. One could tackle this problem with e.g.\ off-diagonal
heat-kernel methods, but then a derivative expansion in momenta and
curvature is necessary \cite{Knorr:2017fus}.  

In this work we
construct an approximate momentum space on a curved background, which
facilitates computations considerably and allows for full momentum and
curvature dependences.  In order to derive the vertex functions, we
first take functional derivatives with respect to the Einstein-Hilbert
action on an arbitrary background. The result is a function depending
on the Laplacian, products of covariant derivatives with respect to
coinciding or different spacetime points and explicit curvature
terms. In the expression for the vertex functions we symmetrise all
covariant derivatives, which produces further $\bar{R}$-terms
\begin{align} \label{eq:symmetrisation}
 \bar{\nabla}^\mu \bar{\nabla}^\nu &=  \012 \{\bar{\nabla}^\mu,\bar{\nabla}^\nu\} + \bar{R}\text{-terms} \,.
\end{align}
In the curved momentum space approximation here, the product of symmetrised 
covariant derivatives acts on the set $\{|\varphi_{\lambda=p^2}\rangle \}$ 
according to
\begin{align}\label{eq:flatproduct}
  \bar{\nabla}_1 \cdot  \bar{\nabla}_2 = p_g\cdot q_g=
  \sqrt{p^2}\sqrt{q^2}\, x\,, \quad {\rm with}\quad
  x=\cos\theta_{\text{flat}}\,,
\end{align}   
with an integration measure $\int \sqrt{1-\cos^2\theta}\,
\mathrm d\cos\theta$. The integration measure is chosen such that in the limit 
$\bar{R}\to 0$ precisely the flat results are obtained.
As a consequence, in this approximation $\sumint$ factorises into an angular 
integration and a sum/integration over the spectral values of $\Delta_{\bar g}$. According to 
\eqref{eq:flatproduct}, external spectral values are described by the angle to 
the internal one and their absolute values, which appear as parameters that can 
be treated as real numbers. 
We emphasise that this curved momentum space approximation has the correct 
flat background limit by construction and is correct for all terms that contain only 
Laplace operators. A comparison of the approximation as a function of the background 
curvature is detailed in App.~\ref{app:approximations}.
With the above approximation associated with covariant 
derivatives, we arrive at a
relatively simple flat-background-type representation of the flow equation in terms 
of angular
integrals and spectral values $p_i^2=\lambda_i$
\begin{align}
  \partial_t \Gamma^{(2)}&(\lambda,\bar R) \notag\\[1ex]  
 = & -\012 \sumint_{\lambda_1}
  \int \! \mathrm d\Omega \,\Gamma^{(4)}(\lambda,\lambda_1,x,\bar R) 
  (G \dot{R}_k G)(\lambda_1,\bar R)\notag\\[1ex]
  & + \sumint_{\lambda_1} \int \!\mathrm d\Omega \, 
  \Gamma^{(3)}(\lambda,\lambda_1,x,\bar{R})
  G(\lambda_1+\lambda+ \sqrt{\lambda\,\lambda_1} x,\bar R) \notag \\[1ex]
  & \times (G \dot{R}_k  G)(\lambda_1,\bar R)
  \Gamma^{(3)}(\lambda,\lambda_1,x,\bar R)\,. 
  \label{eq:flow3fin}
\end{align}
The total $\bar{R}$-dependence of the flow equation enters via the explicit $\bar{R}$-terms 
in the vertex functions, the symmetrised covariant derivatives and the 
spectral values.
The generalisation to flows of higher-order vertex functions is
straightforward.

\subsection{Vertex construction}
\label{sec:vertex_dressing}
The basic ingredients in the flow equations in \autoref{fig:flow_n_point} are the 
vertex functions $\Gamma^{(n)}$. We build on the parameterisation for vertex 
functions introduced in
\cite{Fischer:2009tn,Christiansen:2014raa,Christiansen:2015rva,Denz:2016qks}. 
In contrast to earlier truncations with vertex expansions around a flat 
background, all quantities exhibit explicit $\bar R$-dependence.
Hence, our general ansatz is given by
\begin{align}\label{eq:vertex} 
  & \Gamma^{(\phi_1 \ldots \phi_n)}(\mathbf{p},\bar R) 
  =  S_\text{EH}^{(\phi_1 \ldots
    \phi_n)}(\mathbf{p};G_{n}(\bar R),\Lambda_{n}(\bar R),\bar R) \,, 
\end{align}
where $\mathbf{p}=(p_1,\dots,p_n)$ is the collection of spectral values of the external legs
and $S_{\text{EH}}$ is the gauge-fixed Einstein-Hilbert action
\begin{align}\label{eq:EH_action}
  S_\text{EH} &= \frac{1}{16 \pi G_N} \int \mathrm d^4 x \sqrt{ g}
  \left(2\Lambda - R\right) + S_{\text{gf}} + S_{\text{gh}} \,.
\end{align}
We employ a De-Donder-type linear gauge condition in the Landau
limit, $\alpha = \beta = 0$.

In \eqref{eq:vertex} the Newton's constant and the cosmological
constant of the classical gauge fixed Einstein Hilbert action are
getting replaced with $G_n(\bar R)$ and $\Lambda_n(\bar R)$,
respectively.  They parameterise the gravitational coupling and the
momentum-independent part of the $n$-point function.  Note that the
graviton $n$-point function in \eqref{eq:vertex} is proportional to
$G_n^{n/2-1}$ as well as to $Z_{h}^{n/2}$ due to the rescaling of the
graviton fluctuation field to a field with mass dimension one.  This
is captured with the split
$g_{\mu\nu}= \bar g_{\mu\nu} + \sqrt{Z_h G_N} h_{\mu\nu}$.  The
wave function renormalisation is in general momentum and
background-curvature dependent, $Z_h=Z_h(p^2,\bar R)$.

The propagator is a pure function of $\Delta_{\bar{g}}$ and $\bar R$,
while the vertices with $n>2$ are functions of $\Delta_{\bar{g}}$,
$\bar{\nabla}_\mu$, $\bar R$, $\bar R_{\mu\nu}$ and
$\bar R_{\mu\nu\rho\sigma}$. Restricting ourselves to a background
sphere, the dependence on the Ricci- and the Riemann-tensor reduces to
a dependence on the constant background curvature $\bar R$.  With the
approximation constructed in the last section, we deal with the
covariant derivatives $\bar\nabla_\mu$ in the vertices. We set the
anomalous dimensions
\begin{align}
 \eta_{\phi_i}(p^2,\bar R) = -\partial_t \ln Z_{\phi_i}(p^2,\bar R)\, ,
\end{align}
throughout this work equal to zero.
In the flat computation \cite{Christiansen:2015rva,Denz:2016qks} this 
approximation led to qualitatively reasonable results.
The graviton three-point function is evaluated at the point of symmetric 
spectral values,
\begin{align}
  p&=|p_1|=|p_2|\,, & 	\theta_{\text{flat}}&=2\pi/3\,.
\end{align}
We close the flow equations by setting the higher-order 
couplings to $G_{n\geq4}=G_3=:G$ and $\Lambda_4=\Lambda_3$
as well as $\Lambda_{n\geq5}=0$.
We also introduce the dimensionless variables
\begin{align}
 r &= \bar R\, k^{-2}\,, & g&= G\, k^{2} \,,\notag\\
 \mu&= -2\Lambda_2 k^{-2}\,, & \lambda_3 &= \Lambda_3 k^{-2}\,.
\end{align}
From the graviton two-point function we extract the mass-parameter $\mu(r)$, 
while from the graviton three-point function we extract the gravitational 
coupling $g(r)$ and the coupling of its momentum independent part $\lambda_3 
(r)$. In App.~\ref{app:flow_eq} we give a derivation and display the flow equations.
In summary, the set of couplings in the present truncation is given by
\begin{align}
 \{ g(r),\, \mu(r) ,\, \lambda_3(r) \} \,.
\end{align}

\subsection{Flow equations and trace evaluation}
\label{sec:FlowTrace}
With the construction presented in the last sections, we are left with an 
explicit expression for the flow of the two- and the three-point function.
The flow of the two-point is of the form \eqref{eq:flow3fin} and 
the three-point function has a similar form according to the diagrammatic 
representation in \autoref{fig:flow_n_point}. 
After projection the resulting flow equations take the form \eqref{eq:lam3-dot} 
and \eqref{eq:g-dot}. 
In this work, we are interested in the fixed point equations, which 
are differential equations in
$r$ due to the dependence on the background curvature. 
According to the 
factorisation property of the approximate curved momentum space construction, 
we evaluate the angular integration in a straightforward manner in complete 
analogy to a flat background computation. We are then left with the evaluation of 
traces of the form 
\begin{align}\label{eq:momentum_trace}
\sumint_{\lambda} f(\lambda, r) \,,
\end{align}
for functions of the curvature $r$ and the spectral value $\lambda$ as well as the 
couplings.
To include the effects of the background curvature we perform a 
spectral sum over a four-sphere.
On a four-sphere the spectrum of the scalar Laplacian
is given by 
\begin{align}
\omega(\ell)=  \frac{\ell (3+\ell)}{12} r\,,
\end{align}
with multiplicities 
\begin{align}
m=\frac{(2 \ell+3) (\ell+2)!}{6 \ell!}\,,
\end{align}
with $\ell$ taking integer values $\ell \geq 0 $.
Since we are left with only scalar spectral values we 
replace the spectral values by
\begin{align}
\lambda \to \omega(\ell) \,,
\end{align}
and replace
\begin{align}
\sumint\nolimits_{\lambda}   \to  V^{-1} \sum_{\ell=2}^{\ell_{\max}}   m(\ell) \,,
\end{align}
where the exact sum is achieved for $\ell_{\max} = \infty$ and we
divide by the volume of a four sphere $V = \frac{384 \pi^2}{k^4r^2}$.
Note, that we exclude the zero modes and start the spectral sum at
$\ell=2$.  This does not affect the result for small curvature $r$.
Performing the spectral sums one then obtains the traces.  However, in
most cases, a closed form for the sums cannot be obtained and we have
to resort to cutting the spectral sums off at a finite value
$\ell_{\max}$.  Nonetheless, since each trace involves a regulator
function that cuts modes off at order $\omega(\ell) \approx k^2$ for
non-zero $r$ the spectral sum is only sensitive to the modes
$\omega(\ell) < k^2$, which are finite in number.  However, in the
limit of vanishing curvature, the spectral sum needs to be extended to
infinite order, as all modes are regulator suppressed only for large
$r$ according to $\exp(- \lambda_n r)$, but become important once
$r \approx 1/\lambda_n$. We need the limit $r \to0$ to set the boundary 
conditions of the fixed-point differential
equations. It is obvious that there is only one physical initial
condition that fixes the solution of the fixed point differential
equation uniquely, and that is the initial condition obtained from the
flat background limit. A proper initial condition is also
necessary from a mathematical point of view if one requires a finite
derivative, $g'(r) < \infty$. One infers from \eqref{eq:lam3-dot} and
\eqref{eq:g-dot}, that the derivative of $g(r)$ diverges in the limit
$r \to 0$ if the initial conditions are not chosen appropriately.
However, as argued above, this limit cannot be calculated in practice
with spectral sums as all modes contribute. In the small-curvature
region the trace is evaluated by the early-time heat-kernel expansion
where the leading order gives the flat-background momentum
integrals. In this case we write the Laplace transform
\begin{align}
\label{eq:heat-kernel}
\sumint\nolimits_{\lambda} f(\lambda, r)  = \frac{1}{V}  \int_0^{\infty}\mathrm 
ds \, \tr [e^{-s \Delta_g}]  \, \tilde{f}(s,r)\,,
\end{align}
and one expands the trace of the heat kernel in the scalar curvature $r$ 
and the explicit dependence on $r$ coming from $ \tilde{f}(s,r)$.
For small curvature the  early-time heat-kernel expansion is given by
\begin{align}
\label{eq:heat-kernel_expansion}
\frac{1}{V}  \int_0^{\infty} \!\! \mathrm ds \, \tr[e^{-s \Delta_g}]  \, 
\tilde{f}(s,r) =     \frac{1}{(4 \pi)^2} (Q_{2}[f]   + Q_1[f] \frac{r}{6} + 
...)\,,
\end{align}
where for $n>0$
\begin{align}
Q_n[f] =\frac{1}{\Gamma(n)} \int \mathrm d\lambda \lambda^{n-1} f(\lambda,r)\,.
\end{align}
Using this heat-kernel expansion we translate the physical initial
condition to finite $r$ where we connect to the spectral sum. In
particular we determine the curvature-dependent couplings as
polynomials in the curvature $r$. The heat kernel provides the
asymptotic limit $r \to 0$ which can be reproduced by the spectral sum
in the limit $\ell_{\max} \to \infty$. Thus, while the spectral sum
with finite $\ell_{\max}$ captures the large $r$ behaviour of the
trace, the heat kernel expanded to a finite order in $r$ captures the
small $r$ behaviour. Both connect smoothly for finite but small $r$,
for details see App.~\ref{app:spectralheat}. 

\section{Results}
\label{sec:results}
In this section, we present the results of the given setup.
First, we discuss the fixed point solutions of the beta functions related 
to the fluctuation field couplings. In our approach with curvature-dependent 
couplings, these solutions are fixed-point functions.  
Subsequently, we analyse the background effective potential, which is 
calculated on the solution of the fluctuation field fixed point solution,
with and without SM matter content.
Last we look for solutions of the quantum EoM
and compare to solutions of the background EoM.

%%%%%%%%%%%%%
\begin{figure}[tbp]
\includegraphics[width=\linewidth]{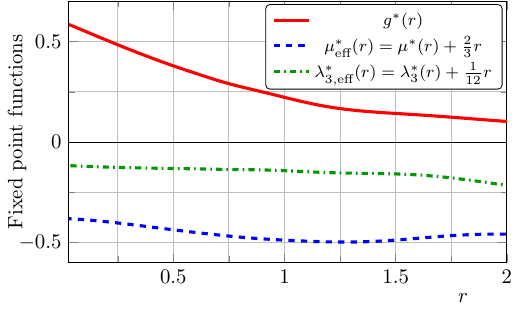}
\caption{Fixed-point functions for the system 
  $(g^*(r),\,\mu_\eff^*(r),\,\lambda^*_{3,\eff}(r))$
  with the boundary condition from the first-order heat kernel.
  The solutions are stable in the whole investigated region.
  Note, that the effective couplings according to \eqref{eq:eff-coupling} 
  are displayed.}
\label{fig:FP-functions}
\end{figure}
%%%%%%%%%%%%%

\subsection{Fixed point solutions}
\label{sec:FP-sol}
The beta functions for a coupling $g_i(r)$ in the present 
framework are partial differential equations. 
Schematically, the equation a coupling $g_i$ takes the form 
\begin{align}
\partial_t g_i(r) = g_i(r) \, A(g_j,\eta_h) + 2 r\, g_i'(r) + \Flow_{g_i} (g_j,\,r) \,, 
\end{align}
with a coefficient $A$ that depends on the other scale-depend parameters $g_j$. 
For explicit expressions we refer to appendix \ref{app:flow_eq}. The fixed point equations are then 
obtained by setting $\partial_t g_i(r) \equiv 0$ and we are left with a system of
ordinary differential equations. The initial condition is imposed at $r=0$ and 
is chosen such that it matches the computation in a flat background 
\cite{Christiansen:2015rva}. For details see \autoref{sec:FlowTrace}.
The UV fixed-point values for the flat background, $g_i(r) = g_{i,0}$, are given by
\begin{align} \label{eq:UV-FP-const}
 (g_{0}^*,\,\lambda_{3,0}^*,\,\mu_{0}^*) = (0.60,\, -0.12,\, -0.38) \,.
\end{align}
with the critical exponents $\theta$, which are the negative eigenvalues of the stability matrix,
\begin{align}\label{eq:crit-exp-const}
 (\theta_{i,0} ) = (-3.7,\,2.0 \pm 2.1 i) \,.
\end{align}
These values differ slightly from the ones in \cite{Christiansen:2015rva} since we 
use the gauge parameter $\beta=0$ and the exponential regulator, see \eqref{eq:exp-reg}.
Taking this difference into account, the agreement is remarkable and highlights 
the insensitivity of our results with respect to the gauge and the 
regulator.

%%%%%%%%%%%%%
\begin{figure*}[tbp]
\includegraphics[width=\linewidth]{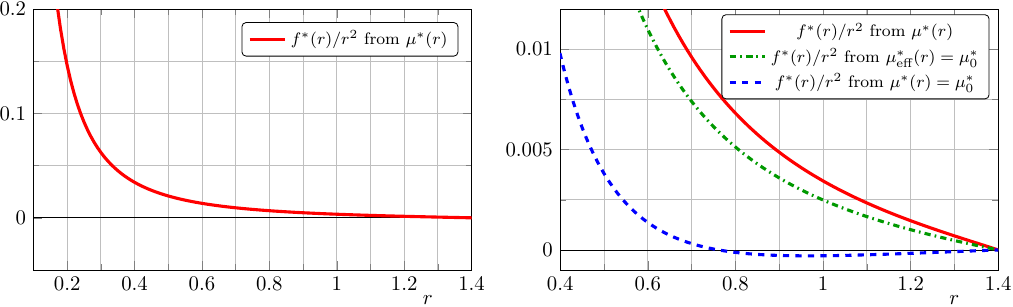}
\caption{Displayed are background potentials $f^*(r)/r^2$ obtained from 
the fixed-point function $\mu^*(r)$ (left and right panel) and from the approximations
$\mu_\eff^*(r)=\mu_0^*$ and $\mu^*(r)=\mu_0^*$ (right panel).
All curves are obtained with the condition $f^*(r=1.4)=0$.
Other conditions just shift the potential $f^*(r)/r^2$ by a constant.
The full solution does not contain a minimum, it becomes asymptotically flat.
The approximation $\mu_\eff^*(r)=\mu_0^*$ is qualitatively very good, see also \autoref{fig:FP-functions}.
The approximation $\mu^*(r)=\mu_0^*$ corresponds to a pure Einstein-Hilbert computation.
Here we find a minimum at $r_0=0.97$.
}
\label{fig:Background-potential}
\end{figure*}
%%%%%%%%%%%%%

To display our results, it is convenient and meaningful to 
introduce effective couplings that include the explicit $r$ 
dependence in the respective graviton $n$-point functions.
According to \eqref{eq:propagator}, \eqref{eq:lam3-dot}, and \eqref{eq:g-dot}, these are given by 
\begin{align}
 g_{\eff} (r) = g (r) \,, \nonumber\\
 \mu_{\eff} (r) = \mu (r) + \023 r \,, \nonumber\\
 \lambda_{3,\eff} (r) = \lambda_3 (r) + \01{12} r \,. 
 \label{eq:eff-coupling}
\end{align}
The interpretation and relevance of these effective couplings can be inferred 
for instance from the graviton two-point function. In terms of $\mu_{\eff} 
(r)$, the transverse-traceless part of the graviton two-point function reads   
\begin{align}
\Gamma^{(0,2)} = \left(\Delta + \mu_{\eff} (r) \right) \,, 
\end{align}
i.e.\ it comprises the non-kinetic part of the correlator. 

The full, $r$-dependent fixed point solutions 
$(g^*(r),\,\mu_\eff^*(r),\,\lambda_{3,\eff}^*(r))$ are displayed in 
\autoref{fig:FP-functions}.
We find a fixed point 
solution with all desired properties. First of all, the fixed point solution 
is characterised by a positive gravitational coupling $g(r)>0$, which decreases 
towards larger background curvatures. To get a feeling for the 
physical meaning of this behaviour, we consider the quantity 
$G(R) R=g(r) r$, i.e.\ the dimensionful Newton's coupling times the curvature. 
As this product is dimensionless, it can in principle be used to define an 
observable. In particular, we expect that this quantity is finite at the fixed 
point, which implies $g^*(r) \sim 1/r$. One might interpret our fixed-point solution 
$g^*(r)$ as an onset of such behaviour. The solutions for the mass-parameter 
$\mu_{\eff}(r)$ and $\lambda_{3,\eff}(r)$ are almost curvature independent, which implies that 
the implicit curvature dependence cancels with the explicit one.

The full solution shown in \autoref{fig:FP-functions} can be
expanded in powers of the dimensionless curvature,
$g^*_i(r) = g^*_{i,0} + g^*_{i,1}\, r + \mathcal{O}(r^2)$. 
The zeroth order is displayed in \eqref{eq:UV-FP-const}
and to linear order in $r$ we find
\begin{align}\label{eq:UV-FP-lin}
 (g_{1}^*,\,\lambda_{3,1}^*,\,\mu_{1}^*) = (-0.43,\, -0.13,\, -0.71) \,.
\end{align}
with the critical exponents $\theta$ given by
\begin{align}\label{eq:crit-exp-lin}
 (\theta_{i,1} ) = (-5.6,\,0.04 \pm 2.2 i) \,.
\end{align}
We find two further UV attractive directions in the linear order of
the background curvature. They are close to marginal and thus 
might easily change their relevance with an increasing truncation.
Further attractive directions of the UV fixed point that are linear 
in the background curvature were found in \cite{Knorr:2017fus}.

%%%%%%%%%%%%%
\begin{figure*}[tbp]
\includegraphics[width=\linewidth]{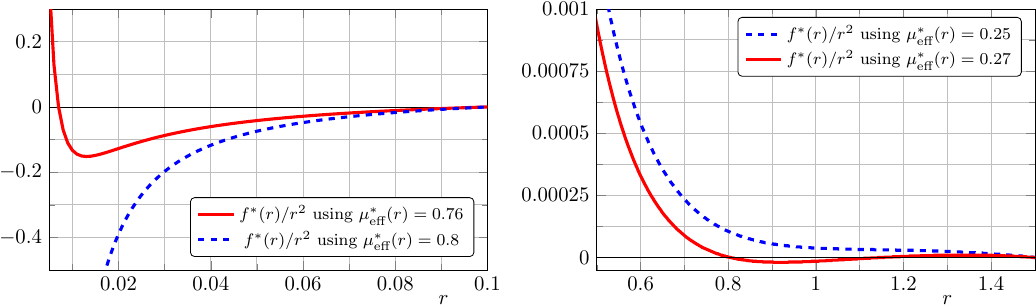}
\caption{Fixed-point background potential for different constant input
  values of $\mu^*_\eff(r)$.  The minimum that
  corresponds to the solution of the background EoM is at $r>0$ for
  $\mu^*_\eff(r) \lesssim 0.77$, while for
  $\mu^*_\eff(r) \gtrsim 0.77$ it is at $r<0$ (left panel).  For
  $\mu^*_\eff(r) \lesssim 0.25$ the minimum vanishes completely, while
  for $\mu^*_\eff(r) = 0.26$ the minimum is located at
  $r_0 = 1.1$ (right panel).  }
\label{fig:Background-pot-variations}
\end{figure*}
%%%%%%%%%%%%%

\subsection{Background potential}
\label{sec:background}
In the previous section, we have presented the fixed point solution for the 
fluctuation field couplings.
All background quantities depend on these dynamical couplings and have to be 
evaluated on the above solution. Along these lines, we calculate a 
background field potential at the fixed point. 
The flow of the background potential is completely determined by the dynamical 
couplings of the two-point function.
In particular, the background flow equation reads
\begin{align}
  \partial_t \Gamma[\bar{g},0] ={}&  \frac{1}{2} \text{Tr}\left[ G\partial_t R_{k} \right]_{hh} 
  - \left.\text{Tr}\left[ G\partial_t R_{k}\right]_{\bar c c} \, \right|_{\phi =0}\,.
  \label{eq:backgroundflow}
\end{align}
On a sphere, the background effective action is given by
\begin{align} \label{eq:ansatz-bg}
\Gamma[\bar{g},0]  = \int \hspace*{-.1cm} \mathrm d^4x \sqrt{\bar{g}}\, k^4 
f(\bar{R}/k^2) = \frac{384 \pi^2}{r^2} f(r) \,.
\end{align}
Denoting the right-hand side of \eqref{eq:backgroundflow} by 
$\mathcal{F}(r, \mu(r))$ 
we obtain a flow equation for the function
$f(r)$ given by 
\begin{align}
\frac{384 \pi^2}{r^2} ( \partial_t f + 4 f(r) - 2 r f'(r)) = 
\mathcal{F}(r, \mu(r))\,.
\end{align}
If we then look at the fixed point for $f^*(r)$ we find
\begin{align}
\frac{384 \pi^2}{r^2} ( 4 f^*(r) - 2 r {f^*}'(r)) = 
\mathcal{F}(r, \mu^*(r))\,.
\end{align}
One then notes that the left-hand side is just the background EoM 
for $f(r)$-gravity on a constant curvature background.
Thus when the function $ \mathcal{F}(r, \mu^*(r))$ vanishes we have a 
solution to the background EoM at the fixed point given by
\begin{align}
\mathcal{F}(r_0, \mu^*(r_0)) = 0\,.
\end{align}
Equivalently we can look for a minimum of the function $f(r)/r^2$.
In \autoref{fig:Background-potential} we plot the background potential $f(r)/r^2$ for our full solution (left panel) 
as well as in comparison with other approximations (right panel). 
There we use $\mu_\eff^*(r) =  \mu_0^*$ and $\mu^*(r) =  \mu_0^*$ as given in \eqref{eq:UV-FP-const}.
The first is seen to be a good approximation from \autoref{fig:FP-functions} 
while the latter reduces our computation to an Einstein-Hilbert approximation.
We observe that there are no solutions to the background EoM
in the full solution and the $\mu_\eff^*(r) =  \mu_0^*$ approximation 
within the investigated curvature regime.
This absence of a constant curvature solution is in agreement with 
studies of $f(R)$ gravity in the background field approximation \cite{Falls:2016wsa},
although solutions have been found in calculations exploiting 
the exponential parameterisation \cite{Ohta:2015fcu,Ohta:2015efa}
and within the geometrical approach \cite{Demmel:2015oqa,Gonzalez-Martin:2017gza}.
For the approximation $\mu^*(r) =  \mu_0^*$, which corresponds 
to a pure Einstein-Hilbert computation, we find a minimum at $r_0=0.97$.
This is again in agreement with computations in the 
background field approximation \cite{Kevin-in-prep,Phd-thesis-Raul}.

In a polynomial expansion around $r=0$ the background potential of the full solution would take the form 
\begin{align}
 f(r) = 0.0065 - 0.0054 \, r + \mathcal{O}(r^2)\,,
\end{align}
and consequently we obtain fixed point values of the background 
Newton's coupling and the background cosmological constant according to
\begin{align}
  \bar g^* &= 3.7 \,, &
  \bar \lambda^* &= 0.60\,.
  \label{eq:backgr-couplings}
\end{align}
Note that $\bar \lambda=\012$ is not a pole in our computation:
the pole is only present in the graviton mass parameter $\mu(r)$.
Surprisingly the fixed-point value of $\bar g^*$ is rather large.
We compare these values with the pure Einstein-Hilbert approximation,
see blue dashed line in \autoref{fig:Background-potential}.
We find 
\begin{align}
 f_\EH(r) = 0.0065 - 0.021 \, r + \mathcal{O}(r^2)\,,
\end{align}
and consequently
\begin{align}
  \bar g_\EH^* &= 0.94 \,, &
  \bar \lambda_\EH^* &= 0.15 \,.
\end{align}
These values are comparable to standard Einstein-Hilbert computations 
in the background field approximation as well as in fluctuation computations.
Thus the large values in \eqref{eq:backgr-couplings} are indeed triggered 
by the non-trivial $r$ dependence of the couplings.

We investigate the stability of the present results by treating
$\mu_\eff^*(r)$ as a free parameter without curvature dependence.  In
this case, $\mu_\eff^*(r)=\mu_0^*$ is a good approximation for our
best solution as discussed above.  By varying this parameter we see
for which values a solution to the background EoM exists. As displayed in
\autoref{fig:Background-pot-variations}, we find that
solutions exist for positive curvature when
$ 0.255 \lesssim \mu^*_\eff \lesssim 0.77$ and for negative curvature
for $\mu^*_\eff \gtrsim 0.77$.  For $\mu^*_\eff \lesssim 0.255$ there
are no solutions.  The transition of the minimum from positive to
negative curvature is depicted in the left panel of
\autoref{fig:Background-pot-variations} while the full disappearance
of the minimum is depicted in the right panel.  The computed value of
$\mu_0^*=-0.38$, see \eqref{eq:UV-FP-const}, is far away from the
value where the solution appears.  Thus we conclude that the absence
of a minimum in the background potential in our full pure gravity
computation is rather stable with respect to changes in the
truncation.

\subsubsection{Dependence on matter}
Matter can potentially have a significant influence on the properties
of the UV fixed point, see e.g.~\cite{Dona:2013qba,Meibohm:2015twa,
  Eichhorn:2016vvy,Eichhorn:2016esv,Christiansen:2017gtg,Christiansen:2017cxa}.
In the present work, matter influences the existence of a minimum in
the background potential in two ways: On the one hand it has an
influence on the fixed point values of the fluctuation couplings,
where in particular the influence on $\mu_\eff^*(r)$ is important.  On
the other hand it has a direct influence on the background potential
via the background matter loops.  Both these effects have been studied
in a fluctuation computation on a flat background, see
\cite{Meibohm:2015twa} for scalars and fermions and
\cite{Christiansen:2017cxa} for gauge bosons.  Consequently, we adapt
the analysis to curved backgrounds under the assumption that the
effective graviton mass parameter $\mu_\eff(r)$ remains almost
curvature independent in these extended systems, similar to the
results displayed in \autoref{sec:FP-sol}.

%%%%%%%%%%%%%
\begin{figure}[tbp]
\includegraphics[width=\linewidth]{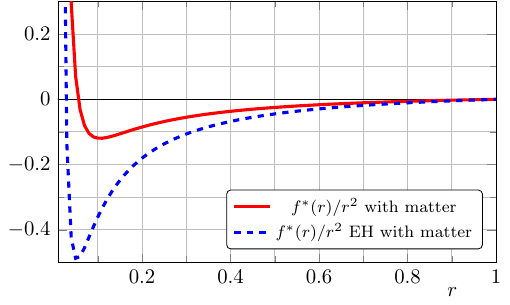}
\caption{
Depicted is the fixed-point background potential if SM matter content is included.
In the full solution as well as in the Einstein-Hilbert solution we find a minimum at small background curvature,
$r_0 = 0.11$ and $r_{0,\EH} = 0.05$, respectively,
which corresponds to the solution of the background EoM.
}
\label{fig:Background-pot-with-matter}
\end{figure}
%%%%%%%%%%%%%

Combining the results of \cite{Meibohm:2015twa} and \cite{Christiansen:2017cxa} 
for SM matter content ($N_s=4$, $N_f=22.5$, and $N_v=12$)
gives a UV fixed point at
\begin{align}
 \label{eq:UV-FP-SM}
 \left(g_{0}^*,\,\lambda_{3,0}^*,\,\mu_{0}^*\right)_\SM = (0.17,\, 0.15,\, -0.71) \,.
\end{align}
For the present analysis
only the value $\mu^*_{0,\SM}$ is important since we now use
$\mu^*_\eff(r)= \mu^*_{0,\SM}$ as an input for the background
potential.  The matter content seemingly pushes $\mu^*_\eff$ in the
wrong direction, cf.~\autoref{fig:Background-pot-variations}.
However, the matter content has also a huge influence on the
background equations.  The combined result is displayed in
\autoref{fig:Background-pot-with-matter}.  Indeed we find a minimum in
the background potential at small curvature, $r_0 = 0.11$.  Also in
the Einstein-Hilbert approximation, i.e.\ $\mu^*(r) = \mu^*_{0,\SM}$,
we find a minimum at $r_{0,\text{EH}} = 0.05$.  With
SM matter content the full solution and the Einstein-Hilbert
approximation are very similar.  This comes as a surprise as the
difference was rather significant without matter content,
cf.~\autoref{fig:Background-potential}.

\subsection{Quantum equation of motion}
In this section, we evaluate the graviton one-point function and thus look for
solutions to the quantum EoM \eqref{eq:quantum-eom}.
As discussed in \autoref{sec:FRG} the solution to this equation leads
to self-consistent backgrounds that improve the convergence of the
Taylor series. Moreover, it has been also argued there that the
quantum EoM in the limit $k\to\infty$ should be seen as an estimate
for the solution of the UV EoM in the physical limit
$k\to 0$ where background and quantum EoM agree due to background
independence. 

Within the present setup, the only invariant linear in the fluctuation field is given by
$f_1(r) h^\Tr$ with some function $f_1$ that is determined by the
fluctuation couplings.  An invariant linear in the transverse
traceless mode does not exist due to our restriction to a spherical
background and thus the absence of terms like
$r^{\mu\nu}h_{\mu\nu}^\TT$.  Consequently, we evaluate
\eqref{eq:quantum-eom} with a projection on the trace mode of the
graviton.

%%%%%%%%%%%%%
\begin{figure}[tbp]
\includegraphics[width=\linewidth]{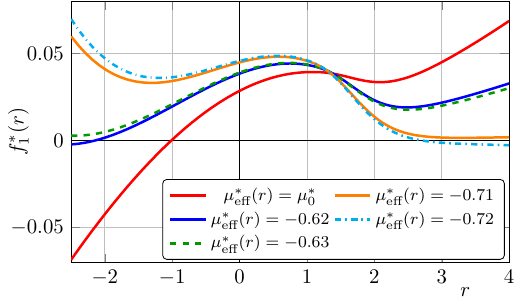}
\caption{
Shown is the fixed-point function $f^*_1(r)$ for different constant input values of $\mu^*_\eff(r)$.
The zeros in these functions correspond to solutions to the quantum EoM \eqref{eq:quantum-eom}.
Our best result $\mu^*_\eff(r)=\mu_0^*= -0.38$ has a solution at negative curvature, $r_0=-1.0$.}
\label{fig:Quantum-eom}
\end{figure}
%%%%%%%%%%%%%

%%%%%%%%%%%%%
\begin{figure*}[tbp]
\includegraphics[width=0.9\linewidth]{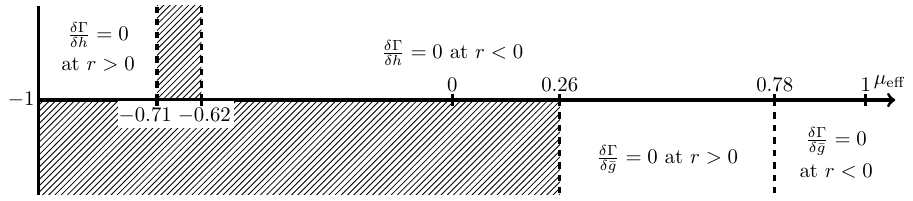} 
\caption{Visualisation of the existence of a solution to the background 
and quantum EoM in dependence on the parameter $\mu_\eff$.
Solutions at positive curvature ($r>0$) and negative curvature ($r<0$)
are distinguished.}
\label{fig:Solutions-eom}
\end{figure*}
%%%%%%%%%%%%%

In straight analogy to the background EoM \eqref{eq:ansatz-bg}, we
parameterise the one-point function by
\begin{align}
\Gamma^{(h_\Tr)}[\bar{g},0]  = \int \hspace*{-.1cm} \mathrm d^4x \sqrt{\bar{g}}\, k^3 
f_1(\bar{R}/k^2) = \frac{384 \pi^2}{k\,r^2} f_1(r) \,.
\end{align}
We denote again the right-hand side by
$\mathcal{F}_1(r,\mu(r))$. Note the difference compared to \eqref{eq:ansatz-bg} due to the different
mass-dimensions of $\bar g$ and $h$.Thus $f_1$ obeys the fixed-point
equation
\begin{align}
\frac{384 \pi^2}{r^2} ( 3 f_1^*(r) - 2 r {f_1^*}'(r)) = \mathcal{F}_1(r, \mu^*(r))\,.
\end{align}
We solve this equation with the initial condition that $f^*_1(r)$
is finite at $r=0$.  Consequently we combine a heat kernel expansion
around $r=0$ up to the order $r^3$ with a spectral sum evaluation for
large, positive curvature. For results at negative
curvature we rely on the heat kernel expansion, but from a comparison
of the heat kernel results with the spectral sum at positive curvature
we can estimate the radius of convergence of the heat kernel.
We estimate the latter by the range where the relative change is in 
the sub per cent regime. We find that the radius of convergence is approximately 
given by $r_\text{conv}\approx1$. The radius of convergence increases 
for larger $\mu^*_\eff(r)$.

The resulting fixed-point functions $f_1^*(r)$ are shown in
\autoref{fig:Quantum-eom}. For our best result
$\mu^*_\eff(r)=\mu_0^*=-0.38$, $f^*_1(r)$ has a root at
negative curvature, $r_0=-1.0$, which corresponds to a solution to the
quantum EoM. The result lies within the radius of convergence of the
heat-kernel expansion and thus we consider it trustworthy.

We again check the stability of the solution by treating
$\mu^*_\eff(r)$ as a constant free input parameter.  For more positive
values, $\mu^*_\eff>\mu_0^*$, the root of $f_1^*(r)$ moves towards larger curvature
but always remains negative. In the limit $\mu^*_\eff\to \infty$ the
root is located at $r_0=-0.42$. For more negative values, $\mu^*_\eff<\mu_0^*$ the
root of $f_1^*(r)$ moves towards smaller curvature and eventually the
root disappears at $\mu^*_\eff=-0.62$, cf.~\autoref{fig:Quantum-eom}.
This result has to be taken very carefully since at $\mu^*_\eff=-0.62$
the root is located at $r_0=-2.2$ and thus lies outside of the radius of
convergence of heat kernel.  At $\mu^*_\eff=-0.71$ a new solution
appears at positive curvature, $r_0=2.7$.  This root remains also for
more negative values of $\mu^*_\eff$ until the pole at $\mu^*_\eff = -1$.  The
roots at positive curvature are obtained with the spectral sum and
thus do not rely on the radius of convergence of the heat kernel.

We have visualised the existence of a solution to the background and
quantum EoM in \autoref{fig:Solutions-eom}.
The quantum EoM has almost always a solution,
only in the range $-0.71<\mu^*_\eff<-0.62$ no solution exists.
This range may even disappear with better truncations or 
an improved computation at large negative curvature.
The background EoM on the other hand 
only allows for a solution for $\mu^*_\eff>0.26$, and thus in a 
region that is very unusual for pure gravity computations.

\section{Summary and Outlook}
\label{sec:summary}
In this work, we have developed an approach to asymptotically safe
gravity with non-trivial backgrounds. As a first application of the
novel approach, we computed the $f(R)$-potential and discussed
solutions of the equations of motion. 

We have also given a discussion of functional approaches to quantum
gravity that take into account the necessary background independence
of the theory. We have discussed, for the first time in quantum
gravity, that background independence and diffeomorphism invariance
can be achieved iteratively in {\it any} approximation scheme, based
on a similar argument in non-Abelian gauge theories, see
\autoref{sec:Motivation}. We have also emphasised the relevance of
aiming for solutions that satisfy all functional relations. We have
argued that this is tightly bound to the question of unitarity.

The approach is based on a vertex expansion of the effective action
about non-trivial backgrounds, which at present are restricted to
constantly curved backgrounds. Our explicit results are based on a
truncation that includes the flow of the graviton two- and three-point
function and thus the couplings $g$, $\lambda_3$, and $\mu$.  The
construction of an approximate momentum space,
cf.~\eqref{eq:flatproduct}, allowed us to evaluate these couplings
without a derivative expansion in momentum $p$ or curvature $r$.  In
this work, we focused on the curvature dependence and thus all
couplings are functions of the curvature, $g(r)$, $\lambda_3(r)$, and
$\mu(r)$.  The flow equations for these coupling functions were
obtained with spectral sums on a sphere. The results are smoothly
connected to known results at vanishing background curvature with 
heat-kernel methods.

As one main result, we found UV fixed point functions that confirm the
asymptotic safety of the present system.  Interestingly, the effective
fixed point couplings, $\lambda^*_{3,\eff}(r)$ and $\mu^*_\eff (r)$,
cf.~\eqref{eq:eff-coupling}, turned out to be almost curvature
independent over the investigated range: the couplings counterbalance
the explicit curvature dependence of the $n$-point functions.

We have also discussed the background and the quantum equation of
motion, \eqref{eq:backgr-eom} and \eqref{eq:quantum-eom}, in
\autoref{sec:genFrame}. At $k=0$, their solutions agree due to
background independence. In turn, at finite $k$ the solutions to
the background and quantum equation of motion differ due to a regulator
contribution to the modified Nielsen identity. This signals the
breaking of background independence in the presence of the cutoff. We
have argued in the present work that at finite cutoff it is the
solution of the quantum equation of motion that relates directly to
the physical solution of the equation of motion at vanishing cutoff. 

We explicitly evaluated both equations of motion with the UV fixed
point functions and indeed found different solutions: The background
equation of motion does not feature a solution. Only with SM
matter content, a solution at small curvature is present.  The
quantum equation of motion exhibits already a solution at negative
curvature without any matter content.  We have checked the stability
of these statements by scanning for solutions in the parameter
$\mu^*_\eff$.  The background equation of motion without matter
features a solution only for very large values of $\mu^*_\eff$, far
away from most values observed in pure quantum gravity truncations.
On the other hand, the quantum equation of motion has a solution for
almost all $\mu^*_\eff$. This indicates that the existence of a
solution seems to be robust with respect to changes in the
truncation. We have visualised this behaviour in
\autoref{fig:Solutions-eom}.

The discussion of the equation of motion leads us directly to a
specific observable: the effective action, evaluated on the equation
of motion. In standard quantum field theories, this is the free energy,
and it is gauge and parameterisation independent. For the present
approach, this is discussed in \autoref{sec:FRG}. Therefore we expect
only a mild dependence on these choices within sensible approximations
to the full effective action. Indeed, this has been observed in the
background field approximation \cite{Benedetti:2011ct,Falls:2014zba}.
It would be interesting to see whether this property also 
holds in the present approach that goes beyond the background field
approximation. At finite cutoff, this investigation can be done by
studying the gauge and parameterisation independence of the effective
action evaluated on the quantum equation of motion. This will be
discussed elsewhere.

Possible improvements of the present work involve the inclusion of
momentum- and curvature-dependent anomalous dimensions as well as the
inclusion of further $R^2$- and $R_{\mu\nu}^2$-tensor structures in
the generating vertices.  It would be very interesting to extend the
present work to more general backgrounds. Moreover, the present
approach also allows us to take the limit $k\to 0$. This allows us,
for the first time, to discuss asymptotically safe physics
directly for the physically relevant cutoff scale $k=0$. Applications 
range from asymptotically safe cosmology with the quantum $f(R)$
potential as well as the UV behaviour and phenomenology of the
asymptotically safe (extensions of the) standard model.  We hope to
report on these applications soon.

\vspace{.3cm}
\noindent {\bf Acknowledgements} We thank B.~Knorr, S.~Lippoldt,
T.~Morris and C.~Wetterich for discussions.  NC acknowledges funding
from the DFG under the Emmy Noether program, grant no.~Ei-1037-1, and
MR from IMPRS-PTFS.  This work is supported by the Helmholtz Alliance
HA216/EMMI and by ERC-AdG-290623. It is part of
and supported by the DFG Collaborative Research Centre ”SFB 1225
(ISOQUANT)”.

\appendix
%%%%%%%%%%%%%
\begin{figure*}[tbp]
\centering
\includegraphics[width=\textwidth]{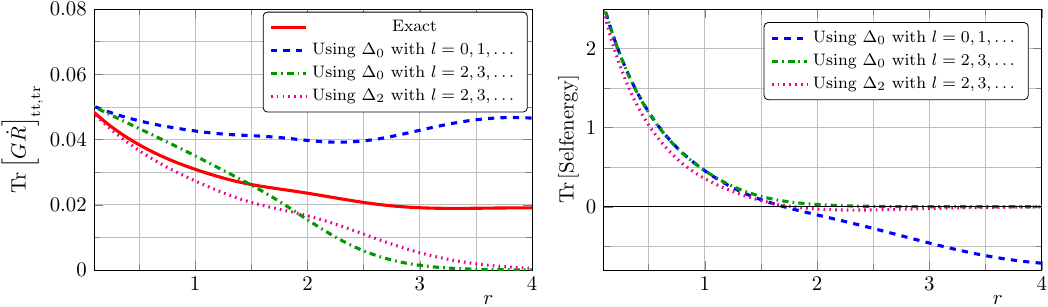}
\caption{Comparison of the trace evaluation using different Laplacians and 
starting with different eigenvalues.
In particular, we compare the spin-two Laplacian $\Delta_{2}$ 
and the spin-zero Laplacian $\Delta_{0}$
and further, we start once from the zero mode and 
once start from the $l=2$ mode.
In the left panel, we display the background flow $\tr\left[G\partial_t R \right]$ 
of the combined transverse traceless and trace mode, where also the exact solution is computed.
In the right panel, we display the self-energy diagram of the two-point function, 
which is the second diagram in \autoref{fig:flow_n_point}.
From these results, we infer that this particular approximation is qualitatively 
reliable in the range $r<2$.}
\label{fig:Comparison}
\end{figure*}
%%%%%%%%%%%%%

\section{Propagator}
\label{app:propagator}
We use the standard York decomposition to invert the two-point functions.
The York-decomposition for the graviton is given by
\begin{align}
 h_{\mu\nu}=  h^{\TT}_{\mu\nu} + \frac{1}{d} \bar g_{\mu\nu} h^{\Tr} + 2 
\bar{\nabla}_{(\mu}\xi_{\nu)}+
 \left(\bar{\nabla}_{\mu}\bar{\nabla}_{\nu}-\frac{\bar 
g_{\mu\nu}}{d}\bar{\nabla}^2\right) \sigma\,. \label{eq:york-decomposition}
\end{align}
and for the ghost by
\begin{align}
 c_\mu = c_\mu^{\text{T}} + \bar{\nabla}_{\mu} \eta\,,
\end{align}
and analogously for the anti ghost.
With the field redefinitions according to 
\cite{Dou:1997fg,Lauscher:2001ya,Gies:2015tca}
\begin{align}
 \xi^\mu &\rightarrow  \01{\sqrt{\bar \Delta - \0{\bar R}4}} \xi^\mu\,, \notag\\
 \sigma &\rightarrow \01{\sqrt{\bar \Delta^2 - \bar \Delta\0{\bar R}3}} \sigma\,, \notag\\
 \eta &\rightarrow \01{\bar \Delta} \eta\,,
\end{align}
we cancel the non-trivial Jacobians and achieve that all field modes have the 
same mass dimension.
We choose the gauge $\alpha=\beta=0$ and choose the regulator proportional to the two-point function
\begin{align}
 R_k = \Gamma_k^{(2)}(\Lambda =0,\,R= 0) \cdot r_k\!\left(p^2\right)\,.
\end{align}
Here and in the following in this appendix, $p^2$ always refers 
to the dimensionless spectral values of the scalar Laplacian.
For the regulator shape function $r_k$, we choose an exponential regulator
\begin{align}
  \label{eq:exp-reg}
  r_k(x) = \0{\E^{-x^2}}x \,.
\end{align}
The propagator has the form 
\begin{align}
 G&= \frac{32\pi}{Z_h} \begin{pmatrix}
 \01{p^2\left(1+r_k\left(p^2\right)\right) + \mu + \023 r}
 & 0 & 0 & 0 \\
 0 & 0 & 0 & 0 \\
 0 & 0 & 
 \0{-\083}{p^2\left(1+r_k\left(p^2\right)\right) + \023 \mu} & 0  \\
 0 & 0 & 0 & 0 \\
\end{pmatrix},
\label{eq:propagator}
\end{align}
where the first entry is the transverse traceless mode and the third entry is the trace mode.
All other modes vanish due to Landau gauge, $\alpha=0$.
Furthermore, we get the following expressions for the background flow of the different graviton modes,
where still the spectral sum/integral or heat-kernel expansion has to be performed,
\begin{align}
 \012 \tr[G\partial_t R]_{h_{\text{tt}}} &= \frac{r^2}{768 \pi^2} 
 \frac{ p^2 \left(\partial_t r_k \left(p^2\right)-\eta_h r_k\left(p^2\right)\right)}
 { p^2 \left(1+ r_k\left(p^2\right)\right)+ \mu+\023 r} \,,\notag\\
 \012 \tr[G\partial_t R]_{\xi} &= \frac{r^2}{768 \pi^2}
 \frac{ p^2 \left(\partial_t r_k\left(p^2\right)-\eta_h r_k\left(p^2\right)\right)}
 {p^2 \left( 1 + r_k\left(p^2\right) \right) - \014 r}  \,,\notag\\
 \012 \tr[G\partial_t R]_{h_{\text{tr}}} &= \frac{r^2}{768 \pi^2}
 \frac{  p^2 \left(\partial_t r_k\left(p^2\right)-\eta_h r_k\left(p^2\right)\right)}
 { p^2\left( 1+ r_k\left(p^2\right)\right)+ \023 \mu} \,,\notag\\
 \012 \tr[G\partial_t R]_{\sigma} &= \frac{r^2}{768 \pi^2}
 \frac{ p^2 \left(\partial_t r_k\left(p^2\right)-\eta_h r_k\left(p^2\right)\right)}
 { p^2 \left( 1+  r_k\left(p^2\right)\right) - \013 r}  \,.
\end{align}
And for the ghosts
\begin{align}
 - \tr[G\partial_t R]_{c} &= -\frac{r^2}{384 \pi^2}
 \frac{ p^2 \left(\partial_t r_k\left(p^2\right)-\eta_c r_k\left(p^2\right)\right)}
 {p^2 \left(r_k\left(p^2\right)+1\right)-\frac{r}{4}}  \,,\notag\\
 - \tr[G\partial_t R]_{\eta} &= -\frac{r^2}{384 \pi^2}
 \frac{ p^2 \left(\partial_t r_k\left(p^2\right)-\eta_c r_k\left(p^2\right)\right)}
 { p^2 \left(r_k\left(p^2\right)+1\right)- \013 r} \,.
\end{align}

\section{Flow equations}
\label{app:flow_eq}
The flow equation for the transverse-traceless part of the graviton two-point 
function is given by
\begin{align} \label{eq:flow-gamma2h}
\01{32\pi} \partial_t \left(Z_h k^2 \left(\mu + p^2 + \023  
r \right)\right) = k^2 Z_h \Flow_\TT^{(2h)}(p^2) \,.
\end{align}
Here we suppressed the dependences of the couplings on e.g.\ background curvature $r$ or
spectral values $p^2$ to improve readability.
All dependences are as in \autoref{sec:vertex_dressing}.
The expression $\Flow$ is used as in \cite{Denz:2016qks} and 
stands here and in the following for the
dimensionless right-hand side of the flow equation divided by
appropriate powers of the wave-function renormalisations.
The superscript specifies the $n$-point function, 
while the subscript refers to the tensor projection.

%%%%%%%%%%%%%
\begin{figure*}[tbp]
\centering
\includegraphics[width=\textwidth]{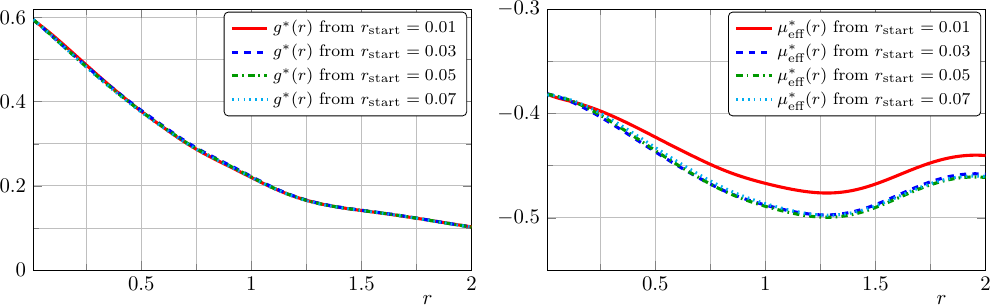}
\caption{
Comparison of fixed-point functions with initial condition
at different curvature values $r_\start \in\{0.01,\,0.03,\,0.05,\,0.07\}$.
In the left panel we compare the fixed point functions of the Newton's coupling 
$g(r)$ and in the right panel the effective graviton mass parameter $\mu_\eff(r)= \mu(r) +\023 r$.
Both fixed-point functions show only a small dependence on the initial condition.
All initial conditions are determined by $g^*_i(r_\start) = g^*_{i,0} + r_\start\, g^*_{i,1}$,
where the zeroth and linear order in $r$ of the couplings are given by \eqref{eq:UV-FP-const} and \eqref{eq:UV-FP-lin}.
}
\label{fig:dependence-startvalues}
\end{figure*}
%%%%%%%%%%%%%

From \eqref{eq:flow-gamma2h} we obtain the flow equation 
for the transverse traceless graviton mass parameter
\begin{align}
 \partial_t \mu  =& (\eta_h - 2) \mu + \023 \eta_h  r + 2  r 
\mu'+ 32\pi \Flow_\TT^{(2h)}(p^2=0) \,, \label{eq:mu}
\end{align}
where the $'$ refers to a derivative with respect to $r$.
The graviton three-point function is projected in straight analogy to the flat 
computation \cite{Christiansen:2015rva}. We focus on the transverse traceless part
and define the two projection operators 
$\Pi_\Lambda$ and $\Pi_G$ as
\begin{align}
 \Pi_\Lambda &= \Pi_\TT^3 \circ S_{\text{EH}}^{(3h)} (\Lambda=1,p^2=0,r=0) \,, \notag\\
 \Pi_G &=  \Pi_\TT^3 \circ S_{\text{EH}}^{(3h)} (\Lambda=0,p^2=1,r=0) \,,
\end{align}
which we use for the projection on $\lambda_3$ and $g$, respectively.
The resulting flow equations are
\begin{widetext}
\begin{align}
 \partial_t \left(Z_h^{3/2} k^2 \sqrt{g} \left( \05{2304} r + 
\05{192}\lambda_3 - \09{4096}p^2  \right)\right) &= k^2 Z_h^{3/2}
\Flow_{\Lambda}^{(3h)}(p^2)\,, \notag\\
 \partial_t \left(Z_h^{3/2} k^2 \sqrt{g} \left( - \03{16384}  r - 
\09{4096}\lambda_3 + \0{171}{32768}p^2 \right)\right) &= k^2 
Z_h^{3/2} \Flow_{G}^{(3h)}(p^2)\,.
\end{align}
\end{widetext}
The flow of $\lambda_3$ is extracted at vanishing spectral value $p^2=0$, 
while the flow of $g$ is extracted with a derivative with
respect to the dimensionless spectral value $p^2$ at $p=0$.
The result is
\begin{widetext}
\begin{align}
 \partial_t \lambda_3 =& -2\lambda_3 + 2  r \lambda_3'  + 
\left(\032\eta_h + \012\0{2g-\partial_t g + 2  r g' }g 
\right)\left(\lambda_3 + \01{12}  r \right) + 
\0{3}{80}\0{(32\pi)^2}{\sqrt{g}k}\text{Flow}_{\Lambda}^{(3h)}(p^2=0) \,,
\label{eq:lam3-dot} \\
 \partial_t g =& 2g + 2  r g' + 3 \eta_h g 
-\0{24}{19}\left(\partial_{p^2} \eta_h \bigg|_{p^2=0}\right)\left(\lambda_3 + 
\01{12}  r \right) g + \0{64}{171} (32\pi)^2\sqrt{g}k 
\partial_{p^2}\text{Flow}^{(3h)}_{G} \bigg|_{p^2=0} \,. \label{eq:g-dot}
\end{align}
\end{widetext}

The derivation of the flow equations in this section 
required contractions of very large tensor structures.
These contractions were computed with the help of the symbolic
manipulation system {\small \emph{FORM}}~\cite{Vermaseren:2000nd,Kuipers:2012rf}. 
We furthermore used the Mathematica packages \emph{xPert}~\cite{xPert}
for the generation of vertex functions,
and the \emph{FormTracer}~\cite{Cyrol:2016zqb} to trace diagrams.

\section{Check of approximations} 
\label{app:approximations}
In \autoref{sec:spectral_decomposition} we have explained that all
vertices in a curved background contain uncontracted covariant
derivatives. We have circumvented this issue by using
the approximation displayed in \eqref{eq:flatproduct}.  This problem
reoccurs during the contraction of the diagrams since the usual
York-decomposition projection operators $\Pi_i$ are needed, with
$i\in\{\TT,\Tr,\dots\}$. The projection operators are functions of the
background Laplacian and the background covariant derivative 
$\Pi_i(\bar\Delta,\bar\nabla)$, where
the latter covariant derivatives are again approximated by
\eqref{eq:symmetrisation} and \eqref{eq:flatproduct}.  This, however,
causes us to mix up the different spin Laplacians, spin-two
$\Delta_{2}$ and spin-zero $\Delta_{0}$.  Other Laplacians do not
occur since the graviton propagator only has a non-vanishing
transverse traceless and trace mode.  In this work, we choose to
use the spin-zero Laplacian.

For the background flow, this mixing of Laplacians does not occur since the
propagator is not a function of the covariant derivative. Hence we
use the background flow to estimate the error of our approximation.
Here we focus on the transverse traceless and the trace part since
these are the relevant modes in the fluctuation computation. The exact
result with our regulator is given by
\begin{align} \label{eq:backgr-exact}
 \tr \left[ G \partial_t R \right]_{\TT,\Tr} =& 
 \sum_{\ell=2}^{\ell_{\max}} m_2(\ell) (G \partial_t R)_\TT (\Delta_{2}(l)) \notag\\ 
 &\quad+ \sum_{\ell=0}^{\ell_{\max}} m_0(\ell) (G \partial_t R)_\Tr (\Delta_{0}(l))\,,
\end{align}
while we compare it to the approximations
\begin{align}
\text{\eqref{eq:backgr-exact}}
 \approx& \sum_{\ell=0}^{\ell_{\max}} m_0(\ell) \left( 5 (G \partial_t R)_\TT  
 +(G \partial_t R)_\Tr\right) \left(\Delta_{0}(l)\right) \label{eq:backgr-approx1} \\
 \approx& \sum_{\ell=2}^{\ell_{\max}} m_0(\ell) \left( 5 (G \partial_t R)_\TT 
 +(G \partial_t R)_\Tr  \right)\left(\Delta_{0}(l)\right) \label{eq:backgr-approx2} \\
  \approx& \sum_{\ell=2}^{\ell_{\max}} m_2(\ell) \left( (G \partial_t R)_\TT 
  + \015(G \partial_t R)_\Tr  \right)\left(\Delta_{2}(l)\right)\,. \label{eq:backgr-approx3}
\end{align}
Here $\ell_{\max}$ is chosen such that the trace is fully converged in the investigated
curvature range and the factors 5 and $\015$ appear due to the five transverse traceless modes
compared to the one trace mode.

The results are shown in \autoref{fig:Comparison} in the left panel.
For small background curvature, all results agree qualitatively well.
For large background curvature, the difference is becoming more significant.
This can be easily understood:
in the exact result \eqref{eq:backgr-exact}, only the trace mode is equipped with a zero mode,
while in the first approximation \eqref{eq:backgr-approx1} all modes are equipped with a zero mode.
In contrast in the second and third approximation, \eqref{eq:backgr-approx2} and \eqref{eq:backgr-approx3},
no mode is equipped with a zero mode.
The zero modes dominate for large curvature and thus it is clear that the approximation fails in this regime.

In other words, the symmetrised products of covariant derivatives in the projectors are 
effectively commuting in our approximation. The transverse-traceless projection  
traces out the degrees of freedom of the transverse-traceless mode and leaves us with a scalar 
quantity. With this approximation, there is an ambiguity related to the 
Laplace operator, which can be chosen as the spin-zero or spin-two Laplacian.
As already mentioned we choose to use the spin-zero Laplacian without zero modes, 
i.e.\ approximation \eqref{eq:backgr-approx2}.

In the right panel of \autoref{fig:Comparison}, we compare these different choices
for one particular diagram of the graviton two-point function,
where the exact result is not available within our truncation.
We observe that the results are almost identical for small curvature, 
i.e.\ $r<2$.
For $r>2$ the results differ qualitatively due to the different treatment of the 
zero modes.
We conclude that the validity of our approximation is bound by $r<2$.

\section{Insensitivity on initial conditions} 
\label{app:spectralheat}
As explained in \autoref{sec:FlowTrace}, we have to give initial conditions 
to the beta function since they are the first-order linear differential equations.
In principle, the initial condition has to be given at vanishing curvature $r=0$ 
since there are the divergences of the differential equations.
However, the spectral sum converges only point-wise and 
the number of modes that have to be included grows exponentially towards $r\to 0$.
Consequently we give the initial conditions at some finite $r_\start$ that should be close to $r=0$.
The value there is obtained by expanding the heat kernel expansion \eqref{eq:heat-kernel_expansion}.
One can then check that the spectral sum and heat kernel agree 
in the small background curvature regime where both methods converge \cite{Kevin-in-prep,Phd-thesis-Raul}.
In this appendix, we discuss the sensitivity of the fixed point functions to the choice of $r_\start$.

The initial condition for some coupling $g_i$ is determined 
from the zero and first order of the heat-kernel expansion around $r=0$, 
i.e.\ by $g^*_i(r_\start) = g^*_{i,0} + r_\start\, g^*_{i,1}$ 
where $g^*_{i,0}$ and $g^*_{i,1}$ are determined by the heat-kernel computation 
and the solutions are displayed in \eqref{eq:UV-FP-const} and \eqref{eq:UV-FP-lin}.
On the one hand, the quality of this initial condition gets worse for large $r_\start$ 
since this is a linear approximation of the curvature dependence of the couplings.
On the other hand, the quality also gets worse for too small $r_\start$ 
since we are too close to the singularity at $r=0$.
Consequently, we have to find a region in between where the fixed point functions 
for the couplings are stable against small variations of $r_\start$.

From the chosen $r_\start$ we integrate the differential equations upwards to large $r$.
Integrating down would quickly run into the singularity at $r=0$.
In \autoref{fig:dependence-startvalues} we display the resulting fixed point functions for
$g^*(r)$ and $\mu^*_\eff(r)$ for different choices of $r_\start\in\{0.01,\,0.03,\,0.05,\,0.07\}$.
We observe that the fixed point functions for $g^*(r)$ (left panel of \autoref{fig:dependence-startvalues})
agree almost perfectly for all chosen start values.
For the fixed point functions of $\mu^*_\eff(r)$ (right panel of \autoref{fig:dependence-startvalues})
we observe larger, but still small deviations.
Only for $r_\start = 0.01$ the deviations are significant.
We conclude that this start value is too close to the singularity at $r=0$.
The results in this work were computed with $r_\start = 0.03$.

\bibliography{flatgravity}

\begin{thebibliography}{129}%
\makeatletter
\providecommand \@ifxundefined [1]{%
 \@ifx{#1\undefined}
}%
\providecommand \@ifnum [1]{%
 \ifnum #1\expandafter \@firstoftwo
 \else \expandafter \@secondoftwo
 \fi
}%
\providecommand \@ifx [1]{%
 \ifx #1\expandafter \@firstoftwo
 \else \expandafter \@secondoftwo
 \fi
}%
\providecommand \natexlab [1]{#1}%
\providecommand \enquote  [1]{``#1''}%
\providecommand \bibnamefont  [1]{#1}%
\providecommand \bibfnamefont [1]{#1}%
\providecommand \citenamefont [1]{#1}%
\providecommand \href@noop [0]{\@secondoftwo}%
\providecommand \href [0]{\begingroup \@sanitize@url \@href}%
\providecommand \@href[1]{\@@startlink{#1}\@@href}%
\providecommand \@@href[1]{\endgroup#1\@@endlink}%
\providecommand \@sanitize@url [0]{\catcode `\\12\catcode `\$12\catcode
  `\&12\catcode `\#12\catcode `\^12\catcode `\_12\catcode `\%12\relax}%
\providecommand \@@startlink[1]{}%
\providecommand \@@endlink[0]{}%
\providecommand \url  [0]{\begingroup\@sanitize@url \@url }%
\providecommand \@url [1]{\endgroup\@href {#1}{\urlprefix }}%
\providecommand \urlprefix  [0]{URL }%
\providecommand \Eprint [0]{\href }%
\providecommand \doibase [0]{http://dx.doi.org/}%
\providecommand \selectlanguage [0]{\@gobble}%
\providecommand \bibinfo  [0]{\@secondoftwo}%
\providecommand \bibfield  [0]{\@secondoftwo}%
\providecommand \translation [1]{[#1]}%
\providecommand \BibitemOpen [0]{}%
\providecommand \bibitemStop [0]{}%
\providecommand \bibitemNoStop [0]{.\EOS\space}%
\providecommand \EOS [0]{\spacefactor3000\relax}%
\providecommand \BibitemShut  [1]{\csname bibitem#1\endcsname}%
\let\auto@bib@innerbib\@empty
%</preamble>
\bibitem [{\citenamefont {Weinberg}(1979)}]{Weinberg:1980gg}%
  \BibitemOpen
  \bibfield  {author} {\bibinfo {author} {\bibfnamefont {S.}~\bibnamefont
  {Weinberg}},\ }\href@noop {} {\bibfield  {journal} {\bibinfo  {journal}
  {General Relativity: An Einstein centenary survey, Eds. Hawking, S.W.,
  Israel, W; Cambridge University Press}\ ,\ \bibinfo {pages} {790}} (\bibinfo
  {year} {1979})}\BibitemShut {NoStop}%
%\%CITATION = INSPIRE-159043;\%\%
\bibitem [{\citenamefont {Reuter}(1998)}]{Reuter:1996cp}%
  \BibitemOpen
  \bibfield  {author} {\bibinfo {author} {\bibfnamefont {M.}~\bibnamefont
  {Reuter}},\ }\href {\doibase 10.1103/PhysRevD.57.971} {\bibfield  {journal}
  {\bibinfo  {journal} {Phys. Rev.}\ }\textbf {\bibinfo {volume} {D57}},\
  \bibinfo {pages} {971} (\bibinfo {year} {1998})},\ \Eprint
  {http://arxiv.org/abs/hep-th/9605030} {arXiv:hep-th/9605030} \BibitemShut
  {NoStop}%
%\%CITATION = HEP-TH/9605030;\%\%
\bibitem [{\citenamefont {Christiansen}\ \emph {et~al.}(2014)\citenamefont
  {Christiansen}, \citenamefont {Litim}, \citenamefont {Pawlowski},\ and\
  \citenamefont {Rodigast}}]{Christiansen:2012rx}%
  \BibitemOpen
  \bibfield  {author} {\bibinfo {author} {\bibfnamefont {N.}~\bibnamefont
  {Christiansen}}, \bibinfo {author} {\bibfnamefont {D.~F.}\ \bibnamefont
  {Litim}}, \bibinfo {author} {\bibfnamefont {J.~M.}\ \bibnamefont
  {Pawlowski}}, \ and\ \bibinfo {author} {\bibfnamefont {A.}~\bibnamefont
  {Rodigast}},\ }\href {\doibase 10.1016/j.physletb.2013.11.025} {\bibfield
  {journal} {\bibinfo  {journal} {Phys.Lett.}\ }\textbf {\bibinfo {volume}
  {B728}},\ \bibinfo {pages} {114} (\bibinfo {year} {2014})},\ \Eprint
  {http://arxiv.org/abs/1209.4038} {arXiv:1209.4038 [hep-th]} \BibitemShut
  {NoStop}%
%\%CITATION = ARXIV:1209.4038;\%\%
\bibitem [{\citenamefont {Codello}\ \emph {et~al.}(2014)\citenamefont
  {Codello}, \citenamefont {D'Odorico},\ and\ \citenamefont
  {Pagani}}]{Codello:2013fpa}%
  \BibitemOpen
  \bibfield  {author} {\bibinfo {author} {\bibfnamefont {A.}~\bibnamefont
  {Codello}}, \bibinfo {author} {\bibfnamefont {G.}~\bibnamefont {D'Odorico}},
  \ and\ \bibinfo {author} {\bibfnamefont {C.}~\bibnamefont {Pagani}},\ }\href
  {\doibase 10.1103/PhysRevD.89.081701} {\bibfield  {journal} {\bibinfo
  {journal} {Phys. Rev.}\ }\textbf {\bibinfo {volume} {D89}},\ \bibinfo {pages}
  {081701} (\bibinfo {year} {2014})},\ \Eprint {http://arxiv.org/abs/1304.4777}
  {arXiv:1304.4777 [gr-qc]} \BibitemShut {NoStop}%
%\%CITATION = ARXIV:1304.4777;\%\%
\bibitem [{\citenamefont {Christiansen}\ \emph {et~al.}(2016)\citenamefont
  {Christiansen}, \citenamefont {Knorr}, \citenamefont {Pawlowski},\ and\
  \citenamefont {Rodigast}}]{Christiansen:2014raa}%
  \BibitemOpen
  \bibfield  {author} {\bibinfo {author} {\bibfnamefont {N.}~\bibnamefont
  {Christiansen}}, \bibinfo {author} {\bibfnamefont {B.}~\bibnamefont {Knorr}},
  \bibinfo {author} {\bibfnamefont {J.~M.}\ \bibnamefont {Pawlowski}}, \ and\
  \bibinfo {author} {\bibfnamefont {A.}~\bibnamefont {Rodigast}},\ }\href
  {\doibase 10.1103/PhysRevD.93.044036} {\bibfield  {journal} {\bibinfo
  {journal} {Phys. Rev.}\ }\textbf {\bibinfo {volume} {D93}},\ \bibinfo {pages}
  {044036} (\bibinfo {year} {2016})},\ \Eprint {http://arxiv.org/abs/1403.1232}
  {arXiv:1403.1232 [hep-th]} \BibitemShut {NoStop}%
%\%CITATION = ARXIV:1403.1232;\%\%
\bibitem [{\citenamefont {Christiansen}\ \emph {et~al.}(2015)\citenamefont
  {Christiansen}, \citenamefont {Knorr}, \citenamefont {Meibohm}, \citenamefont
  {Pawlowski},\ and\ \citenamefont {Reichert}}]{Christiansen:2015rva}%
  \BibitemOpen
  \bibfield  {author} {\bibinfo {author} {\bibfnamefont {N.}~\bibnamefont
  {Christiansen}}, \bibinfo {author} {\bibfnamefont {B.}~\bibnamefont {Knorr}},
  \bibinfo {author} {\bibfnamefont {J.}~\bibnamefont {Meibohm}}, \bibinfo
  {author} {\bibfnamefont {J.~M.}\ \bibnamefont {Pawlowski}}, \ and\ \bibinfo
  {author} {\bibfnamefont {M.}~\bibnamefont {Reichert}},\ }\href {\doibase
  10.1103/PhysRevD.92.121501} {\bibfield  {journal} {\bibinfo  {journal} {Phys.
  Rev.}\ }\textbf {\bibinfo {volume} {D92}},\ \bibinfo {pages} {121501}
  (\bibinfo {year} {2015})},\ \Eprint {http://arxiv.org/abs/1506.07016}
  {arXiv:1506.07016 [hep-th]} \BibitemShut {NoStop}%
%\%CITATION = ARXIV:1506.07016;\%\%
\bibitem [{\citenamefont {Meibohm}\ \emph {et~al.}(2016)\citenamefont
  {Meibohm}, \citenamefont {Pawlowski},\ and\ \citenamefont
  {Reichert}}]{Meibohm:2015twa}%
  \BibitemOpen
  \bibfield  {author} {\bibinfo {author} {\bibfnamefont {J.}~\bibnamefont
  {Meibohm}}, \bibinfo {author} {\bibfnamefont {J.~M.}\ \bibnamefont
  {Pawlowski}}, \ and\ \bibinfo {author} {\bibfnamefont {M.}~\bibnamefont
  {Reichert}},\ }\href {\doibase 10.1103/PhysRevD.93.084035} {\bibfield
  {journal} {\bibinfo  {journal} {Phys. Rev.}\ }\textbf {\bibinfo {volume}
  {D93}},\ \bibinfo {pages} {084035} (\bibinfo {year} {2016})},\ \Eprint
  {http://arxiv.org/abs/1510.07018} {arXiv:1510.07018 [hep-th]} \BibitemShut
  {NoStop}%
%\%CITATION = ARXIV:1510.07018;\%\%
\bibitem [{\citenamefont {Meibohm}\ and\ \citenamefont
  {Pawlowski}(2016)}]{Meibohm:2016mkp}%
  \BibitemOpen
  \bibfield  {author} {\bibinfo {author} {\bibfnamefont {J.}~\bibnamefont
  {Meibohm}}\ and\ \bibinfo {author} {\bibfnamefont {J.~M.}\ \bibnamefont
  {Pawlowski}},\ }\href {\doibase 10.1140/epjc/s10052-016-4132-7} {\bibfield
  {journal} {\bibinfo  {journal} {Eur. Phys. J.}\ }\textbf {\bibinfo {volume}
  {C76}},\ \bibinfo {pages} {285} (\bibinfo {year} {2016})},\ \Eprint
  {http://arxiv.org/abs/1601.04597} {arXiv:1601.04597 [hep-th]} \BibitemShut
  {NoStop}%
%\%CITATION = ARXIV:1601.04597;\%\%
\bibitem [{\citenamefont {Henz}\ \emph {et~al.}(2017)\citenamefont {Henz},
  \citenamefont {Pawlowski},\ and\ \citenamefont {Wetterich}}]{Henz:2016aoh}%
  \BibitemOpen
  \bibfield  {author} {\bibinfo {author} {\bibfnamefont {T.}~\bibnamefont
  {Henz}}, \bibinfo {author} {\bibfnamefont {J.~M.}\ \bibnamefont {Pawlowski}},
  \ and\ \bibinfo {author} {\bibfnamefont {C.}~\bibnamefont {Wetterich}},\
  }\href {\doibase 10.1016/j.physletb.2017.01.057} {\bibfield  {journal}
  {\bibinfo  {journal} {Phys. Lett.}\ }\textbf {\bibinfo {volume} {B769}},\
  \bibinfo {pages} {105} (\bibinfo {year} {2017})},\ \Eprint
  {http://arxiv.org/abs/1605.01858} {arXiv:1605.01858 [hep-th]} \BibitemShut
  {NoStop}%
%\%CITATION = ARXIV:1605.01858;\%\%
\bibitem [{\citenamefont {Christiansen}(2016)}]{Christiansen:2016sjn}%
  \BibitemOpen
  \bibfield  {author} {\bibinfo {author} {\bibfnamefont {N.}~\bibnamefont
  {Christiansen}},\ }\href@noop {} {\  (\bibinfo {year} {2016})},\ \Eprint
  {http://arxiv.org/abs/1612.06223} {arXiv:1612.06223 [hep-th]} \BibitemShut
  {NoStop}%
%\%CITATION = ARXIV:1612.06223;\%\%
\bibitem [{\citenamefont {Denz}\ \emph {et~al.}(2018)\citenamefont {Denz},
  \citenamefont {Pawlowski},\ and\ \citenamefont {Reichert}}]{Denz:2016qks}%
  \BibitemOpen
  \bibfield  {author} {\bibinfo {author} {\bibfnamefont {T.}~\bibnamefont
  {Denz}}, \bibinfo {author} {\bibfnamefont {J.~M.}\ \bibnamefont {Pawlowski}},
  \ and\ \bibinfo {author} {\bibfnamefont {M.}~\bibnamefont {Reichert}},\
  }\href {\doibase 10.1140/epjc/s10052-018-5806-0} {\bibfield  {journal}
  {\bibinfo  {journal} {Eur. Phys. J.}\ }\textbf {\bibinfo {volume} {C78}},\
  \bibinfo {pages} {336} (\bibinfo {year} {2018})},\ \Eprint
  {http://arxiv.org/abs/1612.07315} {arXiv:1612.07315 [hep-th]} \BibitemShut
  {NoStop}%
%%CITATION = ARXIV:1612.07315;%%
\bibitem [{\citenamefont {Knorr}\ and\ \citenamefont
  {Lippoldt}(2017)}]{Knorr:2017fus}%
  \BibitemOpen
  \bibfield  {author} {\bibinfo {author} {\bibfnamefont {B.}~\bibnamefont
  {Knorr}}\ and\ \bibinfo {author} {\bibfnamefont {S.}~\bibnamefont
  {Lippoldt}},\ }\href {\doibase 10.1103/PhysRevD.96.065020} {\bibfield
  {journal} {\bibinfo  {journal} {Phys. Rev.}\ }\textbf {\bibinfo {volume}
  {D96}},\ \bibinfo {pages} {065020} (\bibinfo {year} {2017})},\ \Eprint
  {http://arxiv.org/abs/1707.01397} {arXiv:1707.01397 [hep-th]} \BibitemShut
  {NoStop}%
%\%CITATION = ARXIV:1707.01397;\%\%
\bibitem [{\citenamefont {Christiansen}\ \emph {et~al.}(2018)\citenamefont
  {Christiansen}, \citenamefont {Litim}, \citenamefont {Pawlowski},\ and\
  \citenamefont {Reichert}}]{Christiansen:2017cxa}%
  \BibitemOpen
  \bibfield  {author} {\bibinfo {author} {\bibfnamefont {N.}~\bibnamefont
  {Christiansen}}, \bibinfo {author} {\bibfnamefont {D.~F.}\ \bibnamefont
  {Litim}}, \bibinfo {author} {\bibfnamefont {J.~M.}\ \bibnamefont
  {Pawlowski}}, \ and\ \bibinfo {author} {\bibfnamefont {M.}~\bibnamefont
  {Reichert}},\ }\href {\doibase 10.1103/PhysRevD.97.106012} {\bibfield
  {journal} {\bibinfo  {journal} {Phys. Rev.}\ }\textbf {\bibinfo {volume}
  {D97}},\ \bibinfo {pages} {106012} (\bibinfo {year} {2018})},\ \Eprint
  {http://arxiv.org/abs/1710.04669} {arXiv:1710.04669 [hep-th]} \BibitemShut
  {NoStop}%
%%CITATION = ARXIV:1710.04669;%%
\bibitem [{\citenamefont {Knorr}(2018)}]{Knorr:2017mhu}%
  \BibitemOpen
  \bibfield  {author} {\bibinfo {author} {\bibfnamefont {B.}~\bibnamefont
  {Knorr}},\ }\href {\doibase 10.1088/1361-6382/aabaa0} {\bibfield  {journal}
  {\bibinfo  {journal} {Class. Quant. Grav.}\ }\textbf {\bibinfo {volume}
  {35}},\ \bibinfo {pages} {115005} (\bibinfo {year} {2018})},\ \Eprint
  {http://arxiv.org/abs/1710.07055} {arXiv:1710.07055 [hep-th]} \BibitemShut
  {NoStop}%
%%CITATION = ARXIV:1710.07055;%%
\bibitem [{\citenamefont {Donkin}\ and\ \citenamefont
  {Pawlowski}(2012)}]{Donkin:2012ud}%
  \BibitemOpen
  \bibfield  {author} {\bibinfo {author} {\bibfnamefont {I.}~\bibnamefont
  {Donkin}}\ and\ \bibinfo {author} {\bibfnamefont {J.~M.}\ \bibnamefont
  {Pawlowski}},\ }\href@noop {} {\  (\bibinfo {year} {2012})},\ \Eprint
  {http://arxiv.org/abs/1203.4207} {arXiv:1203.4207 [hep-th]} \BibitemShut
  {NoStop}%
%\%CITATION = ARXIV:1203.4207;\%\%
\bibitem [{\citenamefont {Morris}(2016)}]{Morris:2016spn}%
  \BibitemOpen
  \bibfield  {author} {\bibinfo {author} {\bibfnamefont {T.~R.}\ \bibnamefont
  {Morris}},\ }\href {\doibase 10.1007/JHEP11(2016)160} {\bibfield  {journal}
  {\bibinfo  {journal} {JHEP}\ }\textbf {\bibinfo {volume} {11}},\ \bibinfo
  {pages} {160} (\bibinfo {year} {2016})},\ \Eprint
  {http://arxiv.org/abs/1610.03081} {arXiv:1610.03081 [hep-th]} \BibitemShut
  {NoStop}%
%\%CITATION = ARXIV:1610.03081;\%\%
\bibitem [{\citenamefont {Percacci}\ and\ \citenamefont
  {Vacca}(2017)}]{Percacci:2016arh}%
  \BibitemOpen
  \bibfield  {author} {\bibinfo {author} {\bibfnamefont {R.}~\bibnamefont
  {Percacci}}\ and\ \bibinfo {author} {\bibfnamefont {G.~P.}\ \bibnamefont
  {Vacca}},\ }\href {\doibase 10.1140/epjc/s10052-017-4619-x} {\bibfield
  {journal} {\bibinfo  {journal} {Eur. Phys. J.}\ }\textbf {\bibinfo {volume}
  {C77}},\ \bibinfo {pages} {52} (\bibinfo {year} {2017})},\ \Eprint
  {http://arxiv.org/abs/1611.07005} {arXiv:1611.07005 [hep-th]} \BibitemShut
  {NoStop}%
%\%CITATION = ARXIV:1611.07005;\%\%
\bibitem [{\citenamefont {Manrique}\ and\ \citenamefont
  {Reuter}(2010)}]{Manrique:2009uh}%
  \BibitemOpen
  \bibfield  {author} {\bibinfo {author} {\bibfnamefont {E.}~\bibnamefont
  {Manrique}}\ and\ \bibinfo {author} {\bibfnamefont {M.}~\bibnamefont
  {Reuter}},\ }\href {\doibase 10.1016/j.aop.2009.11.009} {\bibfield  {journal}
  {\bibinfo  {journal} {Annals Phys.}\ }\textbf {\bibinfo {volume} {325}},\
  \bibinfo {pages} {785} (\bibinfo {year} {2010})},\ \Eprint
  {http://arxiv.org/abs/0907.2617} {arXiv:0907.2617 [gr-qc]} \BibitemShut
  {NoStop}%
%\%CITATION = 0907.2617;\%\%
\bibitem [{\citenamefont {Manrique}\ \emph
  {et~al.}(2011{\natexlab{a}})\citenamefont {Manrique}, \citenamefont
  {Reuter},\ and\ \citenamefont {Saueressig}}]{Manrique:2010mq}%
  \BibitemOpen
  \bibfield  {author} {\bibinfo {author} {\bibfnamefont {E.}~\bibnamefont
  {Manrique}}, \bibinfo {author} {\bibfnamefont {M.}~\bibnamefont {Reuter}}, \
  and\ \bibinfo {author} {\bibfnamefont {F.}~\bibnamefont {Saueressig}},\
  }\href {\doibase 10.1016/j.aop.2010.11.003} {\bibfield  {journal} {\bibinfo
  {journal} {Annals Phys.}\ }\textbf {\bibinfo {volume} {326}},\ \bibinfo
  {pages} {440} (\bibinfo {year} {2011}{\natexlab{a}})},\ \Eprint
  {http://arxiv.org/abs/1003.5129} {arXiv:1003.5129 [hep-th]} \BibitemShut
  {NoStop}%
%\%CITATION = 1003.5129;\%\%
\bibitem [{\citenamefont {Manrique}\ \emph
  {et~al.}(2011{\natexlab{b}})\citenamefont {Manrique}, \citenamefont
  {Reuter},\ and\ \citenamefont {Saueressig}}]{Manrique:2010am}%
  \BibitemOpen
  \bibfield  {author} {\bibinfo {author} {\bibfnamefont {E.}~\bibnamefont
  {Manrique}}, \bibinfo {author} {\bibfnamefont {M.}~\bibnamefont {Reuter}}, \
  and\ \bibinfo {author} {\bibfnamefont {F.}~\bibnamefont {Saueressig}},\
  }\href {\doibase 10.1016/j.aop.2010.11.006} {\bibfield  {journal} {\bibinfo
  {journal} {Annals Phys.}\ }\textbf {\bibinfo {volume} {326}},\ \bibinfo
  {pages} {463} (\bibinfo {year} {2011}{\natexlab{b}})},\ \Eprint
  {http://arxiv.org/abs/1006.0099} {arXiv:1006.0099 [hep-th]} \BibitemShut
  {NoStop}%
%\%CITATION = 1006.0099;\%\%
\bibitem [{\citenamefont {Becker}\ and\ \citenamefont
  {Reuter}(2014{\natexlab{a}})}]{Becker:2014qya}%
  \BibitemOpen
  \bibfield  {author} {\bibinfo {author} {\bibfnamefont {D.}~\bibnamefont
  {Becker}}\ and\ \bibinfo {author} {\bibfnamefont {M.}~\bibnamefont
  {Reuter}},\ }\href {\doibase 10.1016/j.aop.2014.07.023} {\bibfield  {journal}
  {\bibinfo  {journal} {Annals Phys.}\ }\textbf {\bibinfo {volume} {350}},\
  \bibinfo {pages} {225} (\bibinfo {year} {2014}{\natexlab{a}})},\ \Eprint
  {http://arxiv.org/abs/1404.4537} {arXiv:1404.4537 [hep-th]} \BibitemShut
  {NoStop}%
%\%CITATION = ARXIV:1404.4537;\%\%
\bibitem [{\citenamefont {Falkenberg}\ and\ \citenamefont
  {Odintsov}(1998)}]{Falkenberg:1996bq}%
  \BibitemOpen
  \bibfield  {author} {\bibinfo {author} {\bibfnamefont {S.}~\bibnamefont
  {Falkenberg}}\ and\ \bibinfo {author} {\bibfnamefont {S.~D.}\ \bibnamefont
  {Odintsov}},\ }\href {\doibase 10.1142/S0217751X98000263} {\bibfield
  {journal} {\bibinfo  {journal} {Int.J.Mod.Phys.}\ }\textbf {\bibinfo {volume}
  {A13}},\ \bibinfo {pages} {607} (\bibinfo {year} {1998})},\ \Eprint
  {http://arxiv.org/abs/hep-th/9612019} {arXiv:hep-th/9612019 [hep-th]}
  \BibitemShut {NoStop}%
%\%CITATION = HEP-TH/9612019;\%\%
\bibitem [{\citenamefont {Reuter}\ and\ \citenamefont
  {Saueressig}(2002)}]{Reuter:2001ag}%
  \BibitemOpen
  \bibfield  {author} {\bibinfo {author} {\bibfnamefont {M.}~\bibnamefont
  {Reuter}}\ and\ \bibinfo {author} {\bibfnamefont {F.}~\bibnamefont
  {Saueressig}},\ }\href {\doibase 10.1103/PhysRevD.65.065016} {\bibfield
  {journal} {\bibinfo  {journal} {Phys. Rev.}\ }\textbf {\bibinfo {volume}
  {D65}},\ \bibinfo {pages} {065016} (\bibinfo {year} {2002})},\ \Eprint
  {http://arxiv.org/abs/hep-th/0110054} {arXiv:hep-th/0110054 [hep-th]}
  \BibitemShut {NoStop}%
%\%CITATION = HEP-TH/0110054;\%\%
\bibitem [{\citenamefont {Lauscher}\ and\ \citenamefont
  {Reuter}(2002{\natexlab{a}})}]{Lauscher:2001ya}%
  \BibitemOpen
  \bibfield  {author} {\bibinfo {author} {\bibfnamefont {O.}~\bibnamefont
  {Lauscher}}\ and\ \bibinfo {author} {\bibfnamefont {M.}~\bibnamefont
  {Reuter}},\ }\href {\doibase 10.1103/PhysRevD.65.025013} {\bibfield
  {journal} {\bibinfo  {journal} {Phys. Rev.}\ }\textbf {\bibinfo {volume}
  {D65}},\ \bibinfo {pages} {025013} (\bibinfo {year} {2002}{\natexlab{a}})},\
  \Eprint {http://arxiv.org/abs/hep-th/0108040} {arXiv:hep-th/0108040}
  \BibitemShut {NoStop}%
%\%CITATION = HEP-TH/0108040;\%\%
\bibitem [{\citenamefont {Lauscher}\ and\ \citenamefont
  {Reuter}(2002{\natexlab{b}})}]{Lauscher:2002sq}%
  \BibitemOpen
  \bibfield  {author} {\bibinfo {author} {\bibfnamefont {O.}~\bibnamefont
  {Lauscher}}\ and\ \bibinfo {author} {\bibfnamefont {M.}~\bibnamefont
  {Reuter}},\ }\href {\doibase 10.1103/PhysRevD.66.025026} {\bibfield
  {journal} {\bibinfo  {journal} {Phys. Rev.}\ }\textbf {\bibinfo {volume}
  {D66}},\ \bibinfo {pages} {025026} (\bibinfo {year} {2002}{\natexlab{b}})},\
  \Eprint {http://arxiv.org/abs/hep-th/0205062} {arXiv:hep-th/0205062}
  \BibitemShut {NoStop}%
%\%CITATION = HEP-TH/0205062;\%\%
\bibitem [{\citenamefont {Litim}(2004)}]{Litim:2003vp}%
  \BibitemOpen
  \bibfield  {author} {\bibinfo {author} {\bibfnamefont {D.~F.}\ \bibnamefont
  {Litim}},\ }\href {\doibase 10.1103/PhysRevLett.92.201301} {\bibfield
  {journal} {\bibinfo  {journal} {Phys.Rev.Lett.}\ }\textbf {\bibinfo {volume}
  {92}},\ \bibinfo {pages} {201301} (\bibinfo {year} {2004})},\ \Eprint
  {http://arxiv.org/abs/hep-th/0312114} {arXiv:hep-th/0312114 [hep-th]}
  \BibitemShut {NoStop}%
%\%CITATION = HEP-TH/0312114;\%\%
\bibitem [{\citenamefont {Fischer}\ and\ \citenamefont
  {Litim}(2006)}]{Fischer:2006fz}%
  \BibitemOpen
  \bibfield  {author} {\bibinfo {author} {\bibfnamefont {P.}~\bibnamefont
  {Fischer}}\ and\ \bibinfo {author} {\bibfnamefont {D.~F.}\ \bibnamefont
  {Litim}},\ }\href {\doibase 10.1016/j.physletb.2006.05.073} {\bibfield
  {journal} {\bibinfo  {journal} {Phys.Lett.}\ }\textbf {\bibinfo {volume}
  {B638}},\ \bibinfo {pages} {497} (\bibinfo {year} {2006})},\ \Eprint
  {http://arxiv.org/abs/hep-th/0602203} {arXiv:hep-th/0602203 [hep-th]}
  \BibitemShut {NoStop}%
%\%CITATION = HEP-TH/0602203;\%\%
\bibitem [{\citenamefont {Codello}\ and\ \citenamefont
  {Percacci}(2006)}]{Codello:2006in}%
  \BibitemOpen
  \bibfield  {author} {\bibinfo {author} {\bibfnamefont {A.}~\bibnamefont
  {Codello}}\ and\ \bibinfo {author} {\bibfnamefont {R.}~\bibnamefont
  {Percacci}},\ }\href {\doibase 10.1103/PhysRevLett.97.221301} {\bibfield
  {journal} {\bibinfo  {journal} {Phys. Rev. Lett.}\ }\textbf {\bibinfo
  {volume} {97}},\ \bibinfo {pages} {221301} (\bibinfo {year} {2006})},\
  \Eprint {http://arxiv.org/abs/hep-th/0607128} {arXiv:hep-th/0607128}
  \BibitemShut {NoStop}%
%\%CITATION = HEP-TH/0607128;\%\%
\bibitem [{\citenamefont {Machado}\ and\ \citenamefont
  {Saueressig}(2008)}]{Machado:2007ea}%
  \BibitemOpen
  \bibfield  {author} {\bibinfo {author} {\bibfnamefont {P.~F.}\ \bibnamefont
  {Machado}}\ and\ \bibinfo {author} {\bibfnamefont {F.}~\bibnamefont
  {Saueressig}},\ }\href {\doibase 10.1103/PhysRevD.77.124045} {\bibfield
  {journal} {\bibinfo  {journal} {Phys. Rev.}\ }\textbf {\bibinfo {volume}
  {D77}},\ \bibinfo {pages} {124045} (\bibinfo {year} {2008})},\ \Eprint
  {http://arxiv.org/abs/0712.0445} {arXiv:0712.0445 [hep-th]} \BibitemShut
  {NoStop}%
%\%CITATION = 0712.0445;\%\%
\bibitem [{\citenamefont {Codello}\ \emph {et~al.}(2009)\citenamefont
  {Codello}, \citenamefont {Percacci},\ and\ \citenamefont
  {Rahmede}}]{Codello:2008vh}%
  \BibitemOpen
  \bibfield  {author} {\bibinfo {author} {\bibfnamefont {A.}~\bibnamefont
  {Codello}}, \bibinfo {author} {\bibfnamefont {R.}~\bibnamefont {Percacci}}, \
  and\ \bibinfo {author} {\bibfnamefont {C.}~\bibnamefont {Rahmede}},\ }\href
  {\doibase 10.1016/j.aop.2008.08.008} {\bibfield  {journal} {\bibinfo
  {journal} {Annals Phys.}\ }\textbf {\bibinfo {volume} {324}},\ \bibinfo
  {pages} {414} (\bibinfo {year} {2009})},\ \Eprint
  {http://arxiv.org/abs/0805.2909} {arXiv:0805.2909 [hep-th]} \BibitemShut
  {NoStop}%
%\%CITATION = 0805.2909;\%\%
\bibitem [{\citenamefont {Eichhorn}\ \emph {et~al.}(2009)\citenamefont
  {Eichhorn}, \citenamefont {Gies},\ and\ \citenamefont
  {Scherer}}]{Eichhorn:2009ah}%
  \BibitemOpen
  \bibfield  {author} {\bibinfo {author} {\bibfnamefont {A.}~\bibnamefont
  {Eichhorn}}, \bibinfo {author} {\bibfnamefont {H.}~\bibnamefont {Gies}}, \
  and\ \bibinfo {author} {\bibfnamefont {M.~M.}\ \bibnamefont {Scherer}},\
  }\href {\doibase 10.1103/PhysRevD.80.104003} {\bibfield  {journal} {\bibinfo
  {journal} {Phys. Rev.}\ }\textbf {\bibinfo {volume} {D80}},\ \bibinfo {pages}
  {104003} (\bibinfo {year} {2009})},\ \Eprint {http://arxiv.org/abs/0907.1828}
  {arXiv:0907.1828 [hep-th]} \BibitemShut {NoStop}%
%\%CITATION = 0907.1828;\%\%
\bibitem [{\citenamefont {Benedetti}\ \emph {et~al.}(2009)\citenamefont
  {Benedetti}, \citenamefont {Machado},\ and\ \citenamefont
  {Saueressig}}]{Benedetti:2009rx}%
  \BibitemOpen
  \bibfield  {author} {\bibinfo {author} {\bibfnamefont {D.}~\bibnamefont
  {Benedetti}}, \bibinfo {author} {\bibfnamefont {P.~F.}\ \bibnamefont
  {Machado}}, \ and\ \bibinfo {author} {\bibfnamefont {F.}~\bibnamefont
  {Saueressig}},\ }\href {\doibase 10.1142/S0217732309031521} {\bibfield
  {journal} {\bibinfo  {journal} {Mod. Phys. Lett.}\ }\textbf {\bibinfo
  {volume} {A24}},\ \bibinfo {pages} {2233} (\bibinfo {year} {2009})},\ \Eprint
  {http://arxiv.org/abs/0901.2984} {arXiv:0901.2984 [hep-th]} \BibitemShut
  {NoStop}%
%\%CITATION = 0901.2984;\%\%
\bibitem [{\citenamefont {Eichhorn}\ and\ \citenamefont
  {Gies}(2010)}]{Eichhorn:2010tb}%
  \BibitemOpen
  \bibfield  {author} {\bibinfo {author} {\bibfnamefont {A.}~\bibnamefont
  {Eichhorn}}\ and\ \bibinfo {author} {\bibfnamefont {H.}~\bibnamefont
  {Gies}},\ }\href {\doibase 10.1103/PhysRevD.81.104010} {\bibfield  {journal}
  {\bibinfo  {journal} {Phys. Rev.}\ }\textbf {\bibinfo {volume} {D81}},\
  \bibinfo {pages} {104010} (\bibinfo {year} {2010})},\ \Eprint
  {http://arxiv.org/abs/1001.5033} {arXiv:1001.5033 [hep-th]} \BibitemShut
  {NoStop}%
%\%CITATION = 1001.5033;\%\%
\bibitem [{\citenamefont {Groh}\ and\ \citenamefont
  {Saueressig}(2010)}]{Groh:2010ta}%
  \BibitemOpen
  \bibfield  {author} {\bibinfo {author} {\bibfnamefont {K.}~\bibnamefont
  {Groh}}\ and\ \bibinfo {author} {\bibfnamefont {F.}~\bibnamefont
  {Saueressig}},\ }\href {\doibase 10.1088/1751-8113/43/36/365403} {\bibfield
  {journal} {\bibinfo  {journal} {J. Phys.}\ }\textbf {\bibinfo {volume}
  {A43}},\ \bibinfo {pages} {365403} (\bibinfo {year} {2010})},\ \Eprint
  {http://arxiv.org/abs/1001.5032} {arXiv:1001.5032 [hep-th]} \BibitemShut
  {NoStop}%
%\%CITATION = 1001.5032;\%\%
\bibitem [{\citenamefont {Manrique}\ \emph
  {et~al.}(2011{\natexlab{c}})\citenamefont {Manrique}, \citenamefont
  {Rechenberger},\ and\ \citenamefont {Saueressig}}]{Manrique:2011jc}%
  \BibitemOpen
  \bibfield  {author} {\bibinfo {author} {\bibfnamefont {E.}~\bibnamefont
  {Manrique}}, \bibinfo {author} {\bibfnamefont {S.}~\bibnamefont
  {Rechenberger}}, \ and\ \bibinfo {author} {\bibfnamefont {F.}~\bibnamefont
  {Saueressig}},\ }\href {\doibase 10.1103/PhysRevLett.106.251302} {\bibfield
  {journal} {\bibinfo  {journal} {Phys.Rev.Lett.}\ }\textbf {\bibinfo {volume}
  {106}},\ \bibinfo {pages} {251302} (\bibinfo {year} {2011}{\natexlab{c}})},\
  \Eprint {http://arxiv.org/abs/1102.5012} {arXiv:1102.5012 [hep-th]}
  \BibitemShut {NoStop}%
%\%CITATION = ARXIV:1102.5012;\%\%
\bibitem [{\citenamefont {Benedetti}\ and\ \citenamefont
  {Caravelli}(2012)}]{Benedetti:2012dx}%
  \BibitemOpen
  \bibfield  {author} {\bibinfo {author} {\bibfnamefont {D.}~\bibnamefont
  {Benedetti}}\ and\ \bibinfo {author} {\bibfnamefont {F.}~\bibnamefont
  {Caravelli}},\ }\href {\doibase 10.1007/JHEP06(2012)017;;;;;;;;;;;;
  10.1007/JHEP10(2012)157} {\bibfield  {journal} {\bibinfo  {journal} {JHEP}\
  }\textbf {\bibinfo {volume} {06}},\ \bibinfo {pages} {017} (\bibinfo {year}
  {2012})},\ \bibinfo {note} {[Erratum: JHEP10,157(2012)]},\ \Eprint
  {http://arxiv.org/abs/1204.3541} {arXiv:1204.3541 [hep-th]} \BibitemShut
  {NoStop}%
%\%CITATION = ARXIV:1204.3541;\%\%
\bibitem [{\citenamefont {Dietz}\ and\ \citenamefont
  {Morris}(2013)}]{Dietz:2012ic}%
  \BibitemOpen
  \bibfield  {author} {\bibinfo {author} {\bibfnamefont {J.~A.}\ \bibnamefont
  {Dietz}}\ and\ \bibinfo {author} {\bibfnamefont {T.~R.}\ \bibnamefont
  {Morris}},\ }\href {\doibase 10.1007/JHEP01(2013)108} {\bibfield  {journal}
  {\bibinfo  {journal} {JHEP}\ }\textbf {\bibinfo {volume} {01}},\ \bibinfo
  {pages} {108} (\bibinfo {year} {2013})},\ \Eprint
  {http://arxiv.org/abs/1211.0955} {arXiv:1211.0955 [hep-th]} \BibitemShut
  {NoStop}%
%\%CITATION = ARXIV:1211.0955;\%\%
\bibitem [{\citenamefont {Falls}\ \emph {et~al.}(2013)\citenamefont {Falls},
  \citenamefont {Litim}, \citenamefont {Nikolakopoulos},\ and\ \citenamefont
  {Rahmede}}]{Falls:2013bv}%
  \BibitemOpen
  \bibfield  {author} {\bibinfo {author} {\bibfnamefont {K.}~\bibnamefont
  {Falls}}, \bibinfo {author} {\bibfnamefont {D.}~\bibnamefont {Litim}},
  \bibinfo {author} {\bibfnamefont {K.}~\bibnamefont {Nikolakopoulos}}, \ and\
  \bibinfo {author} {\bibfnamefont {C.}~\bibnamefont {Rahmede}},\ }\href@noop
  {} {\  (\bibinfo {year} {2013})},\ \Eprint {http://arxiv.org/abs/1301.4191}
  {arXiv:1301.4191 [hep-th]} \BibitemShut {NoStop}%
%\%CITATION = ARXIV:1301.4191;\%\%
\bibitem [{\citenamefont {Falls}\ \emph {et~al.}(2016)\citenamefont {Falls},
  \citenamefont {Litim}, \citenamefont {Nikolakopoulos},\ and\ \citenamefont
  {Rahmede}}]{Falls:2014tra}%
  \BibitemOpen
  \bibfield  {author} {\bibinfo {author} {\bibfnamefont {K.}~\bibnamefont
  {Falls}}, \bibinfo {author} {\bibfnamefont {D.~F.}\ \bibnamefont {Litim}},
  \bibinfo {author} {\bibfnamefont {K.}~\bibnamefont {Nikolakopoulos}}, \ and\
  \bibinfo {author} {\bibfnamefont {C.}~\bibnamefont {Rahmede}},\ }\href
  {\doibase 10.1103/PhysRevD.93.104022} {\bibfield  {journal} {\bibinfo
  {journal} {Phys. Rev.}\ }\textbf {\bibinfo {volume} {D93}},\ \bibinfo {pages}
  {104022} (\bibinfo {year} {2016})},\ \Eprint {http://arxiv.org/abs/1410.4815}
  {arXiv:1410.4815 [hep-th]} \BibitemShut {NoStop}%
%\%CITATION = ARXIV:1410.4815;\%\%
\bibitem [{\citenamefont {Falls}(2015{\natexlab{a}})}]{Falls:2015qga}%
  \BibitemOpen
  \bibfield  {author} {\bibinfo {author} {\bibfnamefont {K.}~\bibnamefont
  {Falls}},\ }\href {\doibase 10.1103/PhysRevD.92.124057} {\bibfield  {journal}
  {\bibinfo  {journal} {Phys. Rev.}\ }\textbf {\bibinfo {volume} {D92}},\
  \bibinfo {pages} {124057} (\bibinfo {year} {2015}{\natexlab{a}})},\ \Eprint
  {http://arxiv.org/abs/1501.05331} {arXiv:1501.05331 [hep-th]} \BibitemShut
  {NoStop}%
%\%CITATION = ARXIV:1501.05331;\%\%
\bibitem [{\citenamefont {Eichhorn}(2015)}]{Eichhorn:2015bna}%
  \BibitemOpen
  \bibfield  {author} {\bibinfo {author} {\bibfnamefont {A.}~\bibnamefont
  {Eichhorn}},\ }\href {\doibase 10.1007/JHEP04(2015)096} {\bibfield  {journal}
  {\bibinfo  {journal} {JHEP}\ }\textbf {\bibinfo {volume} {04}},\ \bibinfo
  {pages} {096} (\bibinfo {year} {2015})},\ \Eprint
  {http://arxiv.org/abs/1501.05848} {arXiv:1501.05848 [gr-qc]} \BibitemShut
  {NoStop}%
%\%CITATION = ARXIV:1501.05848;\%\%
\bibitem [{\citenamefont {Falls}(2015{\natexlab{b}})}]{Falls:2015cta}%
  \BibitemOpen
  \bibfield  {author} {\bibinfo {author} {\bibfnamefont {K.}~\bibnamefont
  {Falls}},\ }\href@noop {} {\  (\bibinfo {year} {2015}{\natexlab{b}})},\
  \Eprint {http://arxiv.org/abs/1503.06233} {arXiv:1503.06233 [hep-th]}
  \BibitemShut {NoStop}%
%\%CITATION = ARXIV:1503.06233;\%\%
\bibitem [{\citenamefont {Demmel}\ \emph {et~al.}(2015)\citenamefont {Demmel},
  \citenamefont {Saueressig},\ and\ \citenamefont {Zanusso}}]{Demmel:2015oqa}%
  \BibitemOpen
  \bibfield  {author} {\bibinfo {author} {\bibfnamefont {M.}~\bibnamefont
  {Demmel}}, \bibinfo {author} {\bibfnamefont {F.}~\bibnamefont {Saueressig}},
  \ and\ \bibinfo {author} {\bibfnamefont {O.}~\bibnamefont {Zanusso}},\ }\href
  {\doibase 10.1007/JHEP08(2015)113} {\bibfield  {journal} {\bibinfo  {journal}
  {JHEP}\ }\textbf {\bibinfo {volume} {08}},\ \bibinfo {pages} {113} (\bibinfo
  {year} {2015})},\ \Eprint {http://arxiv.org/abs/1504.07656} {arXiv:1504.07656
  [hep-th]} \BibitemShut {NoStop}%
%\%CITATION = ARXIV:1504.07656;\%\%
\bibitem [{\citenamefont {Gies}\ \emph {et~al.}(2015)\citenamefont {Gies},
  \citenamefont {Knorr},\ and\ \citenamefont {Lippoldt}}]{Gies:2015tca}%
  \BibitemOpen
  \bibfield  {author} {\bibinfo {author} {\bibfnamefont {H.}~\bibnamefont
  {Gies}}, \bibinfo {author} {\bibfnamefont {B.}~\bibnamefont {Knorr}}, \ and\
  \bibinfo {author} {\bibfnamefont {S.}~\bibnamefont {Lippoldt}},\ }\href
  {\doibase 10.1103/PhysRevD.92.084020} {\bibfield  {journal} {\bibinfo
  {journal} {Phys. Rev.}\ }\textbf {\bibinfo {volume} {D92}},\ \bibinfo {pages}
  {084020} (\bibinfo {year} {2015})},\ \Eprint
  {http://arxiv.org/abs/1507.08859} {arXiv:1507.08859 [hep-th]} \BibitemShut
  {NoStop}%
%\%CITATION = ARXIV:1507.08859;\%\%
\bibitem [{\citenamefont {Gies}\ \emph {et~al.}(2016)\citenamefont {Gies},
  \citenamefont {Knorr}, \citenamefont {Lippoldt},\ and\ \citenamefont
  {Saueressig}}]{Gies:2016con}%
  \BibitemOpen
  \bibfield  {author} {\bibinfo {author} {\bibfnamefont {H.}~\bibnamefont
  {Gies}}, \bibinfo {author} {\bibfnamefont {B.}~\bibnamefont {Knorr}},
  \bibinfo {author} {\bibfnamefont {S.}~\bibnamefont {Lippoldt}}, \ and\
  \bibinfo {author} {\bibfnamefont {F.}~\bibnamefont {Saueressig}},\ }\href
  {\doibase 10.1103/PhysRevLett.116.211302} {\bibfield  {journal} {\bibinfo
  {journal} {Phys. Rev. Lett.}\ }\textbf {\bibinfo {volume} {116}},\ \bibinfo
  {pages} {211302} (\bibinfo {year} {2016})},\ \Eprint
  {http://arxiv.org/abs/1601.01800} {arXiv:1601.01800 [hep-th]} \BibitemShut
  {NoStop}%
%\%CITATION = ARXIV:1601.01800;\%\%
\bibitem [{\citenamefont {Biemans}\ \emph
  {et~al.}(2017{\natexlab{a}})\citenamefont {Biemans}, \citenamefont
  {Platania},\ and\ \citenamefont {Saueressig}}]{Biemans:2016rvp}%
  \BibitemOpen
  \bibfield  {author} {\bibinfo {author} {\bibfnamefont {J.}~\bibnamefont
  {Biemans}}, \bibinfo {author} {\bibfnamefont {A.}~\bibnamefont {Platania}}, \
  and\ \bibinfo {author} {\bibfnamefont {F.}~\bibnamefont {Saueressig}},\
  }\href {\doibase 10.1103/PhysRevD.95.086013} {\bibfield  {journal} {\bibinfo
  {journal} {Phys. Rev.}\ }\textbf {\bibinfo {volume} {D95}},\ \bibinfo {pages}
  {086013} (\bibinfo {year} {2017}{\natexlab{a}})},\ \Eprint
  {http://arxiv.org/abs/1609.04813} {arXiv:1609.04813 [hep-th]} \BibitemShut
  {NoStop}%
%\%CITATION = ARXIV:1609.04813;\%\%
\bibitem [{\citenamefont {Falls}(2017)}]{Falls:2017cze}%
  \BibitemOpen
  \bibfield  {author} {\bibinfo {author} {\bibfnamefont {K.}~\bibnamefont
  {Falls}},\ }\href {\doibase 10.1103/PhysRevD.96.126016} {\bibfield  {journal}
  {\bibinfo  {journal} {Phys. Rev.}\ }\textbf {\bibinfo {volume} {D96}},\
  \bibinfo {pages} {126016} (\bibinfo {year} {2017})},\ \Eprint
  {http://arxiv.org/abs/1702.03577} {arXiv:1702.03577 [hep-th]} \BibitemShut
  {NoStop}%
%%CITATION = ARXIV:1702.03577;%%
\bibitem [{\citenamefont {Hamada}\ and\ \citenamefont
  {Yamada}(2017)}]{Hamada:2017rvn}%
  \BibitemOpen
  \bibfield  {author} {\bibinfo {author} {\bibfnamefont {Y.}~\bibnamefont
  {Hamada}}\ and\ \bibinfo {author} {\bibfnamefont {M.}~\bibnamefont
  {Yamada}},\ }\href {\doibase 10.1007/JHEP08(2017)070} {\bibfield  {journal}
  {\bibinfo  {journal} {JHEP}\ }\textbf {\bibinfo {volume} {08}},\ \bibinfo
  {pages} {070} (\bibinfo {year} {2017})},\ \Eprint
  {http://arxiv.org/abs/1703.09033} {arXiv:1703.09033 [hep-th]} \BibitemShut
  {NoStop}%
%\%CITATION = ARXIV:1703.09033;\%\%
\bibitem [{\citenamefont {Gonzalez-Martin}\ \emph {et~al.}(2017)\citenamefont
  {Gonzalez-Martin}, \citenamefont {Morris},\ and\ \citenamefont
  {Slade}}]{Gonzalez-Martin:2017gza}%
  \BibitemOpen
  \bibfield  {author} {\bibinfo {author} {\bibfnamefont {S.}~\bibnamefont
  {Gonzalez-Martin}}, \bibinfo {author} {\bibfnamefont {T.~R.}\ \bibnamefont
  {Morris}}, \ and\ \bibinfo {author} {\bibfnamefont {Z.~H.}\ \bibnamefont
  {Slade}},\ }\href {\doibase 10.1103/PhysRevD.95.106010} {\bibfield  {journal}
  {\bibinfo  {journal} {Phys. Rev.}\ }\textbf {\bibinfo {volume} {D95}},\
  \bibinfo {pages} {106010} (\bibinfo {year} {2017})},\ \Eprint
  {http://arxiv.org/abs/1704.08873} {arXiv:1704.08873 [hep-th]} \BibitemShut
  {NoStop}%
%\%CITATION = ARXIV:1704.08873;\%\%
\bibitem [{\citenamefont {Becker}\ \emph {et~al.}(2017)\citenamefont {Becker},
  \citenamefont {Ripken},\ and\ \citenamefont {Saueressig}}]{Becker:2017tcx}%
  \BibitemOpen
  \bibfield  {author} {\bibinfo {author} {\bibfnamefont {D.}~\bibnamefont
  {Becker}}, \bibinfo {author} {\bibfnamefont {C.}~\bibnamefont {Ripken}}, \
  and\ \bibinfo {author} {\bibfnamefont {F.}~\bibnamefont {Saueressig}},\
  }\href {\doibase 10.1007/JHEP12(2017)121} {\bibfield  {journal} {\bibinfo
  {journal} {JHEP}\ }\textbf {\bibinfo {volume} {12}},\ \bibinfo {pages} {121}
  (\bibinfo {year} {2017})},\ \Eprint {http://arxiv.org/abs/1709.09098}
  {arXiv:1709.09098 [hep-th]} \BibitemShut {NoStop}%
%%CITATION = ARXIV:1709.09098;%%
\bibitem [{\citenamefont {Dou}\ and\ \citenamefont
  {Percacci}(1998)}]{Dou:1997fg}%
  \BibitemOpen
  \bibfield  {author} {\bibinfo {author} {\bibfnamefont {D.}~\bibnamefont
  {Dou}}\ and\ \bibinfo {author} {\bibfnamefont {R.}~\bibnamefont {Percacci}},\
  }\href {\doibase 10.1088/0264-9381/15/11/011} {\bibfield  {journal} {\bibinfo
   {journal} {Class. Quant. Grav.}\ }\textbf {\bibinfo {volume} {15}},\
  \bibinfo {pages} {3449} (\bibinfo {year} {1998})},\ \Eprint
  {http://arxiv.org/abs/hep-th/9707239} {arXiv:hep-th/9707239 [hep-th]}
  \BibitemShut {NoStop}%
%\%CITATION = HEP-TH/9707239;\%\%
\bibitem [{\citenamefont {Percacci}\ and\ \citenamefont
  {Perini}(2003)}]{Percacci:2002ie}%
  \BibitemOpen
  \bibfield  {author} {\bibinfo {author} {\bibfnamefont {R.}~\bibnamefont
  {Percacci}}\ and\ \bibinfo {author} {\bibfnamefont {D.}~\bibnamefont
  {Perini}},\ }\href {\doibase 10.1103/PhysRevD.67.081503} {\bibfield
  {journal} {\bibinfo  {journal} {Phys. Rev.}\ }\textbf {\bibinfo {volume}
  {D67}},\ \bibinfo {pages} {081503} (\bibinfo {year} {2003})},\ \Eprint
  {http://arxiv.org/abs/hep-th/0207033} {arXiv:hep-th/0207033} \BibitemShut
  {NoStop}%
%\%CITATION = HEP-TH/0207033;\%\%
\bibitem [{\citenamefont {Folkerts}\ \emph {et~al.}(2012)\citenamefont
  {Folkerts}, \citenamefont {Litim},\ and\ \citenamefont
  {Pawlowski}}]{Folkerts:2011jz}%
  \BibitemOpen
  \bibfield  {author} {\bibinfo {author} {\bibfnamefont {S.}~\bibnamefont
  {Folkerts}}, \bibinfo {author} {\bibfnamefont {D.~F.}\ \bibnamefont {Litim}},
  \ and\ \bibinfo {author} {\bibfnamefont {J.~M.}\ \bibnamefont {Pawlowski}},\
  }\href {\doibase 10.1016/j.physletb.2012.02.002} {\bibfield  {journal}
  {\bibinfo  {journal} {Phys.Lett.}\ }\textbf {\bibinfo {volume} {B709}},\
  \bibinfo {pages} {234} (\bibinfo {year} {2012})},\ \Eprint
  {http://arxiv.org/abs/1101.5552} {arXiv:1101.5552 [hep-th]} \BibitemShut
  {NoStop}%
%\%CITATION = ARXIV:1101.5552;\%\%
\bibitem [{\citenamefont {Harst}\ and\ \citenamefont
  {Reuter}(2011)}]{Harst:2011zx}%
  \BibitemOpen
  \bibfield  {author} {\bibinfo {author} {\bibfnamefont {U.}~\bibnamefont
  {Harst}}\ and\ \bibinfo {author} {\bibfnamefont {M.}~\bibnamefont {Reuter}},\
  }\href {\doibase 10.1007/JHEP05(2011)119} {\bibfield  {journal} {\bibinfo
  {journal} {JHEP}\ }\textbf {\bibinfo {volume} {05}},\ \bibinfo {pages} {119}
  (\bibinfo {year} {2011})},\ \Eprint {http://arxiv.org/abs/1101.6007}
  {arXiv:1101.6007 [hep-th]} \BibitemShut {NoStop}%
%\%CITATION = 1101.6007;\%\%
\bibitem [{\citenamefont {Eichhorn}\ and\ \citenamefont
  {Gies}(2011)}]{Eichhorn:2011pc}%
  \BibitemOpen
  \bibfield  {author} {\bibinfo {author} {\bibfnamefont {A.}~\bibnamefont
  {Eichhorn}}\ and\ \bibinfo {author} {\bibfnamefont {H.}~\bibnamefont
  {Gies}},\ }\href {\doibase 10.1088/1367-2630/13/12/125012} {\bibfield
  {journal} {\bibinfo  {journal} {New J. Phys.}\ }\textbf {\bibinfo {volume}
  {13}},\ \bibinfo {pages} {125012} (\bibinfo {year} {2011})},\ \Eprint
  {http://arxiv.org/abs/1104.5366} {arXiv:1104.5366 [hep-th]} \BibitemShut
  {NoStop}%
%\%CITATION = ARXIV:1104.5366;\%\%
\bibitem [{\citenamefont {Eichhorn}(2012)}]{Eichhorn:2012va}%
  \BibitemOpen
  \bibfield  {author} {\bibinfo {author} {\bibfnamefont {A.}~\bibnamefont
  {Eichhorn}},\ }\href {\doibase 10.1103/PhysRevD.86.105021} {\bibfield
  {journal} {\bibinfo  {journal} {Phys. Rev.}\ }\textbf {\bibinfo {volume}
  {D86}},\ \bibinfo {pages} {105021} (\bibinfo {year} {2012})},\ \Eprint
  {http://arxiv.org/abs/1204.0965} {arXiv:1204.0965 [gr-qc]} \BibitemShut
  {NoStop}%
%\%CITATION = ARXIV:1204.0965;\%\%
\bibitem [{\citenamefont {Don{\`a}}\ and\ \citenamefont
  {Percacci}(2013)}]{Dona:2012am}%
  \BibitemOpen
  \bibfield  {author} {\bibinfo {author} {\bibfnamefont {P.}~\bibnamefont
  {Don{\`a}}}\ and\ \bibinfo {author} {\bibfnamefont {R.}~\bibnamefont
  {Percacci}},\ }\href {\doibase 10.1103/PhysRevD.87.045002} {\bibfield
  {journal} {\bibinfo  {journal} {Phys. Rev.}\ }\textbf {\bibinfo {volume}
  {D87}},\ \bibinfo {pages} {045002} (\bibinfo {year} {2013})},\ \Eprint
  {http://arxiv.org/abs/1209.3649} {arXiv:1209.3649 [hep-th]} \BibitemShut
  {NoStop}%
%\%CITATION = ARXIV:1209.3649;\%\%
\bibitem [{\citenamefont {Henz}\ \emph {et~al.}(2013)\citenamefont {Henz},
  \citenamefont {Pawlowski}, \citenamefont {Rodigast},\ and\ \citenamefont
  {Wetterich}}]{Henz:2013oxa}%
  \BibitemOpen
  \bibfield  {author} {\bibinfo {author} {\bibfnamefont {T.}~\bibnamefont
  {Henz}}, \bibinfo {author} {\bibfnamefont {J.~M.}\ \bibnamefont {Pawlowski}},
  \bibinfo {author} {\bibfnamefont {A.}~\bibnamefont {Rodigast}}, \ and\
  \bibinfo {author} {\bibfnamefont {C.}~\bibnamefont {Wetterich}},\ }\href
  {\doibase 10.1016/j.physletb.2013.10.015} {\bibfield  {journal} {\bibinfo
  {journal} {Phys. Lett.}\ }\textbf {\bibinfo {volume} {B727}},\ \bibinfo
  {pages} {298} (\bibinfo {year} {2013})},\ \Eprint
  {http://arxiv.org/abs/1304.7743} {arXiv:1304.7743 [hep-th]} \BibitemShut
  {NoStop}%
%\%CITATION = ARXIV:1304.7743;\%\%
\bibitem [{\citenamefont {Don{\`a}}\ \emph {et~al.}(2014)\citenamefont
  {Don{\`a}}, \citenamefont {Eichhorn},\ and\ \citenamefont
  {Percacci}}]{Dona:2013qba}%
  \BibitemOpen
  \bibfield  {author} {\bibinfo {author} {\bibfnamefont {P.}~\bibnamefont
  {Don{\`a}}}, \bibinfo {author} {\bibfnamefont {A.}~\bibnamefont {Eichhorn}},
  \ and\ \bibinfo {author} {\bibfnamefont {R.}~\bibnamefont {Percacci}},\
  }\href {\doibase 10.1103/PhysRevD.89.084035} {\bibfield  {journal} {\bibinfo
  {journal} {Phys.Rev.}\ }\textbf {\bibinfo {volume} {D89}},\ \bibinfo {pages}
  {084035} (\bibinfo {year} {2014})},\ \Eprint {http://arxiv.org/abs/1311.2898}
  {arXiv:1311.2898 [hep-th]} \BibitemShut {NoStop}%
%\%CITATION = ARXIV:1311.2898;\%\%
\bibitem [{\citenamefont {Percacci}\ and\ \citenamefont
  {Vacca}(2015)}]{Percacci:2015wwa}%
  \BibitemOpen
  \bibfield  {author} {\bibinfo {author} {\bibfnamefont {R.}~\bibnamefont
  {Percacci}}\ and\ \bibinfo {author} {\bibfnamefont {G.~P.}\ \bibnamefont
  {Vacca}},\ }\href {\doibase 10.1140/epjc/s10052-015-3410-0} {\bibfield
  {journal} {\bibinfo  {journal} {Eur. Phys. J.}\ }\textbf {\bibinfo {volume}
  {C75}},\ \bibinfo {pages} {188} (\bibinfo {year} {2015})},\ \Eprint
  {http://arxiv.org/abs/1501.00888} {arXiv:1501.00888 [hep-th]} \BibitemShut
  {NoStop}%
%\%CITATION = ARXIV:1501.00888;\%\%
\bibitem [{\citenamefont {Oda}\ and\ \citenamefont
  {Yamada}(2016)}]{Oda:2015sma}%
  \BibitemOpen
  \bibfield  {author} {\bibinfo {author} {\bibfnamefont {K.-y.}\ \bibnamefont
  {Oda}}\ and\ \bibinfo {author} {\bibfnamefont {M.}~\bibnamefont {Yamada}},\
  }\href {\doibase 10.1088/0264-9381/33/12/125011} {\bibfield  {journal}
  {\bibinfo  {journal} {Class. Quant. Grav.}\ }\textbf {\bibinfo {volume}
  {33}},\ \bibinfo {pages} {125011} (\bibinfo {year} {2016})},\ \Eprint
  {http://arxiv.org/abs/1510.03734} {arXiv:1510.03734 [hep-th]} \BibitemShut
  {NoStop}%
%\%CITATION = ARXIV:1510.03734;\%\%
\bibitem [{\citenamefont {Don{\`a}}\ \emph {et~al.}(2016)\citenamefont
  {Don{\`a}}, \citenamefont {Eichhorn}, \citenamefont {Labus},\ and\
  \citenamefont {Percacci}}]{Dona:2015tnf}%
  \BibitemOpen
  \bibfield  {author} {\bibinfo {author} {\bibfnamefont {P.}~\bibnamefont
  {Don{\`a}}}, \bibinfo {author} {\bibfnamefont {A.}~\bibnamefont {Eichhorn}},
  \bibinfo {author} {\bibfnamefont {P.}~\bibnamefont {Labus}}, \ and\ \bibinfo
  {author} {\bibfnamefont {R.}~\bibnamefont {Percacci}},\ }\href {\doibase
  10.1103/PhysRevD.93.129904;;;;;;;;;;;;;;;;; 10.1103/PhysRevD.93.044049}
  {\bibfield  {journal} {\bibinfo  {journal} {Phys. Rev.}\ }\textbf {\bibinfo
  {volume} {D93}},\ \bibinfo {pages} {044049} (\bibinfo {year} {2016})},\
  \bibinfo {note} {[Erratum: Phys. Rev.D93,no.12,129904(2016)]},\ \Eprint
  {http://arxiv.org/abs/1512.01589} {arXiv:1512.01589 [gr-qc]} \BibitemShut
  {NoStop}%
%\%CITATION = ARXIV:1512.01589;\%\%
\bibitem [{\citenamefont {Eichhorn}\ \emph {et~al.}(2016)\citenamefont
  {Eichhorn}, \citenamefont {Held},\ and\ \citenamefont
  {Pawlowski}}]{Eichhorn:2016esv}%
  \BibitemOpen
  \bibfield  {author} {\bibinfo {author} {\bibfnamefont {A.}~\bibnamefont
  {Eichhorn}}, \bibinfo {author} {\bibfnamefont {A.}~\bibnamefont {Held}}, \
  and\ \bibinfo {author} {\bibfnamefont {J.~M.}\ \bibnamefont {Pawlowski}},\
  }\href {\doibase 10.1103/PhysRevD.94.104027} {\bibfield  {journal} {\bibinfo
  {journal} {Phys. Rev.}\ }\textbf {\bibinfo {volume} {D94}},\ \bibinfo {pages}
  {104027} (\bibinfo {year} {2016})},\ \Eprint
  {http://arxiv.org/abs/1604.02041} {arXiv:1604.02041 [hep-th]} \BibitemShut
  {NoStop}%
%\%CITATION = ARXIV:1604.02041;\%\%
\bibitem [{\citenamefont {Eichhorn}\ and\ \citenamefont
  {Lippoldt}(2017)}]{Eichhorn:2016vvy}%
  \BibitemOpen
  \bibfield  {author} {\bibinfo {author} {\bibfnamefont {A.}~\bibnamefont
  {Eichhorn}}\ and\ \bibinfo {author} {\bibfnamefont {S.}~\bibnamefont
  {Lippoldt}},\ }\href {\doibase 10.1016/j.physletb.2017.01.064} {\bibfield
  {journal} {\bibinfo  {journal} {Phys. Lett.}\ }\textbf {\bibinfo {volume}
  {B767}},\ \bibinfo {pages} {142} (\bibinfo {year} {2017})},\ \Eprint
  {http://arxiv.org/abs/1611.05878} {arXiv:1611.05878 [gr-qc]} \BibitemShut
  {NoStop}%
%\%CITATION = ARXIV:1611.05878;\%\%
\bibitem [{\citenamefont {Christiansen}\ and\ \citenamefont
  {Eichhorn}(2017)}]{Christiansen:2017gtg}%
  \BibitemOpen
  \bibfield  {author} {\bibinfo {author} {\bibfnamefont {N.}~\bibnamefont
  {Christiansen}}\ and\ \bibinfo {author} {\bibfnamefont {A.}~\bibnamefont
  {Eichhorn}},\ }\href {\doibase 10.1016/j.physletb.2017.04.047} {\bibfield
  {journal} {\bibinfo  {journal} {Phys. Lett.}\ }\textbf {\bibinfo {volume}
  {B770}},\ \bibinfo {pages} {154} (\bibinfo {year} {2017})},\ \Eprint
  {http://arxiv.org/abs/1702.07724} {arXiv:1702.07724 [hep-th]} \BibitemShut
  {NoStop}%
%\%CITATION = ARXIV:1702.07724;\%\%
\bibitem [{\citenamefont {Eichhorn}\ and\ \citenamefont
  {Held}(2017)}]{Eichhorn:2017eht}%
  \BibitemOpen
  \bibfield  {author} {\bibinfo {author} {\bibfnamefont {A.}~\bibnamefont
  {Eichhorn}}\ and\ \bibinfo {author} {\bibfnamefont {A.}~\bibnamefont
  {Held}},\ }\href {\doibase 10.1103/PhysRevD.96.086025} {\bibfield  {journal}
  {\bibinfo  {journal} {Phys. Rev.}\ }\textbf {\bibinfo {volume} {D96}},\
  \bibinfo {pages} {086025} (\bibinfo {year} {2017})},\ \Eprint
  {http://arxiv.org/abs/1705.02342} {arXiv:1705.02342 [gr-qc]} \BibitemShut
  {NoStop}%
%\%CITATION = ARXIV:1705.02342;\%\%
\bibitem [{\citenamefont {Biemans}\ \emph
  {et~al.}(2017{\natexlab{b}})\citenamefont {Biemans}, \citenamefont
  {Platania},\ and\ \citenamefont {Saueressig}}]{Biemans:2017zca}%
  \BibitemOpen
  \bibfield  {author} {\bibinfo {author} {\bibfnamefont {J.}~\bibnamefont
  {Biemans}}, \bibinfo {author} {\bibfnamefont {A.}~\bibnamefont {Platania}}, \
  and\ \bibinfo {author} {\bibfnamefont {F.}~\bibnamefont {Saueressig}},\
  }\href {\doibase 10.1007/JHEP05(2017)093} {\bibfield  {journal} {\bibinfo
  {journal} {JHEP}\ }\textbf {\bibinfo {volume} {05}},\ \bibinfo {pages} {093}
  (\bibinfo {year} {2017}{\natexlab{b}})},\ \Eprint
  {http://arxiv.org/abs/1702.06539} {arXiv:1702.06539 [hep-th]} \BibitemShut
  {NoStop}%
%\%CITATION = ARXIV:1702.06539;\%\%
\bibitem [{\citenamefont {Eichhorn}\ and\ \citenamefont
  {Held}(2018)}]{Eichhorn:2017ylw}%
  \BibitemOpen
  \bibfield  {author} {\bibinfo {author} {\bibfnamefont {A.}~\bibnamefont
  {Eichhorn}}\ and\ \bibinfo {author} {\bibfnamefont {A.}~\bibnamefont
  {Held}},\ }\href {\doibase 10.1016/j.physletb.2017.12.040} {\bibfield
  {journal} {\bibinfo  {journal} {Phys. Lett.}\ }\textbf {\bibinfo {volume}
  {B777}},\ \bibinfo {pages} {217} (\bibinfo {year} {2018})},\ \Eprint
  {http://arxiv.org/abs/1707.01107} {arXiv:1707.01107 [hep-th]} \BibitemShut
  {NoStop}%
%%CITATION = ARXIV:1707.01107;%%
\bibitem [{\citenamefont {Eichhorn}\ and\ \citenamefont
  {Versteegen}(2018)}]{Eichhorn:2017lry}%
  \BibitemOpen
  \bibfield  {author} {\bibinfo {author} {\bibfnamefont {A.}~\bibnamefont
  {Eichhorn}}\ and\ \bibinfo {author} {\bibfnamefont {F.}~\bibnamefont
  {Versteegen}},\ }\href {\doibase 10.1007/JHEP01(2018)030} {\bibfield
  {journal} {\bibinfo  {journal} {JHEP}\ }\textbf {\bibinfo {volume} {01}},\
  \bibinfo {pages} {030} (\bibinfo {year} {2018})},\ \Eprint
  {http://arxiv.org/abs/1709.07252} {arXiv:1709.07252 [hep-th]} \BibitemShut
  {NoStop}%
%%CITATION = ARXIV:1709.07252;%%
\bibitem [{\citenamefont {Eichhorn}\ \emph
  {et~al.}(2018{\natexlab{a}})\citenamefont {Eichhorn}, \citenamefont
  {Lippoldt},\ and\ \citenamefont {Skrinjar}}]{Eichhorn:2017sok}%
  \BibitemOpen
  \bibfield  {author} {\bibinfo {author} {\bibfnamefont {A.}~\bibnamefont
  {Eichhorn}}, \bibinfo {author} {\bibfnamefont {S.}~\bibnamefont {Lippoldt}},
  \ and\ \bibinfo {author} {\bibfnamefont {V.}~\bibnamefont {Skrinjar}},\
  }\href {\doibase 10.1103/PhysRevD.97.026002} {\bibfield  {journal} {\bibinfo
  {journal} {Phys. Rev.}\ }\textbf {\bibinfo {volume} {D97}},\ \bibinfo {pages}
  {026002} (\bibinfo {year} {2018}{\natexlab{a}})},\ \Eprint
  {http://arxiv.org/abs/1710.03005} {arXiv:1710.03005 [hep-th]} \BibitemShut
  {NoStop}%
%%CITATION = ARXIV:1710.03005;%%
\bibitem [{\citenamefont {Niedermaier}\ and\ \citenamefont
  {Reuter}(2006)}]{Niedermaier:2006wt}%
  \BibitemOpen
  \bibfield  {author} {\bibinfo {author} {\bibfnamefont {M.}~\bibnamefont
  {Niedermaier}}\ and\ \bibinfo {author} {\bibfnamefont {M.}~\bibnamefont
  {Reuter}},\ }\href@noop {} {\bibfield  {journal} {\bibinfo  {journal} {Living
  Rev.Rel.}\ }\textbf {\bibinfo {volume} {9}},\ \bibinfo {pages} {5} (\bibinfo
  {year} {2006})}\BibitemShut {NoStop}%
%\%CITATION = 00222,9,5;\%\%
\bibitem [{\citenamefont {Percacci}(2007)}]{Percacci:2007sz}%
  \BibitemOpen
  \bibfield  {author} {\bibinfo {author} {\bibfnamefont {R.}~\bibnamefont
  {Percacci}},\ }\href@noop {} {\bibfield  {journal} {\bibinfo  {journal} {In
  *Oriti, D. (ed.): Approaches to quantum gravity* 111-128}\ } (\bibinfo {year}
  {2007})},\ \Eprint {http://arxiv.org/abs/0709.3851} {arXiv:0709.3851
  [hep-th]} \BibitemShut {NoStop}%
%\%CITATION = ARXIV:0709.3851;\%\%
\bibitem [{\citenamefont {Litim}(2011)}]{Litim:2011cp}%
  \BibitemOpen
  \bibfield  {author} {\bibinfo {author} {\bibfnamefont {D.~F.}\ \bibnamefont
  {Litim}},\ }\href@noop {} {\bibfield  {journal} {\bibinfo  {journal}
  {Phil.Trans.Roy.Soc.Lond.}\ }\textbf {\bibinfo {volume} {A369}},\ \bibinfo
  {pages} {2759} (\bibinfo {year} {2011})},\ \Eprint
  {http://arxiv.org/abs/1102.4624} {arXiv:1102.4624 [hep-th]} \BibitemShut
  {NoStop}%
%\%CITATION = ARXIV:1102.4624;\%\%
\bibitem [{\citenamefont {Reuter}\ and\ \citenamefont
  {Saueressig}(2012)}]{Reuter:2012id}%
  \BibitemOpen
  \bibfield  {author} {\bibinfo {author} {\bibfnamefont {M.}~\bibnamefont
  {Reuter}}\ and\ \bibinfo {author} {\bibfnamefont {F.}~\bibnamefont
  {Saueressig}},\ }\href {\doibase 10.1088/1367-2630/14/5/055022} {\bibfield
  {journal} {\bibinfo  {journal} {New J. Phys.}\ }\textbf {\bibinfo {volume}
  {14}},\ \bibinfo {pages} {055022} (\bibinfo {year} {2012})},\ \Eprint
  {http://arxiv.org/abs/1202.2274} {arXiv:1202.2274 [hep-th]} \BibitemShut
  {NoStop}%
%\%CITATION = ARXIV:1202.2274;\%\%
\bibitem [{\citenamefont {Bonanno}\ and\ \citenamefont
  {Saueressig}(2017)}]{Bonanno:2017pkg}%
  \BibitemOpen
  \bibfield  {author} {\bibinfo {author} {\bibfnamefont {A.}~\bibnamefont
  {Bonanno}}\ and\ \bibinfo {author} {\bibfnamefont {F.}~\bibnamefont
  {Saueressig}},\ }\href {\doibase 10.1016/j.crhy.2017.02.002} {\bibfield
  {journal} {\bibinfo  {journal} {Comptes Rendus Physique}\ }\textbf {\bibinfo
  {volume} {18}},\ \bibinfo {pages} {254} (\bibinfo {year} {2017})},\ \Eprint
  {http://arxiv.org/abs/1702.04137} {arXiv:1702.04137 [hep-th]} \BibitemShut
  {NoStop}%
%\%CITATION = ARXIV:1702.04137;\%\%
\bibitem [{\citenamefont {Eichhorn}(2017)}]{Eichhorn:2017egq}%
  \BibitemOpen
  \bibfield  {author} {\bibinfo {author} {\bibfnamefont {A.}~\bibnamefont
  {Eichhorn}},\ }in\ \href
  {http://inspirehep.net/record/1623009/files/arXiv:1709.03696.pdf} {\emph
  {\bibinfo {booktitle} {{Black Holes, Gravitational Waves and Spacetime
  Singularities Rome, Italy, May 9-12, 2017}}}}\ (\bibinfo {year} {2017})\
  \Eprint {http://arxiv.org/abs/1709.03696} {arXiv:1709.03696 [gr-qc]}
  \BibitemShut {NoStop}%
%\%CITATION = ARXIV:1709.03696;\%\%
\bibitem [{\citenamefont {Wetterich}(1993)}]{Wetterich:1992yh}%
  \BibitemOpen
  \bibfield  {author} {\bibinfo {author} {\bibfnamefont {C.}~\bibnamefont
  {Wetterich}},\ }\href {\doibase 10.1016/0370-2693(93)90726-X} {\bibfield
  {journal} {\bibinfo  {journal} {Phys. Lett.}\ }\textbf {\bibinfo {volume}
  {B301}},\ \bibinfo {pages} {90} (\bibinfo {year} {1993})},\ \Eprint
  {http://arxiv.org/abs/1710.05815} {arXiv:1710.05815 [hep-th]} \BibitemShut
  {NoStop}%
%%CITATION = ARXIV:1710.05815;%%
\bibitem [{\citenamefont {Ellwanger}(1994)}]{Ellwanger:1993mw}%
  \BibitemOpen
  \bibfield  {author} {\bibinfo {author} {\bibfnamefont {U.}~\bibnamefont
  {Ellwanger}},\ }\bibfield  {booktitle} {\emph {\bibinfo {booktitle}
  {{Proceedings, Workshop on Quantum field theoretical aspects of high energy
  physics: Bad Frankenhausen, Germany, September 20-24, 1993}}},\ }\href
  {\doibase 10.1007/BF01555911} {\bibfield  {journal} {\bibinfo  {journal} {Z.
  Phys.}\ }\textbf {\bibinfo {volume} {C62}},\ \bibinfo {pages} {503} (\bibinfo
  {year} {1994})},\ \Eprint {http://arxiv.org/abs/hep-ph/9308260}
  {arXiv:hep-ph/9308260 [hep-ph]} \BibitemShut {NoStop}%
%\%CITATION = HEP-PH/9308260;\%\%
\bibitem [{\citenamefont {Morris}(1994)}]{Morris:1993qb}%
  \BibitemOpen
  \bibfield  {author} {\bibinfo {author} {\bibfnamefont {T.~R.}\ \bibnamefont
  {Morris}},\ }\href {\doibase 10.1142/S0217751X94000972} {\bibfield  {journal}
  {\bibinfo  {journal} {Int. J. Mod. Phys.}\ }\textbf {\bibinfo {volume}
  {A9}},\ \bibinfo {pages} {2411} (\bibinfo {year} {1994})},\ \Eprint
  {http://arxiv.org/abs/hep-ph/9308265} {arXiv:hep-ph/9308265} \BibitemShut
  {NoStop}%
%\%CITATION = HEP-PH/9308265;\%\%
\bibitem [{\citenamefont {Litim}\ and\ \citenamefont
  {Pawlowski}(2002{\natexlab{a}})}]{Litim:2002ce}%
  \BibitemOpen
  \bibfield  {author} {\bibinfo {author} {\bibfnamefont {D.~F.}\ \bibnamefont
  {Litim}}\ and\ \bibinfo {author} {\bibfnamefont {J.~M.}\ \bibnamefont
  {Pawlowski}},\ }\href {\doibase 10.1088/1126-6708/2002/09/049} {\bibfield
  {journal} {\bibinfo  {journal} {JHEP}\ }\textbf {\bibinfo {volume} {09}},\
  \bibinfo {pages} {049} (\bibinfo {year} {2002}{\natexlab{a}})},\ \Eprint
  {http://arxiv.org/abs/hep-th/0203005} {arXiv:hep-th/0203005 [hep-th]}
  \BibitemShut {NoStop}%
%%CITATION = HEP-TH/0203005;%%
\bibitem [{\citenamefont {Litim}\ and\ \citenamefont
  {Pawlowski}(2002{\natexlab{b}})}]{Litim:2002hj}%
  \BibitemOpen
  \bibfield  {author} {\bibinfo {author} {\bibfnamefont {D.~F.}\ \bibnamefont
  {Litim}}\ and\ \bibinfo {author} {\bibfnamefont {J.~M.}\ \bibnamefont
  {Pawlowski}},\ }\href {\doibase 10.1016/S0370-2693(02)02693-X} {\bibfield
  {journal} {\bibinfo  {journal} {Phys.Lett.}\ }\textbf {\bibinfo {volume}
  {B546}},\ \bibinfo {pages} {279} (\bibinfo {year} {2002}{\natexlab{b}})},\
  \Eprint {http://arxiv.org/abs/hep-th/0208216} {arXiv:hep-th/0208216 [hep-th]}
  \BibitemShut {NoStop}%
%\%CITATION = HEP-TH/0208216;\%\%
\bibitem [{\citenamefont {Pawlowski}(2003)}]{Pawlowski:2003sk}%
  \BibitemOpen
  \bibfield  {author} {\bibinfo {author} {\bibfnamefont {J.~M.}\ \bibnamefont
  {Pawlowski}},\ }\href@noop {} {\  (\bibinfo {year} {2003})},\ \Eprint
  {http://arxiv.org/abs/hep-th/0310018} {arXiv:hep-th/0310018 [hep-th]}
  \BibitemShut {NoStop}%
%\%CITATION = HEP-TH/0310018;\%\%
\bibitem [{\citenamefont {Pawlowski}(2007)}]{Pawlowski:2005xe}%
  \BibitemOpen
  \bibfield  {author} {\bibinfo {author} {\bibfnamefont {J.~M.}\ \bibnamefont
  {Pawlowski}},\ }\href {\doibase 10.1016/j.aop.2007.01.007} {\bibfield
  {journal} {\bibinfo  {journal} {Annals Phys.}\ }\textbf {\bibinfo {volume}
  {322}},\ \bibinfo {pages} {2831} (\bibinfo {year} {2007})},\ \Eprint
  {http://arxiv.org/abs/hep-th/0512261} {arXiv:hep-th/0512261 [hep-th]}
  \BibitemShut {NoStop}%
%\%CITATION = HEP-TH/0512261;\%\%
\bibitem [{\citenamefont {Safari}(2016)}]{Safari:2015dva}%
  \BibitemOpen
  \bibfield  {author} {\bibinfo {author} {\bibfnamefont {M.}~\bibnamefont
  {Safari}},\ }\href {\doibase 10.1140/epjc/s10052-016-4036-6} {\bibfield
  {journal} {\bibinfo  {journal} {Eur. Phys. J.}\ }\textbf {\bibinfo {volume}
  {C76}},\ \bibinfo {pages} {201} (\bibinfo {year} {2016})},\ \Eprint
  {http://arxiv.org/abs/1508.06244} {arXiv:1508.06244 [hep-th]} \BibitemShut
  {NoStop}%
%\%CITATION = ARXIV:1508.06244;\%\%
\bibitem [{\citenamefont {Bridle}\ \emph {et~al.}(2014)\citenamefont {Bridle},
  \citenamefont {Dietz},\ and\ \citenamefont {Morris}}]{Bridle:2013sra}%
  \BibitemOpen
  \bibfield  {author} {\bibinfo {author} {\bibfnamefont {I.~H.}\ \bibnamefont
  {Bridle}}, \bibinfo {author} {\bibfnamefont {J.~A.}\ \bibnamefont {Dietz}}, \
  and\ \bibinfo {author} {\bibfnamefont {T.~R.}\ \bibnamefont {Morris}},\
  }\href {\doibase 10.1007/JHEP03(2014)093} {\bibfield  {journal} {\bibinfo
  {journal} {JHEP}\ }\textbf {\bibinfo {volume} {03}},\ \bibinfo {pages} {093}
  (\bibinfo {year} {2014})},\ \Eprint {http://arxiv.org/abs/1312.2846}
  {arXiv:1312.2846 [hep-th]} \BibitemShut {NoStop}%
%\%CITATION = ARXIV:1312.2846;\%\%
\bibitem [{\citenamefont {Dietz}\ and\ \citenamefont
  {Morris}(2015)}]{Dietz:2015owa}%
  \BibitemOpen
  \bibfield  {author} {\bibinfo {author} {\bibfnamefont {J.~A.}\ \bibnamefont
  {Dietz}}\ and\ \bibinfo {author} {\bibfnamefont {T.~R.}\ \bibnamefont
  {Morris}},\ }\href {\doibase 10.1007/JHEP04(2015)118} {\bibfield  {journal}
  {\bibinfo  {journal} {JHEP}\ }\textbf {\bibinfo {volume} {04}},\ \bibinfo
  {pages} {118} (\bibinfo {year} {2015})},\ \Eprint
  {http://arxiv.org/abs/1502.07396} {arXiv:1502.07396 [hep-th]} \BibitemShut
  {NoStop}%
%\%CITATION = ARXIV:1502.07396;\%\%
\bibitem [{\citenamefont {Safari}\ and\ \citenamefont
  {Vacca}(2016)}]{Safari:2016gtj}%
  \BibitemOpen
  \bibfield  {author} {\bibinfo {author} {\bibfnamefont {M.}~\bibnamefont
  {Safari}}\ and\ \bibinfo {author} {\bibfnamefont {G.~P.}\ \bibnamefont
  {Vacca}},\ }\href {\doibase 10.1007/JHEP11(2016)139} {\bibfield  {journal}
  {\bibinfo  {journal} {JHEP}\ }\textbf {\bibinfo {volume} {11}},\ \bibinfo
  {pages} {139} (\bibinfo {year} {2016})},\ \Eprint
  {http://arxiv.org/abs/1607.07074} {arXiv:1607.07074 [hep-th]} \BibitemShut
  {NoStop}%
%\%CITATION = ARXIV:1607.07074;\%\%
\bibitem [{\citenamefont {Nieto}\ \emph {et~al.}(2017)\citenamefont {Nieto},
  \citenamefont {Percacci},\ and\ \citenamefont {Skrinjar}}]{Nieto:2017ddk}%
  \BibitemOpen
  \bibfield  {author} {\bibinfo {author} {\bibfnamefont {C.~M.}\ \bibnamefont
  {Nieto}}, \bibinfo {author} {\bibfnamefont {R.}~\bibnamefont {Percacci}}, \
  and\ \bibinfo {author} {\bibfnamefont {V.}~\bibnamefont {Skrinjar}},\ }\href
  {\doibase 10.1103/PhysRevD.96.106019} {\bibfield  {journal} {\bibinfo
  {journal} {Phys. Rev.}\ }\textbf {\bibinfo {volume} {D96}},\ \bibinfo {pages}
  {106019} (\bibinfo {year} {2017})},\ \Eprint
  {http://arxiv.org/abs/1708.09760} {arXiv:1708.09760 [gr-qc]} \BibitemShut
  {NoStop}%
%%CITATION = ARXIV:1708.09760;%%
\bibitem [{\citenamefont {Braun}\ \emph
  {et~al.}(2010{\natexlab{a}})\citenamefont {Braun}, \citenamefont {Gies},\
  and\ \citenamefont {Pawlowski}}]{Braun:2007bx}%
  \BibitemOpen
  \bibfield  {author} {\bibinfo {author} {\bibfnamefont {J.}~\bibnamefont
  {Braun}}, \bibinfo {author} {\bibfnamefont {H.}~\bibnamefont {Gies}}, \ and\
  \bibinfo {author} {\bibfnamefont {J.~M.}\ \bibnamefont {Pawlowski}},\ }\href
  {\doibase 10.1016/j.physletb.2010.01.009} {\bibfield  {journal} {\bibinfo
  {journal} {Phys.Lett.}\ }\textbf {\bibinfo {volume} {B684}},\ \bibinfo
  {pages} {262} (\bibinfo {year} {2010}{\natexlab{a}})},\ \Eprint
  {http://arxiv.org/abs/0708.2413} {arXiv:0708.2413 [hep-th]} \BibitemShut
  {NoStop}%
\bibitem [{\citenamefont {Braun}\ \emph {et~al.}(2011)\citenamefont {Braun},
  \citenamefont {Haas}, \citenamefont {Marhauser},\ and\ \citenamefont
  {Pawlowski}}]{Braun:2009gm}%
  \BibitemOpen
  \bibfield  {author} {\bibinfo {author} {\bibfnamefont {J.}~\bibnamefont
  {Braun}}, \bibinfo {author} {\bibfnamefont {L.~M.}\ \bibnamefont {Haas}},
  \bibinfo {author} {\bibfnamefont {F.}~\bibnamefont {Marhauser}}, \ and\
  \bibinfo {author} {\bibfnamefont {J.~M.}\ \bibnamefont {Pawlowski}},\ }\href
  {\doibase 10.1103/PhysRevLett.106.022002} {\bibfield  {journal} {\bibinfo
  {journal} {Phys.Rev.Lett.}\ }\textbf {\bibinfo {volume} {106}},\ \bibinfo
  {pages} {022002} (\bibinfo {year} {2011})},\ \Eprint
  {http://arxiv.org/abs/0908.0008} {arXiv:0908.0008 [hep-ph]} \BibitemShut
  {NoStop}%
\bibitem [{\citenamefont {Braun}\ \emph
  {et~al.}(2010{\natexlab{b}})\citenamefont {Braun}, \citenamefont {Eichhorn},
  \citenamefont {Gies},\ and\ \citenamefont {Pawlowski}}]{Braun:2010cy}%
  \BibitemOpen
  \bibfield  {author} {\bibinfo {author} {\bibfnamefont {J.}~\bibnamefont
  {Braun}}, \bibinfo {author} {\bibfnamefont {A.}~\bibnamefont {Eichhorn}},
  \bibinfo {author} {\bibfnamefont {H.}~\bibnamefont {Gies}}, \ and\ \bibinfo
  {author} {\bibfnamefont {J.~M.}\ \bibnamefont {Pawlowski}},\ }\href {\doibase
  10.1140/epjc/s10052-010-1485-1} {\bibfield  {journal} {\bibinfo  {journal}
  {Eur.Phys.J.}\ }\textbf {\bibinfo {volume} {C70}},\ \bibinfo {pages} {689}
  (\bibinfo {year} {2010}{\natexlab{b}})},\ \Eprint
  {http://arxiv.org/abs/1007.2619} {arXiv:1007.2619 [hep-ph]} \BibitemShut
  {NoStop}%
%\%CITATION = ARXIV:1007.2619;\%\%
\bibitem [{\citenamefont {Fister}\ and\ \citenamefont
  {Pawlowski}(2013)}]{Fister:2013bh}%
  \BibitemOpen
  \bibfield  {author} {\bibinfo {author} {\bibfnamefont {L.}~\bibnamefont
  {Fister}}\ and\ \bibinfo {author} {\bibfnamefont {J.~M.}\ \bibnamefont
  {Pawlowski}},\ }\href {\doibase 10.1103/PhysRevD.88.045010} {\bibfield
  {journal} {\bibinfo  {journal} {Phys.Rev.}\ }\textbf {\bibinfo {volume}
  {D88}},\ \bibinfo {pages} {045010} (\bibinfo {year} {2013})},\ \Eprint
  {http://arxiv.org/abs/1301.4163} {arXiv:1301.4163 [hep-ph]} \BibitemShut
  {NoStop}%
%\%CITATION = ARXIV:1301.4163;\%\%
\bibitem [{\citenamefont {Reinosa}\ \emph {et~al.}(2015)\citenamefont
  {Reinosa}, \citenamefont {Serreau}, \citenamefont {Tissier},\ and\
  \citenamefont {Wschebor}}]{Reinosa:2014ooa}%
  \BibitemOpen
  \bibfield  {author} {\bibinfo {author} {\bibfnamefont {U.}~\bibnamefont
  {Reinosa}}, \bibinfo {author} {\bibfnamefont {J.}~\bibnamefont {Serreau}},
  \bibinfo {author} {\bibfnamefont {M.}~\bibnamefont {Tissier}}, \ and\
  \bibinfo {author} {\bibfnamefont {N.}~\bibnamefont {Wschebor}},\ }\href
  {\doibase 10.1016/j.physletb.2015.01.006} {\bibfield  {journal} {\bibinfo
  {journal} {Phys. Lett.}\ }\textbf {\bibinfo {volume} {B742}},\ \bibinfo
  {pages} {61} (\bibinfo {year} {2015})},\ \Eprint
  {http://arxiv.org/abs/1407.6469} {arXiv:1407.6469 [hep-ph]} \BibitemShut
  {NoStop}%
%\%CITATION = ARXIV:1407.6469;\%\%
\bibitem [{\citenamefont {Bonanno}\ and\ \citenamefont
  {Reuter}(2002{\natexlab{a}})}]{Bonanno:2001xi}%
  \BibitemOpen
  \bibfield  {author} {\bibinfo {author} {\bibfnamefont {A.}~\bibnamefont
  {Bonanno}}\ and\ \bibinfo {author} {\bibfnamefont {M.}~\bibnamefont
  {Reuter}},\ }\href {\doibase 10.1103/PhysRevD.65.043508} {\bibfield
  {journal} {\bibinfo  {journal} {Phys. Rev.}\ }\textbf {\bibinfo {volume}
  {D65}},\ \bibinfo {pages} {043508} (\bibinfo {year} {2002}{\natexlab{a}})},\
  \Eprint {http://arxiv.org/abs/hep-th/0106133} {arXiv:hep-th/0106133}
  \BibitemShut {NoStop}%
%\%CITATION = HEP-TH/0106133;\%\%
\bibitem [{\citenamefont {Bonanno}\ and\ \citenamefont
  {Reuter}(2002{\natexlab{b}})}]{Bonanno:2001hi}%
  \BibitemOpen
  \bibfield  {author} {\bibinfo {author} {\bibfnamefont {A.}~\bibnamefont
  {Bonanno}}\ and\ \bibinfo {author} {\bibfnamefont {M.}~\bibnamefont
  {Reuter}},\ }\href {\doibase 10.1016/S0370-2693(01)01522-2} {\bibfield
  {journal} {\bibinfo  {journal} {Phys. Lett.}\ }\textbf {\bibinfo {volume}
  {B527}},\ \bibinfo {pages} {9} (\bibinfo {year} {2002}{\natexlab{b}})},\
  \Eprint {http://arxiv.org/abs/astro-ph/0106468} {arXiv:astro-ph/0106468}
  \BibitemShut {NoStop}%
%\%CITATION = ASTRO-PH/0106468;\%\%
\bibitem [{\citenamefont {Bentivegna}\ \emph {et~al.}(2004)\citenamefont
  {Bentivegna}, \citenamefont {Bonanno},\ and\ \citenamefont
  {Reuter}}]{Bentivegna:2003rr}%
  \BibitemOpen
  \bibfield  {author} {\bibinfo {author} {\bibfnamefont {E.}~\bibnamefont
  {Bentivegna}}, \bibinfo {author} {\bibfnamefont {A.}~\bibnamefont {Bonanno}},
  \ and\ \bibinfo {author} {\bibfnamefont {M.}~\bibnamefont {Reuter}},\ }\href
  {\doibase 10.1088/1475-7516/2004/01/001} {\bibfield  {journal} {\bibinfo
  {journal} {JCAP}\ }\textbf {\bibinfo {volume} {0401}},\ \bibinfo {pages}
  {001} (\bibinfo {year} {2004})},\ \Eprint
  {http://arxiv.org/abs/astro-ph/0303150} {arXiv:astro-ph/0303150} \BibitemShut
  {NoStop}%
%\%CITATION = ASTRO-PH/0303150;\%\%
\bibitem [{\citenamefont {Reuter}\ and\ \citenamefont
  {Saueressig}(2005)}]{Reuter:2005kb}%
  \BibitemOpen
  \bibfield  {author} {\bibinfo {author} {\bibfnamefont {M.}~\bibnamefont
  {Reuter}}\ and\ \bibinfo {author} {\bibfnamefont {F.}~\bibnamefont
  {Saueressig}},\ }\href {\doibase 10.1088/1475-7516/2005/09/012} {\bibfield
  {journal} {\bibinfo  {journal} {JCAP}\ }\textbf {\bibinfo {volume} {0509}},\
  \bibinfo {pages} {012} (\bibinfo {year} {2005})},\ \Eprint
  {http://arxiv.org/abs/hep-th/0507167} {arXiv:hep-th/0507167} \BibitemShut
  {NoStop}%
%\%CITATION = HEP-TH/0507167;\%\%
\bibitem [{\citenamefont {Bonanno}\ and\ \citenamefont
  {Reuter}(2007)}]{Bonanno:2007wg}%
  \BibitemOpen
  \bibfield  {author} {\bibinfo {author} {\bibfnamefont {A.}~\bibnamefont
  {Bonanno}}\ and\ \bibinfo {author} {\bibfnamefont {M.}~\bibnamefont
  {Reuter}},\ }\href {\doibase 10.1088/1475-7516/2007/08/024} {\bibfield
  {journal} {\bibinfo  {journal} {JCAP}\ }\textbf {\bibinfo {volume} {0708}},\
  \bibinfo {pages} {024} (\bibinfo {year} {2007})},\ \Eprint
  {http://arxiv.org/abs/0706.0174} {arXiv:0706.0174 [hep-th]} \BibitemShut
  {NoStop}%
%\%CITATION = ARXIV:0706.0174;\%\%
\bibitem [{\citenamefont {Weinberg}(2010)}]{Weinberg:2009wa}%
  \BibitemOpen
  \bibfield  {author} {\bibinfo {author} {\bibfnamefont {S.}~\bibnamefont
  {Weinberg}},\ }\href {\doibase 10.1103/PhysRevD.81.083535} {\bibfield
  {journal} {\bibinfo  {journal} {Phys. Rev.}\ }\textbf {\bibinfo {volume}
  {D81}},\ \bibinfo {pages} {083535} (\bibinfo {year} {2010})},\ \Eprint
  {http://arxiv.org/abs/0911.3165} {arXiv:0911.3165 [hep-th]} \BibitemShut
  {NoStop}%
%\%CITATION = 0911.3165;\%\%
\bibitem [{\citenamefont {Bonanno}(2011)}]{Bonanno:2009nj}%
  \BibitemOpen
  \bibfield  {author} {\bibinfo {author} {\bibfnamefont {A.}~\bibnamefont
  {Bonanno}},\ }\bibfield  {booktitle} {\emph {\bibinfo {booktitle}
  {{Proceedings, Workshop on Continuum and lattice approaches to quantum
  gravity (CLAQG08): Brighton, UK, September 17-19, 2008}}},\ }\href {\doibase
  10.22323/1.079.0008} {\bibfield  {journal} {\bibinfo  {journal} {PoS}\
  }\textbf {\bibinfo {volume} {CLAQG08}},\ \bibinfo {pages} {008} (\bibinfo
  {year} {2011})},\ \Eprint {http://arxiv.org/abs/0911.2727} {arXiv:0911.2727
  [hep-th]} \BibitemShut {NoStop}%
%%CITATION = ARXIV:0911.2727;%%
\bibitem [{\citenamefont {Bonanno}\ and\ \citenamefont
  {Reuter}(2011)}]{Bonanno:2010mk}%
  \BibitemOpen
  \bibfield  {author} {\bibinfo {author} {\bibfnamefont {A.}~\bibnamefont
  {Bonanno}}\ and\ \bibinfo {author} {\bibfnamefont {M.}~\bibnamefont
  {Reuter}},\ }\href {\doibase 10.3390/e13010274} {\bibfield  {journal}
  {\bibinfo  {journal} {Entropy}\ }\textbf {\bibinfo {volume} {13}},\ \bibinfo
  {pages} {274} (\bibinfo {year} {2011})},\ \Eprint
  {http://arxiv.org/abs/1011.2794} {arXiv:1011.2794 [hep-th]} \BibitemShut
  {NoStop}%
%\%CITATION = ARXIV:1011.2794;\%\%
\bibitem [{\citenamefont {Koch}\ and\ \citenamefont
  {Ramirez}(2011)}]{Koch:2010nn}%
  \BibitemOpen
  \bibfield  {author} {\bibinfo {author} {\bibfnamefont {B.}~\bibnamefont
  {Koch}}\ and\ \bibinfo {author} {\bibfnamefont {I.}~\bibnamefont {Ramirez}},\
  }\href {\doibase 10.1088/0264-9381/28/5/055008} {\bibfield  {journal}
  {\bibinfo  {journal} {Class. Quant. Grav.}\ }\textbf {\bibinfo {volume}
  {28}},\ \bibinfo {pages} {055008} (\bibinfo {year} {2011})},\ \Eprint
  {http://arxiv.org/abs/1010.2799} {arXiv:1010.2799 [gr-qc]} \BibitemShut
  {NoStop}%
%\%CITATION = 1010.2799;\%\%
\bibitem [{\citenamefont {Casadio}\ \emph {et~al.}(2011)\citenamefont
  {Casadio}, \citenamefont {Hsu},\ and\ \citenamefont
  {Mirza}}]{Casadio:2010fw}%
  \BibitemOpen
  \bibfield  {author} {\bibinfo {author} {\bibfnamefont {R.}~\bibnamefont
  {Casadio}}, \bibinfo {author} {\bibfnamefont {S.~D.~H.}\ \bibnamefont {Hsu}},
  \ and\ \bibinfo {author} {\bibfnamefont {B.}~\bibnamefont {Mirza}},\ }\href
  {\doibase 10.1016/j.physletb.2010.10.060} {\bibfield  {journal} {\bibinfo
  {journal} {Phys. Lett.}\ }\textbf {\bibinfo {volume} {B695}},\ \bibinfo
  {pages} {317} (\bibinfo {year} {2011})},\ \Eprint
  {http://arxiv.org/abs/1008.2768} {arXiv:1008.2768 [gr-qc]} \BibitemShut
  {NoStop}%
%\%CITATION = 1008.2768;\%\%
\bibitem [{\citenamefont {Contillo}(2011)}]{Contillo:2010ju}%
  \BibitemOpen
  \bibfield  {author} {\bibinfo {author} {\bibfnamefont {A.}~\bibnamefont
  {Contillo}},\ }\href {\doibase 10.1103/PhysRevD.83.085016} {\bibfield
  {journal} {\bibinfo  {journal} {Phys. Rev.}\ }\textbf {\bibinfo {volume}
  {D83}},\ \bibinfo {pages} {085016} (\bibinfo {year} {2011})},\ \Eprint
  {http://arxiv.org/abs/1011.4618} {arXiv:1011.4618 [gr-qc]} \BibitemShut
  {NoStop}%
%\%CITATION = 1011.4618;\%\%
\bibitem [{\citenamefont {Bonanno}\ \emph {et~al.}(2011)\citenamefont
  {Bonanno}, \citenamefont {Contillo},\ and\ \citenamefont
  {Percacci}}]{Bonanno:2010bt}%
  \BibitemOpen
  \bibfield  {author} {\bibinfo {author} {\bibfnamefont {A.}~\bibnamefont
  {Bonanno}}, \bibinfo {author} {\bibfnamefont {A.}~\bibnamefont {Contillo}}, \
  and\ \bibinfo {author} {\bibfnamefont {R.}~\bibnamefont {Percacci}},\ }\href
  {\doibase 10.1088/0264-9381/28/14/145026} {\bibfield  {journal} {\bibinfo
  {journal} {Class. Quant. Grav.}\ }\textbf {\bibinfo {volume} {28}},\ \bibinfo
  {pages} {145026} (\bibinfo {year} {2011})},\ \Eprint
  {http://arxiv.org/abs/1006.0192} {arXiv:1006.0192 [gr-qc]} \BibitemShut
  {NoStop}%
%\%CITATION = 1006.0192;\%\%
\bibitem [{\citenamefont {Hindmarsh}\ \emph {et~al.}(2011)\citenamefont
  {Hindmarsh}, \citenamefont {Litim},\ and\ \citenamefont
  {Rahmede}}]{Hindmarsh:2011hx}%
  \BibitemOpen
  \bibfield  {author} {\bibinfo {author} {\bibfnamefont {M.}~\bibnamefont
  {Hindmarsh}}, \bibinfo {author} {\bibfnamefont {D.}~\bibnamefont {Litim}}, \
  and\ \bibinfo {author} {\bibfnamefont {C.}~\bibnamefont {Rahmede}},\ }\href
  {\doibase 10.1088/1475-7516/2011/07/019} {\bibfield  {journal} {\bibinfo
  {journal} {JCAP}\ }\textbf {\bibinfo {volume} {1107}},\ \bibinfo {pages}
  {019} (\bibinfo {year} {2011})},\ \Eprint {http://arxiv.org/abs/1101.5401}
  {arXiv:1101.5401 [gr-qc]} \BibitemShut {NoStop}%
%\%CITATION = ARXIV:1101.5401;\%\%
\bibitem [{\citenamefont {Cai}\ and\ \citenamefont
  {Easson}(2011)}]{Cai:2011kd}%
  \BibitemOpen
  \bibfield  {author} {\bibinfo {author} {\bibfnamefont {Y.-F.}\ \bibnamefont
  {Cai}}\ and\ \bibinfo {author} {\bibfnamefont {D.~A.}\ \bibnamefont
  {Easson}},\ }\href {\doibase 10.1103/PhysRevD.84.103502} {\bibfield
  {journal} {\bibinfo  {journal} {Phys. Rev.}\ }\textbf {\bibinfo {volume}
  {D84}},\ \bibinfo {pages} {103502} (\bibinfo {year} {2011})},\ \Eprint
  {http://arxiv.org/abs/1107.5815} {arXiv:1107.5815 [hep-th]} \BibitemShut
  {NoStop}%
%\%CITATION = ARXIV:1107.5815;\%\%
\bibitem [{\citenamefont {Bonanno}(2012)}]{Bonanno:2012jy}%
  \BibitemOpen
  \bibfield  {author} {\bibinfo {author} {\bibfnamefont {A.}~\bibnamefont
  {Bonanno}},\ }\href {\doibase 10.1103/PhysRevD.85.081503} {\bibfield
  {journal} {\bibinfo  {journal} {Phys. Rev.}\ }\textbf {\bibinfo {volume}
  {D85}},\ \bibinfo {pages} {081503} (\bibinfo {year} {2012})},\ \Eprint
  {http://arxiv.org/abs/1203.1962} {arXiv:1203.1962 [hep-th]} \BibitemShut
  {NoStop}%
%\%CITATION = ARXIV:1203.1962;\%\%
\bibitem [{\citenamefont {Hindmarsh}\ and\ \citenamefont
  {Saltas}(2012)}]{Hindmarsh:2012rc}%
  \BibitemOpen
  \bibfield  {author} {\bibinfo {author} {\bibfnamefont {M.}~\bibnamefont
  {Hindmarsh}}\ and\ \bibinfo {author} {\bibfnamefont {I.~D.}\ \bibnamefont
  {Saltas}},\ }\href {\doibase 10.1103/PhysRevD.86.064029} {\bibfield
  {journal} {\bibinfo  {journal} {Phys. Rev.}\ }\textbf {\bibinfo {volume}
  {D86}},\ \bibinfo {pages} {064029} (\bibinfo {year} {2012})},\ \Eprint
  {http://arxiv.org/abs/1203.3957} {arXiv:1203.3957 [gr-qc]} \BibitemShut
  {NoStop}%
%\%CITATION = ARXIV:1203.3957;\%\%
\bibitem [{\citenamefont {Bonanno}\ and\ \citenamefont
  {Reuter}(2013)}]{Bonanno:2013dja}%
  \BibitemOpen
  \bibfield  {author} {\bibinfo {author} {\bibfnamefont {A.}~\bibnamefont
  {Bonanno}}\ and\ \bibinfo {author} {\bibfnamefont {M.}~\bibnamefont
  {Reuter}},\ }\href {\doibase 10.1103/PhysRevD.87.084019} {\bibfield
  {journal} {\bibinfo  {journal} {Phys. Rev.}\ }\textbf {\bibinfo {volume}
  {D87}},\ \bibinfo {pages} {084019} (\bibinfo {year} {2013})},\ \Eprint
  {http://arxiv.org/abs/1302.2928} {arXiv:1302.2928 [hep-th]} \BibitemShut
  {NoStop}%
%\%CITATION = ARXIV:1302.2928;\%\%
\bibitem [{\citenamefont {Copeland}\ \emph {et~al.}(2015)\citenamefont
  {Copeland}, \citenamefont {Rahmede},\ and\ \citenamefont
  {Saltas}}]{Copeland:2013vva}%
  \BibitemOpen
  \bibfield  {author} {\bibinfo {author} {\bibfnamefont {E.~J.}\ \bibnamefont
  {Copeland}}, \bibinfo {author} {\bibfnamefont {C.}~\bibnamefont {Rahmede}}, \
  and\ \bibinfo {author} {\bibfnamefont {I.~D.}\ \bibnamefont {Saltas}},\
  }\href {\doibase 10.1103/PhysRevD.91.103530} {\bibfield  {journal} {\bibinfo
  {journal} {Phys. Rev.}\ }\textbf {\bibinfo {volume} {D91}},\ \bibinfo {pages}
  {103530} (\bibinfo {year} {2015})},\ \Eprint {http://arxiv.org/abs/1311.0881}
  {arXiv:1311.0881 [gr-qc]} \BibitemShut {NoStop}%
%\%CITATION = ARXIV:1311.0881;\%\%
\bibitem [{\citenamefont {Becker}\ and\ \citenamefont
  {Reuter}(2014{\natexlab{b}})}]{Becker:2014jua}%
  \BibitemOpen
  \bibfield  {author} {\bibinfo {author} {\bibfnamefont {D.}~\bibnamefont
  {Becker}}\ and\ \bibinfo {author} {\bibfnamefont {M.}~\bibnamefont
  {Reuter}},\ }\href {\doibase 10.1007/JHEP12(2014)025} {\bibfield  {journal}
  {\bibinfo  {journal} {JHEP}\ }\textbf {\bibinfo {volume} {12}},\ \bibinfo
  {pages} {025} (\bibinfo {year} {2014}{\natexlab{b}})},\ \Eprint
  {http://arxiv.org/abs/1407.5848} {arXiv:1407.5848 [hep-th]} \BibitemShut
  {NoStop}%
%\%CITATION = ARXIV:1407.5848;\%\%
\bibitem [{\citenamefont {Saltas}(2016)}]{Saltas:2015vsc}%
  \BibitemOpen
  \bibfield  {author} {\bibinfo {author} {\bibfnamefont {I.~D.}\ \bibnamefont
  {Saltas}},\ }\href {\doibase 10.1088/1475-7516/2016/02/048} {\bibfield
  {journal} {\bibinfo  {journal} {JCAP}\ }\textbf {\bibinfo {volume} {1602}},\
  \bibinfo {pages} {048} (\bibinfo {year} {2016})},\ \Eprint
  {http://arxiv.org/abs/1512.06134} {arXiv:1512.06134 [hep-th]} \BibitemShut
  {NoStop}%
%\%CITATION = ARXIV:1512.06134;\%\%
\bibitem [{\citenamefont {Nielsen}\ \emph {et~al.}(2015)\citenamefont
  {Nielsen}, \citenamefont {Sannino},\ and\ \citenamefont
  {Svendsen}}]{Nielsen:2015una}%
  \BibitemOpen
  \bibfield  {author} {\bibinfo {author} {\bibfnamefont {N.~G.}\ \bibnamefont
  {Nielsen}}, \bibinfo {author} {\bibfnamefont {F.}~\bibnamefont {Sannino}}, \
  and\ \bibinfo {author} {\bibfnamefont {O.}~\bibnamefont {Svendsen}},\ }\href
  {\doibase 10.1103/PhysRevD.91.103521} {\bibfield  {journal} {\bibinfo
  {journal} {Phys. Rev.}\ }\textbf {\bibinfo {volume} {D91}},\ \bibinfo {pages}
  {103521} (\bibinfo {year} {2015})},\ \Eprint
  {http://arxiv.org/abs/1503.00702} {arXiv:1503.00702 [hep-ph]} \BibitemShut
  {NoStop}%
%\%CITATION = ARXIV:1503.00702;\%\%
\bibitem [{\citenamefont {Bonanno}\ and\ \citenamefont
  {Platania}(2015)}]{Bonanno:2015fga}%
  \BibitemOpen
  \bibfield  {author} {\bibinfo {author} {\bibfnamefont {A.}~\bibnamefont
  {Bonanno}}\ and\ \bibinfo {author} {\bibfnamefont {A.}~\bibnamefont
  {Platania}},\ }\href {\doibase 10.1016/j.physletb.2015.10.005} {\bibfield
  {journal} {\bibinfo  {journal} {Phys. Lett.}\ }\textbf {\bibinfo {volume}
  {B750}},\ \bibinfo {pages} {638} (\bibinfo {year} {2015})},\ \Eprint
  {http://arxiv.org/abs/1507.03375} {arXiv:1507.03375 [gr-qc]} \BibitemShut
  {NoStop}%
%\%CITATION = ARXIV:1507.03375;\%\%
\bibitem [{\citenamefont {Falls}\ \emph {et~al.}(2018)\citenamefont {Falls},
  \citenamefont {Litim}, \citenamefont {Nikolakopoulos},\ and\ \citenamefont
  {Rahmede}}]{Falls:2016wsa}%
  \BibitemOpen
  \bibfield  {author} {\bibinfo {author} {\bibfnamefont {K.}~\bibnamefont
  {Falls}}, \bibinfo {author} {\bibfnamefont {D.~F.}\ \bibnamefont {Litim}},
  \bibinfo {author} {\bibfnamefont {K.}~\bibnamefont {Nikolakopoulos}}, \ and\
  \bibinfo {author} {\bibfnamefont {C.}~\bibnamefont {Rahmede}},\ }\href
  {\doibase 10.1088/1361-6382/aac440} {\bibfield  {journal} {\bibinfo
  {journal} {Class. Quant. Grav.}\ }\textbf {\bibinfo {volume} {35}},\ \bibinfo
  {pages} {135006} (\bibinfo {year} {2018})},\ \Eprint
  {http://arxiv.org/abs/1607.04962} {arXiv:1607.04962 [gr-qc]} \BibitemShut
  {NoStop}%
%%CITATION = ARXIV:1607.04962;%%
\bibitem [{\citenamefont {Eichhorn}\ \emph
  {et~al.}(2018{\natexlab{b}})\citenamefont {Eichhorn}, \citenamefont {Labus},
  \citenamefont {Pawlowski},\ and\ \citenamefont
  {Reichert}}]{Eichhorn:2018akn}%
  \BibitemOpen
  \bibfield  {author} {\bibinfo {author} {\bibfnamefont {A.}~\bibnamefont
  {Eichhorn}}, \bibinfo {author} {\bibfnamefont {P.}~\bibnamefont {Labus}},
  \bibinfo {author} {\bibfnamefont {J.~M.}\ \bibnamefont {Pawlowski}}, \ and\
  \bibinfo {author} {\bibfnamefont {M.}~\bibnamefont {Reichert}},\ }\href
  {\doibase 10.21468/SciPostPhys.5.4.031} {\bibfield  {journal} {\bibinfo
  {journal} {SciPost Phys.}\ }\textbf {\bibinfo {volume} {5}},\ \bibinfo
  {pages} {031} (\bibinfo {year} {2018}{\natexlab{b}})},\ \Eprint
  {http://arxiv.org/abs/1804.00012} {arXiv:1804.00012 [hep-th]} \BibitemShut
  {NoStop}%
%%CITATION = ARXIV:1804.00012;%%
\bibitem [{\citenamefont {Ohta}\ \emph {et~al.}(2016)\citenamefont {Ohta},
  \citenamefont {Percacci},\ and\ \citenamefont {Vacca}}]{Ohta:2015fcu}%
  \BibitemOpen
  \bibfield  {author} {\bibinfo {author} {\bibfnamefont {N.}~\bibnamefont
  {Ohta}}, \bibinfo {author} {\bibfnamefont {R.}~\bibnamefont {Percacci}}, \
  and\ \bibinfo {author} {\bibfnamefont {G.~P.}\ \bibnamefont {Vacca}},\ }\href
  {\doibase 10.1140/epjc/s10052-016-3895-1} {\bibfield  {journal} {\bibinfo
  {journal} {Eur. Phys. J.}\ }\textbf {\bibinfo {volume} {C76}},\ \bibinfo
  {pages} {46} (\bibinfo {year} {2016})},\ \Eprint
  {http://arxiv.org/abs/1511.09393} {arXiv:1511.09393 [hep-th]} \BibitemShut
  {NoStop}%
%\%CITATION = ARXIV:1511.09393;\%\%
\bibitem [{\citenamefont {Ohta}\ \emph {et~al.}(2015)\citenamefont {Ohta},
  \citenamefont {Percacci},\ and\ \citenamefont {Vacca}}]{Ohta:2015efa}%
  \BibitemOpen
  \bibfield  {author} {\bibinfo {author} {\bibfnamefont {N.}~\bibnamefont
  {Ohta}}, \bibinfo {author} {\bibfnamefont {R.}~\bibnamefont {Percacci}}, \
  and\ \bibinfo {author} {\bibfnamefont {G.~P.}\ \bibnamefont {Vacca}},\ }\href
  {\doibase 10.1103/PhysRevD.92.061501} {\bibfield  {journal} {\bibinfo
  {journal} {Phys. Rev.}\ }\textbf {\bibinfo {volume} {D92}},\ \bibinfo {pages}
  {061501} (\bibinfo {year} {2015})},\ \Eprint
  {http://arxiv.org/abs/1507.00968} {arXiv:1507.00968 [hep-th]} \BibitemShut
  {NoStop}%
%\%CITATION = ARXIV:1507.00968;\%\%
\bibitem [{\citenamefont {Fischer}\ \emph {et~al.}(2009)\citenamefont
  {Fischer}, \citenamefont {Maas},\ and\ \citenamefont
  {Pawlowski}}]{Fischer:2008uz}%
  \BibitemOpen
  \bibfield  {author} {\bibinfo {author} {\bibfnamefont {C.~S.}\ \bibnamefont
  {Fischer}}, \bibinfo {author} {\bibfnamefont {A.}~\bibnamefont {Maas}}, \
  and\ \bibinfo {author} {\bibfnamefont {J.~M.}\ \bibnamefont {Pawlowski}},\
  }\href {\doibase 10.1016/j.aop.2009.07.009} {\bibfield  {journal} {\bibinfo
  {journal} {Annals Phys.}\ }\textbf {\bibinfo {volume} {324}},\ \bibinfo
  {pages} {2408} (\bibinfo {year} {2009})},\ \Eprint
  {http://arxiv.org/abs/0810.1987} {arXiv:0810.1987 [hep-ph]} \BibitemShut
  {NoStop}%
%\%CITATION = 0810.1987;\%\%
\bibitem [{\citenamefont {Fischer}\ and\ \citenamefont
  {Pawlowski}(2009)}]{Fischer:2009tn}%
  \BibitemOpen
  \bibfield  {author} {\bibinfo {author} {\bibfnamefont {C.~S.}\ \bibnamefont
  {Fischer}}\ and\ \bibinfo {author} {\bibfnamefont {J.~M.}\ \bibnamefont
  {Pawlowski}},\ }\href {\doibase 10.1103/PhysRevD.80.025023} {\bibfield
  {journal} {\bibinfo  {journal} {Phys. Rev.}\ }\textbf {\bibinfo {volume}
  {D80}},\ \bibinfo {pages} {025023} (\bibinfo {year} {2009})},\ \Eprint
  {http://arxiv.org/abs/0903.2193} {arXiv:0903.2193 [hep-th]} \BibitemShut
  {NoStop}%
%\%CITATION = 0903.2193;\%\%
\bibitem [{\citenamefont {Cuesta}\ \emph {et~al.}()\citenamefont {Cuesta},
  \citenamefont {Falls},\ and\ \citenamefont {Litim}}]{Kevin-in-prep}%
  \BibitemOpen
  \bibfield  {author} {\bibinfo {author} {\bibfnamefont {R.}~\bibnamefont
  {Cuesta}}, \bibinfo {author} {\bibfnamefont {K.}~\bibnamefont {Falls}}, \
  and\ \bibinfo {author} {\bibfnamefont {D.~F.}\ \bibnamefont {Litim}},\
  }\href@noop {} {\bibinfo  {journal} {in preparation}\ }\BibitemShut {NoStop}%
\bibitem [{\citenamefont {{Cuesta Ramos}}(2016)}]{Phd-thesis-Raul}%
  \BibitemOpen
\bibfield  {journal} {  }\bibfield  {author} {\bibinfo {author} {\bibfnamefont
  {R.~A.}\ \bibnamefont {{Cuesta Ramos}}},\ }\emph {\bibinfo {title} {{Quantum
  gravity and the renormalisation group: from the UV to the IR}}},\ \href
  {http://sro.sussex.ac.uk/65582} {Ph.D. thesis},\ \bibinfo  {school}
  {University of Sussex} (\bibinfo {year} {2016})\BibitemShut {NoStop}%
\bibitem [{\citenamefont {Benedetti}(2012)}]{Benedetti:2011ct}%
  \BibitemOpen
  \bibfield  {author} {\bibinfo {author} {\bibfnamefont {D.}~\bibnamefont
  {Benedetti}},\ }\href {\doibase 10.1088/1367-2630/14/1/015005} {\bibfield
  {journal} {\bibinfo  {journal} {New J. Phys.}\ }\textbf {\bibinfo {volume}
  {14}},\ \bibinfo {pages} {015005} (\bibinfo {year} {2012})},\ \Eprint
  {http://arxiv.org/abs/1107.3110} {arXiv:1107.3110 [hep-th]} \BibitemShut
  {NoStop}%
%\%CITATION = ARXIV:1107.3110;\%\%
\bibitem [{\citenamefont {Falls}(2016)}]{Falls:2014zba}%
  \BibitemOpen
  \bibfield  {author} {\bibinfo {author} {\bibfnamefont {K.}~\bibnamefont
  {Falls}},\ }\href {\doibase 10.1007/JHEP01(2016)069} {\bibfield  {journal}
  {\bibinfo  {journal} {JHEP}\ }\textbf {\bibinfo {volume} {01}},\ \bibinfo
  {pages} {069} (\bibinfo {year} {2016})},\ \Eprint
  {http://arxiv.org/abs/1408.0276} {arXiv:1408.0276 [hep-th]} \BibitemShut
  {NoStop}%
%\%CITATION = ARXIV:1408.0276;\%\%
\bibitem [{\citenamefont {Vermaseren}(2000)}]{Vermaseren:2000nd}%
  \BibitemOpen
  \bibfield  {author} {\bibinfo {author} {\bibfnamefont {J.~A.~M.}\
  \bibnamefont {Vermaseren}},\ }\href@noop {} {\  (\bibinfo {year} {2000})},\
  \Eprint {http://arxiv.org/abs/math-ph/0010025} {arXiv:math-ph/0010025
  [math-ph]} \BibitemShut {NoStop}%
%\%CITATION = MATH-PH/0010025;\%\%
\bibitem [{\citenamefont {Kuipers}\ \emph {et~al.}(2013)\citenamefont
  {Kuipers}, \citenamefont {Ueda}, \citenamefont {Vermaseren},\ and\
  \citenamefont {Vollinga}}]{Kuipers:2012rf}%
  \BibitemOpen
  \bibfield  {author} {\bibinfo {author} {\bibfnamefont {J.}~\bibnamefont
  {Kuipers}}, \bibinfo {author} {\bibfnamefont {T.}~\bibnamefont {Ueda}},
  \bibinfo {author} {\bibfnamefont {J.~A.~M.}\ \bibnamefont {Vermaseren}}, \
  and\ \bibinfo {author} {\bibfnamefont {J.}~\bibnamefont {Vollinga}},\ }\href
  {\doibase 10.1016/j.cpc.2012.12.028} {\bibfield  {journal} {\bibinfo
  {journal} {Comput. Phys. Commun.}\ }\textbf {\bibinfo {volume} {184}},\
  \bibinfo {pages} {1453} (\bibinfo {year} {2013})},\ \Eprint
  {http://arxiv.org/abs/1203.6543} {arXiv:1203.6543 [cs.SC]} \BibitemShut
  {NoStop}%
%\%CITATION = ARXIV:1203.6543;\%\%
\bibitem [{\citenamefont {Brizuela}\ \emph {et~al.}(2008)\citenamefont
  {Brizuela}, \citenamefont {Martin-Garcia},\ and\ \citenamefont
  {Marugan}}]{xPert}%
  \BibitemOpen
  \bibfield  {author} {\bibinfo {author} {\bibfnamefont {D.}~\bibnamefont
  {Brizuela}}, \bibinfo {author} {\bibfnamefont {J.~M.}\ \bibnamefont
  {Martin-Garcia}}, \ and\ \bibinfo {author} {\bibfnamefont {G.~A.~M.}\
  \bibnamefont {Marugan}},\ }\href {\doibase 10.1007/s10714-009-0773-2} {\
  (\bibinfo {year} {2008}),\ 10.1007/s10714-009-0773-2},\ \Eprint
  {http://arxiv.org/abs/arXiv:0807.0824} {arXiv:0807.0824} \BibitemShut
  {NoStop}%
\bibitem [{\citenamefont {Cyrol}\ \emph {et~al.}(2017)\citenamefont {Cyrol},
  \citenamefont {Mitter},\ and\ \citenamefont {Strodthoff}}]{Cyrol:2016zqb}%
  \BibitemOpen
  \bibfield  {author} {\bibinfo {author} {\bibfnamefont {A.~K.}\ \bibnamefont
  {Cyrol}}, \bibinfo {author} {\bibfnamefont {M.}~\bibnamefont {Mitter}}, \
  and\ \bibinfo {author} {\bibfnamefont {N.}~\bibnamefont {Strodthoff}},\
  }\href {\doibase 10.1016/j.cpc.2017.05.024} {\bibfield  {journal} {\bibinfo
  {journal} {Comput. Phys. Commun.}\ }\textbf {\bibinfo {volume} {219}},\
  \bibinfo {pages} {346} (\bibinfo {year} {2017})},\ \Eprint
  {http://arxiv.org/abs/1610.09331} {arXiv:1610.09331 [hep-ph]} \BibitemShut
  {NoStop}%
%\%CITATION = ARXIV:1610.09331;\%\%
\end{thebibliography}%
\end{document}